\documentclass[onecolumn,aps,prd,preprintnumbers,showpacs,superscriptaddress,nofootinbib,amsmath,amssymb,floats,floatfix,showkeys,notitlepage,longbibliography]{revtex4-1}

\usepackage{orcidlink}
\usepackage{lipsum}
\usepackage{graphicx}
\usepackage{subfigure}
\usepackage[utf8]{inputenc}
\usepackage{sans}
\usepackage{changes}
\usepackage{hyperref}
\hypersetup{colorlinks=true,linkcolor=blue,urlcolor=blue,citecolor=blue}
\usepackage[toc,page]{appendix}
\usepackage[normalem]{ulem}
\usepackage{adjustbox}
\usepackage{latexsym}
\usepackage{amsmath}
\usepackage{amssymb}
\usepackage{amsfonts}
\usepackage{dcolumn}
\usepackage{bm}
\usepackage{tikz}
\usepackage{bigints}
\usepackage{array,tabularx,multirow,booktabs}
\usepackage[tracking=true]{microtype}
\SetTracking{}{500}
\SetTracking{encoding={*}, shape=sc}{40}
\UseRawInputEncoding %for inputenc error%
\allowdisplaybreaks

\begin{document} \sloppy

\title{An Effective Model for the Quantum Schwarzschild Black Hole: 
\\
Weak Deflection Angle, Quasinormal Modes and Bounding of Greybody Factor}

\author{\'Angel Rinc\'on
\orcidlink{0000-0001-8069-9162}
}
\email{angel.rincon@ua.es}
\affiliation{Departamento de Física Aplicada, Universidad de Alicante, Campus de San Vicente del Raspeig, E-03690 Alicante, Spain}
\affiliation{Sede Esmeralda, Universidad de Tarapac\'a, Avda. Luis Emilio Recabarren 2477, Iquique, Chile}

\author{Ali \"Ovg\"un
\orcidlink{0000-0002-9889-342X}
}
\email{ali.ovgun@emu.edu.tr}

\affiliation{Physics Department, Eastern Mediterranean
University, Famagusta, 99628 North Cyprus, via Mersin 10, Turkey}

\author{Reggie C. Pantig
\orcidlink{0000-0002-3101-8591}
}
\email{rcpantig@mapua.edu.ph}
\affiliation{Physics Department, Map\'ua University, 658 Muralla St., Intramuros, Manila 1002, Philippines}

\date{\today}
\begin{abstract}
In this paper, we thoroughly explore two crucial aspects of a quantum Schwarzschild black solution within four-dimensional space-time: i) the weak deflection angle, ii) the rigorous greybody factor \textcolor{black}{and, iii) the Dirac quasinormal modes}. Our investigation involves employing the Gauss-Bonnet theorem to precisely compute the deflection angle and establishing its correlation with the Einstein ring. Additionally, we derive the rigorous bounds for greybody factors through the utilization of general bounds for reflection and transmission coefficients in the context of Schrodinger-like one-dimensional potential scattering. 
\textcolor{black}{We also compute the corresponding Dirac quasinormal modes using the WKB approximation. We reduce the Dirac equation to a Schrodinger-like differential equation and solve it with appropriate boundary conditions to obtain the quasinormal frequencies.}
To visually underscore the quantum effect, we present figures that illustrate the impact of varying the parameter $r_0$, or more specifically, in terms of the parameter $\alpha$. 
This comprehensive examination enhances our understanding of the quantum characteristics inherent in the Schwarzschild black solution, shedding light on both the deflection angle and greybody factors in a four-dimensional space-time framework.
\end{abstract}

\keywords{General relativity; Black holes; Weak deflection angle; Greybody factor; Quasinormal modes.}

\pacs{95.30.Sf, 04.70.-s, 97.60.Lf, 04.50.+h}

\maketitle
%\tableofcontents
\section{Introduction} \label{intro}
%-Brief Introduction

As predicted by Einstein's general theory of relativity and other theories of gravity, black holes are fundamental objects in the universe. They are of great importance in classical and quantum gravity. 
Black holes are the perfect arena to combine general relativity with quantum mechanics in light of several well-known concepts where classical and quantum features must coexist. In this sense, the concept of Hawking radiation \cite{hawking1,hawking2}, although not yet observed, has captivated the scientific community. 
However, regardless of their relevance, black holes have several disadvantageous properties, for example, they contain a curvature singularity, which means that the black hole spacetime is geodesically incomplete. Also, unless their spin is exactly zero, black holes contain a Cauchy horizon, which makes it impossible to predict the future evolution of physical quantities \cite{Eichhorn:2022bgu}.
To gain a deeper understanding of black hole physics, it is useful to study black holes in alternative theories of gravity, especially those in which quantum features can emerge naturally.
Black holes and their properties can offer insights into the foundations of general relativity and beyond. They assume essential roles in both classical and quantum regimes, which is why they are regarded as ideal candidates for investigating alternative theories of gravity.

The study of black holes is well motivated from at least two complementary perspectives. The first motivation is the observational one.
The direct detection of gravitational waves from black hole mergers  \cite{ligo1,ligo2,ligo3,ligo4,ligo5} and the groundbreaking image of a supermassive black hole \cite{L1,L2,L3,L4,L5,L6} at the center of galaxy Messier 87 by the Event Horizon Telescope project \cite{project} have intensified research into various aspects of black hole physics. These recent advancements have sparked heightened interest and exploration in understanding black holes.
Thus, the relevance of black holes is supported not only by observational evidence but also by theoretical arguments such as the "no-hair theorem" or the "no-hair conjecture" \cite{heusler_1996}. 
Such a conjecture states that, regardless of how the
how the BH is formed, it can be parameterized by only three
quantities and these are: 
i) the mass
ii) the electric charge, and finally 
iii) the angular momentum \cite{Hawking:2016msc}.
However, in some special cases, black holes can have more than three quantities that are useful to describe their properties (see \cite{Babichev:2013cya,Antoniou:2017acq,Sotiriou:2013qea} and references therein).
From the vast amount of black hole solutions in four-dimensional spacetime, we can highlight certain solutions in light of their importance. The first solutions in vacuum, with electric charge and rotation, are:
i) The Schwarzschild solution \cite{Schwarzschild:1916uq},
ii) the Reissner-Nordstr\"om solution \cite{1916AnP...355..106R,1918KNAB...20.1238N},
iii) the Kerr solution \cite{Kerr:1963ud} and, finally,
iv) the Kerr-Newman solution \cite{Newman:1965my}.
Such solutions are often denoted as the "black-hole" of general relativity and include the three basic parameters, i.e., $\{ M, Q, J \}$.
From the last four works, the Schwarzschild black hole (a non-rotating, spherically symmetric solution to general relativity with a singularity at its centre, surrounded by an event horizon), albeit simple, is the most conventional "toy model" or background, and it has been significantly used merge quantum features into a classical background.
Be aware and notice that a consistent quantum theory of gravity remains an open task in modern theoretical physics. Thus, there are a significant number of approaches that attempt to obtain a well-defined theory of quantum gravity, for instance, we can mention \cite{Jacobson:1995ab,Connes:1996gi,Reuter:1996cp,Rovelli:1997yv,Gambini:2004vz,Ashtekar:2004vs,Nicolini:2008aj,Horava:2009uw,Verlinde:2010hp} and references therein.
In particular, we can highlight several approaches, as are, for instance, the asymptotic safety scenario which is based on the renormalization group flow that controls coupling constants \cite{Reuter:1996cp}. The formalism has been implemented in several contexts (in particular, in black hole physics) producing good results consistent in some limits with GR (see \cite{Platania:2023srt,Saueressig:2015xua,Koch:2014cqa,Borissova:2022mgd} and references therein).
An interesting application to black hole physics is the work of Bonanno and Reuter \cite{Bonanno:2000ep} (usually referred to as the "quantum improved" black hole formalism), where they studied the implications of asymptotically safe gravity on the Schwarzschild metric, where the ordinary Newton constant in the classical metric component is replaced by the running gravitational constant obtained from the renormalization group equation. As in the case of black holes in asymptotically safe gravity, formalism has been used to analyze how quantum features perturb classical backgrounds (see \cite{Ishibashi:2021kmf,Chen:2023wdg,Konoplya:2023bpf,Ruiz:2021qfp,Chen:2023wdg,Rincon:2020iwy} and references therein).
In addition, a related approach is the scale-dependent scenario, where not only the Newton constant but also all other coupling constants appearing in the action are replaced by functions of a given energy scale. In such a formalism, the need to establish a direct link between the energy scale and the physical coordinate is avoided (in problems with a very high degree of symmetry). For further details see \cite{Rincon:2018sgd,Rincon:2017goj,Koch:2016uso,Contreras:2019cmf,Rincon:2018dsq,Contreras:2017eza,Rincon:2018lyd,Panotopoulos:2020mii,Fathi:2019jid,Panotopoulos:2020zqa,Ovgun:2023ego,Rincon:2021hjj,Rincon:2020cpz,Alvarez:2022mlf,Rincon:2022hpy,Rincon:2023kun,Fathi:2023qyl} and references therein.
In both cases, the running Newton coupling acquires a certain scale dependence and evolves. In particular, for black holes, Newton's coupling evolves with the radial coordinate. As a consequence, the global structure of the quantum black hole solution differs significantly from its classical solution. 
In the case of the improved Schwarzschild black hole, such a solution admits an inner (or Cauchy) horizon in addition to the black hole event horizon. Alternatively, in the case of the scale-dependent Schwarzschild black hole, the event horizon was found to be slightly modified. These two examples suggest that the horizon, and hence its thermodynamic properties, may be perturbed in quantum black holes. 

Black holes can be investigated in light of Loop quantum gravity \cite{Perez:2017cmj}, one of the best-known (and also accepted) theories of quantum gravity \cite{Ashtekar:2021kfp}.
One of the main motivations for revisiting black holes within the framework of loop quantum gravity is the discretization of geometric properties at the Planck scale. This quantum discreteness arises from the canonical quantization procedure when applied to the action of general relativity, which is suitable for describing the coupling of gravity with gauge fields and fermions.
More precisely, loop quantum gravity is a mathematically well-defined, background-independent, and non-perturbative quantization of general relativity. Some of the most important outcomes are:
i) the calculation of the physical spectra of the geometric quantities such as area and volume, leading to quantitative predictions of Planck-scale physics,  
ii) a derivation of the black hole entropy method proposed by Bekenstein and Hawking,
iii) a non-trivial scientific representation of quantum physical space’s micro-structure that is distinguished by Planck scale discreteness, among others. 
In light of previous remarks, the loop quantum gravity theory has provided a path to parameterize quantum properties of spacetime revealed by a black hole.

\textcolor{black}{
Last but not least, the so-called quasinormal modes (QNMs) are important features of the late-time response of a black hole to external perturbations. The study of black hole perturbations has gained significant importance in the last decade, especially after the first detection of gravitational waves. Dominant QNMs are observed in the late-time gravitational wave signals of black holes (and other compact objects). In particular, collaborations such as LIGO and VIRGO have confirmed the detection of QNMs \cite{LIGOScientific:2016aoc}.
QNMs are also an essential tool for testing the stability of black holes. If the imaginary part of the quasinormal frequencies is positive, then the black hole is unstable to perturbations; otherwise, it remains stable. For a detailed explanation, see \cite{Kokkotas:1999bd}.
Let us focus on the Dirac quasinormal modes: they are particularly relevant because they provide crucial insights into the behavior of fermionic fields, such as spin-1/2 particles, near black holes. The study of these modes completes the topic of quasinormal modes, which usually includes only i)  scalar, ii) electromagnetic, and iii) gravitational perturbations. 
}

The structure of our work is as follows: In the next section, we summarize the fundamental aspects of a novel model for a quantum Schwarzschild black hole. Moving on to the third section, we calculate the deflection angle by comparing the solution with the Schwarzschild case. Subsequently, in the fourth section, we compute the Rigorous Bounds of Greybody factors for a massless scalar field.
\textcolor{black}{After that, in section five, we compute the Dirac QNMs and compare our results with the classical Schwarzschild case $\alpha =0$.}
Finally, we conclude our work with a summary and concluding remarks in section six. Throughout the study, we adopt the mostly positive metric signature $(-,+,+,+)$ and employ geometrical units where the universal constants are set to unity ($c=1=G_0$).

%\textcolor{purple}{Angel is here}

%

%

%%%%%%%%%%%%%%%%%%%%%%%%%%%%%%%%
\section{An effective model for the quantum Schwarzschild black hole}
%%%%%%%%%%%%%%%%%%%%%%%%%%%%%%%%

Let us start by considering a static region recently investigated in the context of LQG \cite{Alonso-Bardaji:2021yls,Alonso-Bardaji:2023niu}. 
To construct such a space, we first should take advantage of the Ashtekar-Barbero variables and then rewrite the corresponding Poisson algebra. Also, for an adequate effective descriptions a polymerization procedure is typically required. Doing so, an discrete parameter $\lambda$ accounts the quantum spacetime. Given that the concrete construction of this spacetime have been recently shown in \cite{Alonso-Bardaji:2021yls,Alonso-Bardaji:2023niu} we will focus on the study of some properties of the static region (exterior), in particular: 
 i) the deflection angle and 
ii) the rigorous bounds of Greybody Factors. Furthermore, we only considered specializing in static and spherically symmetric case for conservative reasons since (1) there was no consensus across
different estimates of Sgr A*'s rotation, and (2) the effect of the spin parameter $a$ on the shadow is small for the non-extremal case. Although there is a considerable effect relative to an observer's inclination position, it is still small (See the complete discussion in \cite{Vagnozzi:2022moj}).

\textcolor{black}{
In order to keep the paper self-contained, we will briefly summarize the more relevant equations needed to understand how this solution is found.
The first step is to write a spherically symmetric spacetime using the new variables of Ashtekar (see for instance \cite{Boehmer:2007ket} and references therein).   
The set of triads, labeled $E_i^a$, are the canonical variables in LQG, and $A^i_a$ are their corresponding SO(3) connections. 
In the spherically symmetric case, only three pairs of canonical variables remain, namely $\{ \eta, P^{\eta}, A_{\phi}, E^{\phi}, A_x, E^x \}$. 
As originally described in \cite{Gambini:2008dy}, a polar set of variables is chosen and that $x$ is used because it is not necessarily described by the standard radial coordinate.
Note that it is always possible to define the gauge invariant variables $K_i$ as a function of the connections $A_i$ and $\eta$, in other words, a more suitable variable can be defined. This means that $E^x, K_x$ and $E^\phi, K_\phi$ both represent the canonically conjugated pairs.
In the context of General Relativity, diffeomorphism invariance is based on four constraints. The first one, the Hamiltonian $\mathcal{H}$, and the remaining three come from the diffeomorphism constraint $\mathcal{D}$. 
For this concrete case, we can write down the relevant quantities as follows:
\begin{align}
\begin{split}
    \mathcal{H}  = & -  \frac{\tilde{E}^\phi}{2\sqrt{\tilde{E}^x}} \left( 1+\tilde{K}_{\phi}^2 \right)
    - 2\sqrt{\tilde{E}^x} \tilde{K}_{x}\tilde{K}_{\phi} \ +
     \frac{1}{8\sqrt{\tilde{E}^x} \tilde{E}^\phi} \left( \left( \tilde{E}^x \right)^\prime \right)^2 
    \\
    &     
    - 
    \frac{\sqrt{\tilde{E}^x}}{2\left( \tilde{E}^\phi \right)^2} \left( \tilde{E}^x \right)^\prime 
    %\\
    %& 
    \left( \tilde{E}^\phi \right)^\prime +
    \frac{\sqrt{\tilde{E}^x}}{2 \tilde{E}^\phi } \left( \tilde{E}^x \right)^{\prime \prime},
\end{split}
    \\
    \mathcal{D}   =  &\ \ \  \left( \tilde{E}^x \right)^\prime \tilde{K}_{x}+ \tilde{E}^\phi \left( \tilde{K}_{\phi}\right)^\prime.
\end{align}
Here the notation and symbols are clarified as follows:
i) the prime denotes the derivative with respect to $x$, 
ii) $\tilde{E}^i$ are the symmetry-reduced triad components, and finally, 
iii) $\tilde{K}_i$ means the conjugated momenta with $i=\left\lbrace x, \phi \right\rbrace$.
Given that the holonomies of the connection have well-defined operators in loop quantum gravity, a polymerization procedure is needed (see \cite{Gambini:2021uzf} for clarifications).
Thus, the basic idea is to replace the variables by an exponential form. 
In the case where real variables are used, a suitable replacement is of the form 
\begin{align}
\tilde{x} &\rightarrow \frac{\sin\left( \lambda x \right)}{\lambda},
\end{align}
where the last parameters are defined as follows: 
 i) $x$ is the variable and 
ii) $\lambda$ is the polymerization parameter. 
Note that the classical theory is valid in the limit
$\lambda \to 0$. Moreover, the polymerization parameter $\lambda$ is directly related to the length of the loop along which the holonomy is computed, since it is responsible for the space-time discretization.
It is important to note that there may be anomalies in the last replacement since the modified constraint algebra is typically not closed.
Alternatively, both $K_\phi$ and $E^\phi$ can be replaced by $\tilde{K}_\phi$ and $\tilde{E}^\phi$ using the expressions:
\begin{align}
    \tilde{K}_\phi & \rightarrow \frac{\sin\left( \lambda K_\phi \right)}{\lambda},
    \\
    \tilde{E}^\phi & \rightarrow \frac{E^\phi}{\cos \left( \lambda K_\phi \right)},
\end{align}
and thus the theory becomes free of anomalies \cite{Alonso-Bardaji:2021yls}.
The canonical transformation is bijective 
while $\cos \left( \lambda K_\phi \right) \neq 0$ and the dynamical content of the theory is equivalent to that obtained in GR.
The case $\cos \left( \lambda K_\phi \right) =0$ could be crucial because it could be a sign of new physics. Due to the fact that the Hamiltonian constraint diverges there, a regularization is required.
As can be viewed in \cite{Alonso-Bardaji:2021yls}, using a simplified map 
$\{t,x\} = \{\tilde{t}, \tilde{r}\}$ and choosing $E^x = \tilde{r}^2$ and $K_{\phi} = 0$ we get the corresponding spacetime.
Finally, taking into account the last ingredients, we can now introduce 
an effective quantum-corrected Schwarzschild space-time background ~\cite{Alonso-Bardaji:2021yls,Alonso-Bardaji:2022ear}. As the discussion is too involved from a mathematical point of view, we will not go into the technical details here and refer the interested reader to ~\cite{Alonso-Bardaji:2021yls,Alonso-Bardaji:2022ear}. 
In addition, this spacetime, along with similar ones, has been partially studied, and related calculations have been done independently during the submission of this paper. Please see \cite{Soares:2023uup,Junior:2023xgl,Jha:2023rem} for other studies that have used similar ideas.
}

%%%%%%%%%%%%%%%%%%%%%%%%%%%%%%%%%
The spacetime metric for the quantum Schwarzschild black hole is:
\begin{equation}\label{deformedschwarzsmetric}
  \!\!\!ds^2\!=\!-\!\bigg(\!1-\frac{{2M}}{r}\!\bigg){d t}^2 +\bigg(\!1-\frac{r_0}{r}\!\bigg)^{\!\!-1}\!\bigg(\!1-\frac{{2M}}{r}\!\bigg)^{\!\!-1}\!\!\!{dr}^2 \!+r^2d\Omega^2,
\end{equation}
with $r\in(2M,\infty)$. This region is asymptotically flat, and will describe one exterior domain.
Also, notice that the parameter $r_0$ contains the quantum effect via the following relation
\begin{equation}\label{r0}
r_0 = 2M 
\Bigg( 
\frac{\lambda^2}{1 + \lambda^2}
\Bigg)
\equiv \alpha r_H ,
\end{equation}
being $r_0$ smaller than the Schwarzschild horizon (which is $r_s = 2M$). The event horizon, $r_H$, obtained demanding $g^{rr}=0$, is not modified and then, the horizon is the same that its classical counterpart (i.e. $r_s = 2M = r_H$). 
The black hole mass, $M$, and the auxiliary parameter, $r_0$, can also be written with help of three well-known definitions of mass in spherical systems.
First, notice that the black hole mass, is not represented by $M$, statement originally explained in \cite{Alonso-Bardaji:2022ear} and reinforced in \cite{Moreira:2023cxy}. 
Commonly we have three well-known masses. They are
  i) the Komar mass $M_{\text{K}}$
 ii) the ADM mass $M_{\text{ADM}}$ and finally
iii) the Misner-Sharp mass $M_{\text{MS}}$.
These quantities are given by 
%
%\begin{subequations}
\begin{align}
\label{Komar}
    M_{\text{K}}=M\sqrt{1-\frac{r_0}{r}} 
    &= 
    \frac{1}{2}r_H \Bigg(1 - \alpha \frac{r_H}{r} \Bigg)^{1/2},\\
\label{ADM}
    M_{\text{ADM}}=M+\frac{r_0}{2} 
    &= 
    \frac{1}{2} r_H\Bigl( 1 + \alpha \Bigl) ,\\
\label{MS}
    M_{\text{MS}}=M+\frac{r_0}{2}-\frac{M r_0}{r} 
    &=
     \frac{1}{2}r_H \Bigg(1 + \alpha \Bigg(1 - \frac{r_H}{r} \Bigg) \Bigg)
    .
\end{align}
%\end{subequations}
%
Be aware and notice that two of them, i.e.,  
the Komar and Misner-Sharp masses,
can be different each other, because of the modified spacetime is not a solution of the Einstein's equations \cite{BEIG1978153}. 
In addition, for large values of $r$ and for spherically symmetric and asymptotically flat spacetimes, the ADM and Misner-Sharp masses must be equal \cite{Hayward:1994bu}, as can be checked by simple inspection. Thus, the black hole mass, $M$, and the "quantum" correction, $r_0$, can be interpreted in term of the different masses according to:
%\begin{subequations}
\begin{align}
&M=\lim\limits_{r\rightarrow \infty}M_{\text{K}},\\
&r_0=2\lim\limits_{r\rightarrow \infty}(M_{\text{MS}}-M_{\text{K}}).
\end{align}
%\end{subequations}

\textcolor{black}{In the next sections, we aim to study the deflection angle and the greybody bounds of the quantum Schwarzschild black hole described by Eq. \eqref{deformedschwarzsmetric}. For instance, light deflection near a black hole has a role in gravitational lensing, which is a powerful tool for probing both the strong-field and weak-field regime of gravity. By studying how the deflection angle is modified in the quantum Schwarzschild scenario, we can gain a deeper understanding of how quantum corrections affect photon trajectories. Greybody factors, on the other hand, influence the spectrum of Hawking radiation. Rigorous bounds on these factors in the quantum Schwarzschild context can reveal how quantum effects modify the black hole evaporation process. This has implications for understanding the end state of black hole evaporation and the information loss paradox. The polymerization parameter $\lambda$ arises from a specific quantization approach in Loop Quantum Gravity (LQG), and by studying the deflection angle and greybody factors, we can gain some insights into how quantum gravity might alter observable phenomena around black holes, as these effects could be observed with current or future astronomical instruments.}

%%%%%%%%%%%%%%%%%%%%%%%%%%%%%%%%%%%%%%%%%%%%%%%%%%%%%%%%%%%%%%%%%
%\section{Shadow of the EQSCH black hole}
%REGGIE

\section{Deflection Angle}
It was shown by \cite{Li:2020wvn} that in a static and spherically symmetric (SSS) spacetime with no asymptotic flatness, The Gauss-Bonnet theorem (GBT) can be written as
\begin{equation} \label{eIshi}
    \hat{\alpha} = \iint_{D}KdS + \phi_{\text{RS}},
\end{equation}
where $r_\text{ps}$ is the radius of the particle's circular orbit, and S and R are the radial positions of the source and receiver respectively. These are the integration domains, and we note that the infinitesimal curve surface $dS$ is given by
\begin{equation}
    dS = \sqrt{g}drd\phi.
\end{equation}
Furthermore, $\phi_\text{RS}$ is the coordinate position angle between the source and the receiver defined as $\phi_\text{RS} = \phi_\text{R}-\phi_\text{S}$, which can be found through the iterative solution of
\begin{align}
    F(u) = \left(\frac{du}{d\phi}\right)^2  %\nonumber\\
    = \frac{C(u)^2u^4}{A(u)B(u)}\Bigg[\left(\frac{E}{J}\right)^2-A(u)\left(\frac{1}{J^2}+\frac{1}{C(u)}\right)\Bigg],
\end{align}
where we have used the substitution $r = 1/u$ and the corresponding angular momentum and energy of the massive particle given the impact parameter $b$ as
\begin{equation}
    J = \frac{\mu v b}{\sqrt{1-v^2}}, \quad E = \frac{\mu}{\sqrt{1-v^2}}.
\end{equation}
we find
\begin{align}
    F(u) = \frac{1}{b^2}-u^2-\frac{\left[1+\left(b^{2} u^{2}-1\right) v^{2}\right] r_H u}{v^{2} b^{2}}+\frac{a u \left(b^{2} u^{2}-1\right) r_H}{b^{2}}.
\end{align}
The above enables one to solve for the azimuthal separation angle $\phi$ as
\begin{align} \label{ephi}
    \phi = \arcsin(bu)+\frac{r_H\left[v^{2}\left(b^{2}u^{2}-1\right)-1\right]}{2bv^{2}\sqrt{1-b^{2}u^{2}}}, %- \frac{X}{8 v^{2} \sqrt{-b^{2} u^{2}+1}} %\nonumber \\
    %+ \frac{\left[b^{3} u^{3} v^{2}+2 u^{2} v^{2} b^{2}+\left(-v^{2}+1\right) u b -2 v^{2}\right] X M}{8 \left(-b^{2} u^{2}+1\right)^{\frac{3}{2}} b \,v^{4}},
\end{align}
which is the also the direct expression for $\phi_S$ as $u$ is replaced by $u_S$. Meanwhile, the expression for the receiver is $\phi_R = \pi - \phi_S$ where $u_S$ should be replaced by $u_R$.

Leaving the angle $\phi$ for a while, the Gaussian curvature $K$ in terms of connection coefficients can be calculated as \cite{Gibbons:2008rj}
\begin{align}
    K=\frac{1}{\sqrt{g}}\left[\frac{\partial}{\partial\phi}\left(\frac{\sqrt{g}}{g_{rr}}\Gamma_{rr}^{\phi}\right)-\frac{\partial}{\partial r}\left(\frac{\sqrt{g}}{g_{rr}}\Gamma_{r\phi}^{\phi}\right)\right] %\nonumber \\
    =-\frac{1}{\sqrt{g}}\left[\frac{\partial}{\partial r}\left(\frac{\sqrt{g}}{g_{rr}}\Gamma_{r\phi}^{\phi}\right)\right],
\end{align}
since $\Gamma_{rr}^{\phi} = 0$. If in a certain spacetime there is an analytical solution for the $r_\text{ps}$, then we have the relation
\begin{equation} \label{gct}
    \int_{r_\text{ps}}^{r(\phi)} K\sqrt{g}dr = -\frac{A(r)\left(E^{2}-A(r)\right)C'-E^{2}C(r)A(r)'}{2A(r)\left(E^{2}-A(r)\right)\sqrt{B(r)C(r)}}\bigg|_{r = r(\phi)},
\end{equation}
since
\begin{equation}
    \left[\int K\sqrt{g}dr\right]\bigg|_{r=r_\text{ps}} = 0.
\end{equation}
The prime denotes differentiation with respect to $r$. The weak deflection angle for non-asymptotic spacetime is then \cite{Ishihara:2016vdc,Li:2020wvn},
\begin{align} \label{ewda}
    \hat{\alpha} = \int^{\phi_\text{R}}_{\phi_\text{S}} \left[-\frac{A(r)\left(E^{2}-A(r)\right)C'-E^{2}C(r)A(r)'}{2A(r)\left(E^{2}-A(r)\right)\sqrt{B(r)C(r)}}\bigg|_{r = r(\phi)}\right] %\times \nonumber\\
    d\phi + \phi_\text{RS}.
\end{align}
 we find
\begin{align} \label{gct2}
    &\left[\int K\sqrt{g}dr\right]\bigg|_{r=r_\phi} = -\phi_\text{RS} -\frac{(\cos \phi_R - \cos \phi_S) \left(v^{2}+1\right) r_H}{2v^{2} b} - \frac{(\cos \phi_R - \cos \phi_S) \alpha r_H}{2 b}. %\nonumber\\
    %&+\frac{X M}{8bv^4} \left[\phi_\text{RS}(1+v^2) + (\cos \phi_R - \cos \phi_S) (v^2 + 3v^4 + 2) \right].
\end{align}
With the above expression, Eq. \eqref{ephi} is needed. One should note that if $\phi_S$ is given, $\phi_\text{RS} = \pi - 2\phi_S$. The cosine of $\phi$ is then
\begin{align} \label{cs}
    \cos\phi = \sqrt{1-b^{2}u^{2}}-\frac{r_H u\left[v^{2}\left(b^{2}u^{2}-1\right)-1\right]}{2\sqrt{v^{2}\left(1-b^{2}u^{2}\right)}}, %\nonumber\\
    %+ \frac{b u X}{8 v^{2} \sqrt{-b^{2} u^{2}+1}} - \frac{\left[1+\left(2 b^{4} u^{4}+2 b^{3} u^{3}-3 b^{2} u^{2}-2 b u +1\right) v^{2}\right] X M}{8 \left(-b^{2} u^{2}+1\right)^{\frac{3}{2}} v^{4} b}
\end{align}
which should be applied to the source and the receiver. Using the above expression to Eq. \eqref{gct2}, we get the final analytic expression for the weak deflection angle that accommodates both time-like particles and finite distance as
\begin{align} \label{wda1}
    \Theta &= \left[\frac{r_H}{2b} \left( 1+ \frac{1}{v^2} + \alpha \right) \right] \left(\sqrt{1-b^{2}u_\text{S}^{2}}+\sqrt{1-b^{2}u_\text{S}^{2}}\right). %+ \frac{3}{8}X\left[\pi-2(\sin^{-1}(bu_\text{S})+\sin^{-1}(bu_\text{R}))\right] \nonumber \\
    %&+\frac{MX}{8bv^{4}}\left\{ \left(v^{2}+1\right)\left[\pi-2(\sin^{-1}(bu_\text{S})+\sin^{-1}(bu_\text{S}))\right]-2\left(3v^{4}+v^{2}+2\right)\left(\sqrt{1-b^{2}u_\text{S}^{2}}+\sqrt{1-b^{2}u_\text{R}^{2}}\right)\right\}.
\end{align}
Assuming that $u_R = u_S$, and these are distant from the black hole ($u \rightarrow 0$),
\begin{equation} \label{wda2}
    \Theta = \frac{r_H}{b} \left( 1+ \frac{1}{v^2} + \alpha \right).
\end{equation}
Finally, when $v = 1$,
\begin{equation} \label{wda3}
    \Theta_\text{null} = \frac{r_H}{b}(2 + \alpha).
\end{equation}
\begin{figure*}
    \centering
    \includegraphics[width=0.48\textwidth]{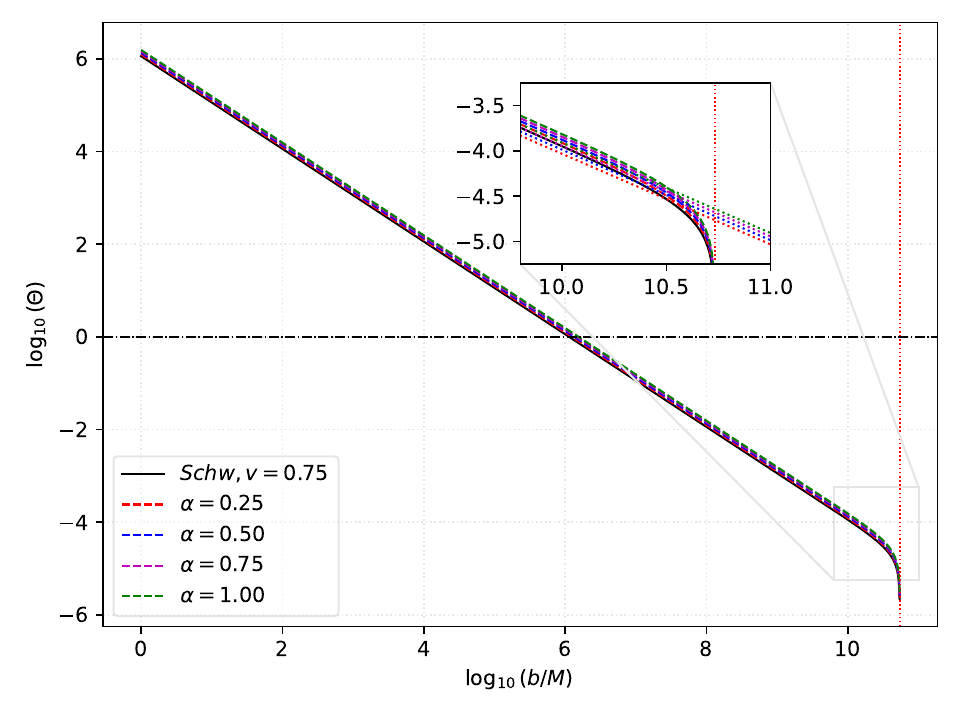}
    \includegraphics[width=0.48\textwidth]{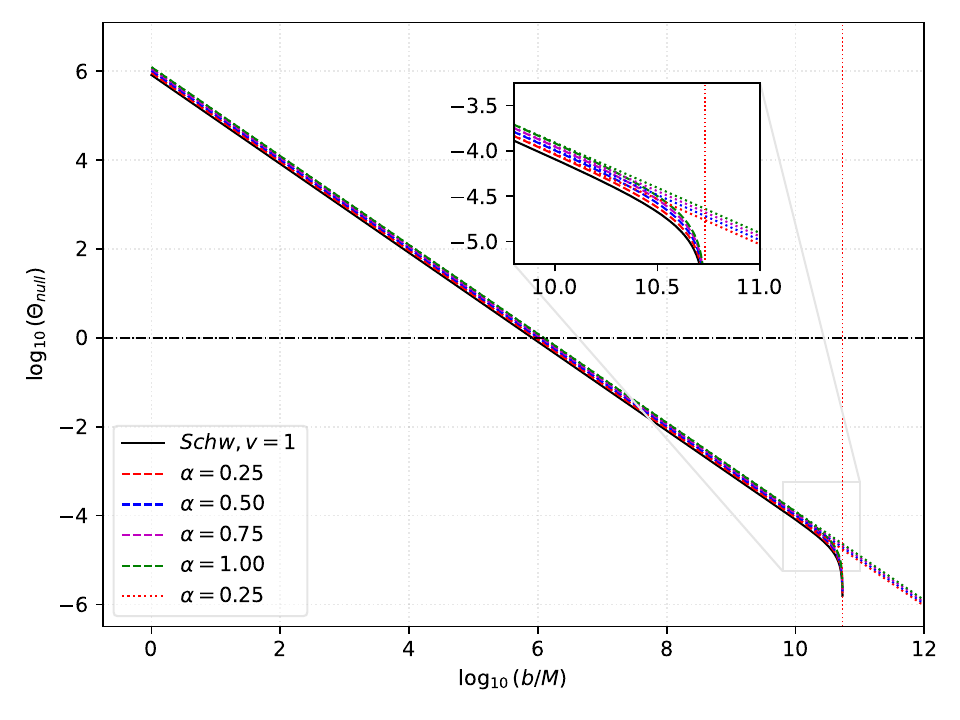}
    \caption{
    Weak deflection angle (in $\mu$as) using M87* parameters. 
    The left panel shows the behavior of the deflection angle of timelike particles according to Eq. \eqref{wda1} as finite distance is considered while $\alpha$ varies.  In contrast, the right panel shows how $\Theta$ behaves for photons. The case without finite distance is depicted by the dotted lines. We also added the Schwarzschild deflection angle for comparison as shown. The vertical dotted line represents our location from M87* SMBH, which is $16.8$ Mpc.}
    \label{fig_wda}
\end{figure*}

We plot the results in Fig. \ref{fig_wda}. Since the actual parameters are used for the finite distance $1/u = 16.8$ Mpc, we need to use a log-log plot to see the changes. The general observation is that the parameter $\alpha$ increases the deflection angle relative to the Schwarzschild case. Also, timelike particles provide a higher deflection angle than photons. The inset plot on the right panel shows the difference between the approximation $r\rightarrow \infty$ and not ignoring the finite distance. The difference only occurs when the impact parameter of photons is so large (comparable to $16.8$ Mpc).

The weak deflection angle has direct application in a phenomenon called the Einstein ring. Here, we will determine the effects of the $\alpha$ parameter on the Einstein ring. To begin with, let us define the distance of the source and the receiver with respect to the lensing object as $d_\text{S}$ and, $d_\text{R}$ respectively. Through the thin lens equation, we then have $d_\text{RS}=d_\text{S}+d_\text{R}$ so that the position of the weak field images is given by \cite{Bozza:2008zr}
\begin{equation}
    d_\text{RS}\tan\beta=\frac{d_\text{R}\sin\theta-d_\text{S}\sin(\hat{\alpha}-\theta)}{\cos(\hat{\alpha}-\theta)}.
\end{equation}
It is well known that an Einstein ring is formed when $\beta=0$, and the above equation leads then to the angular radius  \cite{Virbhadra:1999nm,Virbhadra:2002ju,Bozza:2001xd,Bozza:2002zj,Hasse:2001by,Perlick:2003vg,Atamurotov:2023rye,Abdujabbarov:2017pfw,Atamurotov:2022ahu,Atamurotov:2023tff}
\begin{equation}
\label{approx-wda}
    \theta_E \approx \frac{d_\text{S}}{d_\text{RS}}\hat{\alpha}.
\end{equation}
In addition, since the Einstein ring is assumed to be small it is safe to take relation $b=d_\text{R}\sin\theta \sim d_\text{R}\theta$. Then,  the weak deflection angle (\ref{approx-wda}) takes the explicit form
\begin{equation}
    \theta_E = \frac{1}{v}\sqrt{\frac{2r_H d_\text{S}}{d_\text{R}}} \sqrt{\frac{1 + v^2 (1+ \alpha)}{d_\text{S} +  d_\text{R}}}.
\end{equation}
\begin{figure}
    \centering
    \includegraphics[width=0.48\textwidth]{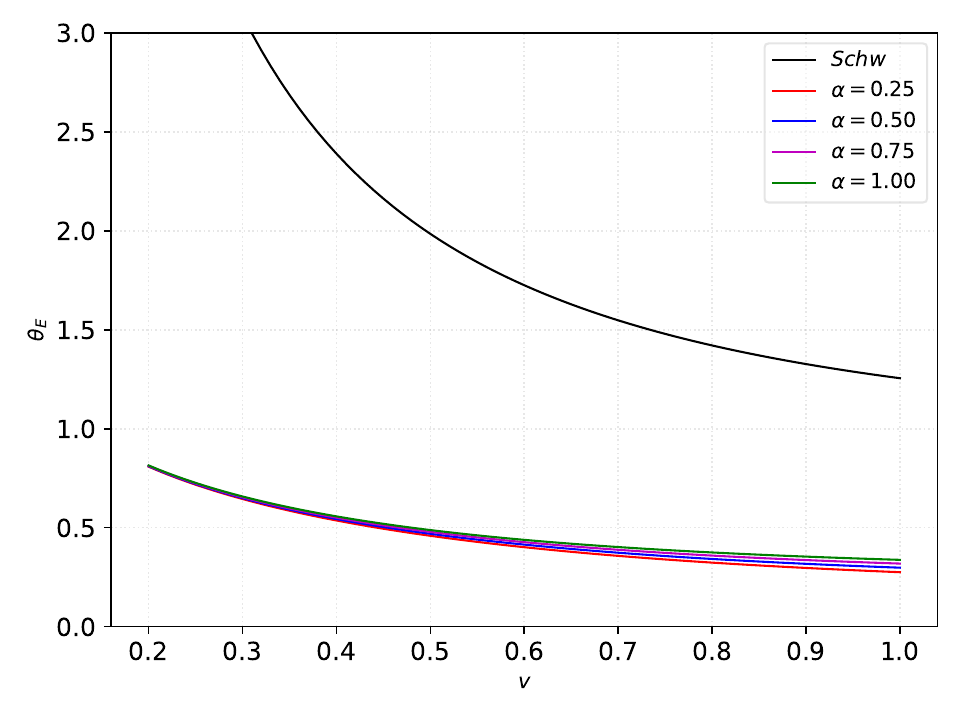}
    \caption{Einstein ring (in $\mu$as) using M87* parameters.}
    \label{fig_einsring}
\end{figure}
This angular radius is depicted in Figure \ref{fig_einsring} as a function of the impact parameter for different values of $\alpha$ This plot corresponds to the right panel in Figure \ref{wda1}, where we can see that it produces a greater value for the angular radius. Let us now consider M87*. Our location is $d_{R}\approx 16.8$ Mpc from the galactic center. 
\textcolor{black}{
Some additional works where the weak deflection angle is calculated are \cite{Hoshimov:2023tlz,Parbin:2023zik,Mushtaq:2024utq,Al-Badawi:2024dzc,QiQi:2023nex} and references therein.
}

%%%%%%%%%%%%%%%%%%%%%%%%%%%%%%%%%%%%%%%%%%%%%%%%%%%%%%%%%%%%%%%%%%
\section{Rigorous Bounds of Greybody factors} \label{sec5}
%%%%%%%%%%%%%%%%%%%%%%%%%%%%%%%%%%%%%%%%%%%%%%%%%%%%%%%%%%%%%%%%%%%
%ALI
In what follows we will focus on bounds for the greybody factors 
\cite{Boonserm:2017qcq,Boonserm:2019mon,Liu:2022ygf,Yang:2022ifo,Gray:2015xig,Boonserm:2014fja,Boonserm:2014rma,Boonserm:2013dua,Ngampitipan:2012dq,Boonserm:2009zba,Boonserm:2009mi,Kumaran:2023brp,Baruah:2023rhd,Javed:2022bdi,Al-Badawi:2023emj}
of a quantum Schwarzschild black hole inspired by loop quantum gravity as follows: 

\begin{equation}
    \label{eq: N and F}
d s^{2}=-\left|g_{t t}\right| d t^{2}+g_{r r} d r^{2}+r^{2}\left(d \theta^{2}+\sin ^{2} \theta d \phi^{2}\right)
\end{equation}
where lapse functions are 
\begin{align}
    g_{tt} &= \bigg(\!1-\frac{{2M}}{r}\!\bigg), 
    \\
g_{rr} &= \frac{1}{\bigg(\!1-\frac{r_0}{r}\!\bigg) \!\bigg(\!1-\frac{{2M}}{r}\!\bigg)}.
\end{align}

Here, we consider the Klein-Gordon equation for the massless scalar field and Maxwell equation for electromagnetic field as follows:

\begin{eqnarray} \frac{1}{\sqrt{-g}} \partial_{\mu}\left(\sqrt{-g} g^{\mu \nu} \partial_{\nu} \Phi\right)-m^{2} \Phi & =0 \\ \frac{1}{\sqrt{-g}} \partial_{\mu}\left(\sqrt{-g} g^{\sigma \mu} g^{\rho \nu} F_{\rho \sigma}\right) & =0.\end{eqnarray}

where \(F_{\rho \sigma}=\partial_{\rho} A_{\sigma}-\partial_{\sigma} A_{\rho}\) and \(A_{\mu}\) is the vector potential.

In the case of a spherically symmetric background,  with the metric functions given in \eqref{eq: N and F}.
As usual, we proceed by considering the obeying the separated variable form
\begin{equation}
 \Phi(t, r, \theta, \phi)=\sum_{l, m} \frac{\psi_{l}(r)}{r} Y_{l m}(\theta, \phi) e^{-i \omega t}
\end{equation}
where $Y_{l m}(\theta,\phi)$ are the spherical harmonic functions and   $l$ is the angular momentum quantum number \cite{Cardoso:2001bb, Konoplya:2006rv,Konoplya:2022tvv,Konoplya:2019hml}.

In order to simplify this equation, we can transform the radial coordinate $r$ to the ``tortoise coordinate'' $r^*$ through the variable transformation
\begin{equation}
   \frac{d r_{*}}{d r}=\sqrt{g_{r r}\left|g_{t t}^{-1}\right|},
\end{equation}
where $r_{*}$ is the tortoise coordinate defined in Schwarzschild metric \cite{Cai:2015fia}. We mention that this \textcolor{black}{tortoise coordinate is defined only outside the event horizon}. In this way, the radial part of the scalar-field equation of motion can be written as a Schr$\ddot{\text{o}}$dinger-like equation
\begin{equation}
    \label{eq: Schrodinger-like equation}
    \partial_{r_{*}}^{2} \psi_{l}\left(r_{*}\right)+\omega^{2} \psi_{l}\left(r_{*}\right)=V_{i}(r) \psi_{l}\left(r_{*}\right)
\end{equation}
with the effective potential $V_i$ in the form for scalar ($i = s$) and electromagnetic ($i = e$)

\begin{eqnarray}\label{eq: effective potential}  V_{s}  =\frac{g_{r r} g_{t t}^{\prime}-g_{t t} g_{r r}^{\prime}}{2 r g_{r r}^{2}}+g_{t t} \frac{l(l+1)}{r^{2}} \\ V_{e} =g_{t t} \frac{l(l+1)}{r^{2}}\end{eqnarray}

For the  quantum Schwarzschild black hole, the effective potential for scalar field ($i = s$)is calculated as follows:
\begin{equation} \label{V0}
    V_{s} =
\frac{(2 M-r) \left(-l (l+1) r^2+6 \alpha  M^2-(\alpha +2) M r\right)}{r^5}, \end{equation}
and for $i=e$ is calculated as follows:
\begin{equation} \label{V1}
    V_{e}=-\frac{l (l+1) (2 M-r)}{r^3},
\end{equation} which is same with Schwarzschild case so for there is not any quantum effect from spacetime.

\begin{figure}
    \centering
\includegraphics[width=0.45\linewidth]{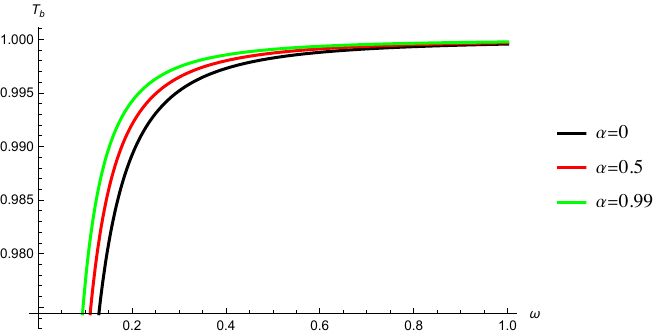}
\includegraphics[width=0.45\linewidth]{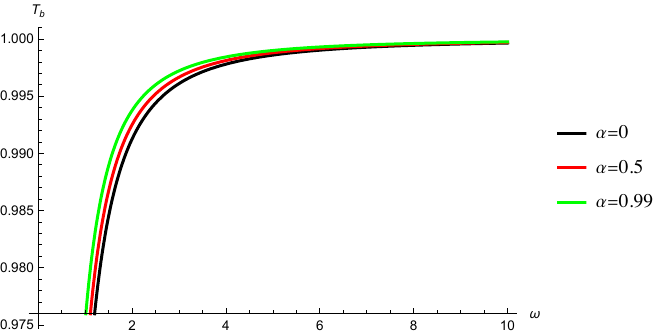}
   \caption{Greybody transmission probability for $M=1$, Left: $l=0$ and Right: $l=1$}
    \label{T01}
\end{figure}

Now, with help of the effective potential, $V(r)$, we study the lower rigorous bound for the greybody factor of the quantum black hole to probe the impact of $\alpha$ on the bound. 
The formula to derive the rigorous bound of greybody factor is given as follows \cite{Visser:1998ke,Boonserm:2008zg}:
\begin{equation} \label{bound}
T \geq \operatorname{sech}^{2}\left(\frac{1}{2 \omega} \int_{-\infty}^{\infty}\left|V_i\right| \frac{d r}{(\sqrt{g_{r r}\left|g_{t t}^{-1}\right|}} \right),
\end{equation}
and the boundary of the above formula is slightly modified when the cosmological constant is included \cite{Boonserm:2019mon} as follows:
\begin{equation}
T \geq T_{b}=\operatorname{sech}^{2}\left(\frac{1}{2 \omega} \int_{r_{H}}^{\infty} \frac{|V_i|}{(\sqrt{g_{r r}\left|g_{t t}^{-1}\right|}}  d r\right)=\operatorname{sech}^{2}\left(\frac{A_{\ell}}{2 \omega}\right),
\end{equation}

where the factor $A_{\ell}$ is defined according to the following expression:
\begin{equation}
A_{\ell}=\int_{r_{H}}^{\infty} \frac{|V_i|}{(\sqrt{g_{r r}\left|g_{t t}^{-1}\right|}}  d r.
\end{equation}
Hence, the bounds of the greybody factor of bosons are numerically plotted in Figs. 
Fig.s \ref{T01} for $\ell=0$ and  for $\ell=1$. The graph shows that when the parameter of $\alpha$ increases, the bound of the
greybody factor of bosons also increases. It is found that the quantum Schwarzschild black hole behave as good barriers.

\begin{figure}
\includegraphics[width=0.4\linewidth]{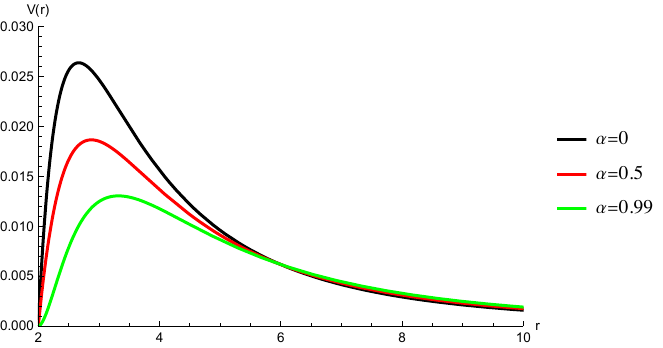}
\includegraphics[width=0.4\linewidth]{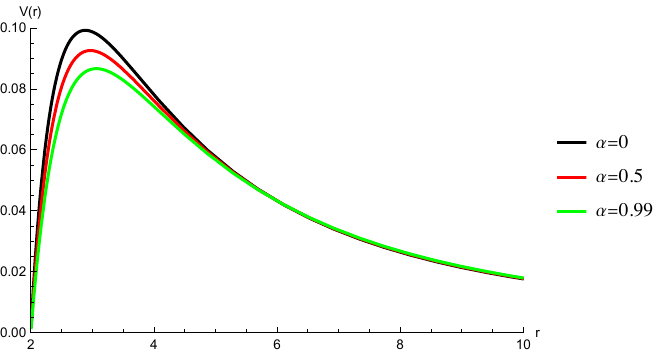}
   \caption{Effective potential for $M=1$, Left: $l=0$ and Right: $l=1$}
    \label{V01}
\end{figure}

% \section{QNMs: Dirac perturbations}

% To complete our analysis, we will investigate non-charged massless fermionic perturbations in the background of 4-dimensional quantum Schwarzschild black hole.
% Let us start by considering the general covariant Dirac equation (see for instance \cite{Brill:1957fx}):
% %
% \begin{equation}
% \gamma^\alpha \left(\frac{\partial}{\partial x^\alpha} - \Gamma^\alpha \right)\Psi = 0, 
% \end{equation}
% %
% where $\gamma^\alpha$ are gamma matrices, and $\Gamma^\alpha$ are spin connections in the tetrad formalism.
% %
% This type of equation is susceptible to be decoupled by using the method of separation of variables. 
% %
% We will discuss the radial part only, because that is the relevant part. The angular contribution appears via the angular numbers which are accounted in the radial differential equation.
% %
% Thus, the corresponding  wave equation can be represented in a Schrodinger-like differential equation (see, for instance, \cite{Konoplya:2006rv,Zinhailo:2018ska} and references therein):
% %
% \begin{equation}
% \frac{d^2 \Psi}{dr_{*}^2} + \left( \omega^2 - V_i(r) \right) \Psi = 0,
% \end{equation}
% %
% where as in the last section was mentioned, $r_{*}$ defines the “tortoise coordinate”.
% %
% The effective potentials of  Dirac ($i = d$) fields in the general background originally introduced can be written as follows:

%%%%%%%%%%%%%%%%%%%%%%%%%%%%%%%%%%%%
\section{QNMs: Dirac Perturbations}
%%%%%%%%%%%%%%%%%%%%%%%%%%%%%%%%%%%%

\textcolor{black}{
In the following section, we will outline the fundamental expression for calculating Quasinormal Modes (QNMs) for neutral Dirac particles. To maintain the discussion as general as we can, we will also consider a spherically symmetric background in a $d$-dimensional space-time and finally, we will consider the particular case $d=4$ at the end of equations (see \cite{Cho:2007zi} for details). The metric is then given by the following line element
\begin{align}
    \mathrm{d} s^{2}=-\left|g_{t t}\right| \mathrm{d} t^{2}+g_{r r} d r^{2} + r^{2}\mathrm{d}\Omega^{2}_{d-2}, 
\end{align}
where $d\Omega^{2}_{d-2}$ represents the metric for the $(d-2)$-dimensional sphere and the two metric potentials are then given as usual in this case, i.e., 
\begin{align}
g_{tt} &= \bigg(\!1-\frac{{2M}}{r}\!\bigg) 
\equiv 
f(r), 
\\
g_{rr} &= \frac{1}{\bigg(\!1-\frac{r_0}{r}\!\bigg) \!\bigg(\!1-\frac{{2M}}{r}\!\bigg)} 
\equiv 
\bigg( 1 - \frac{r_0}{r}\bigg)^{-1}f(r)^{-1}.
\end{align}
Let us perform a conformal transformation (see \cite{Das:1996we,Gibbons:1993hg} and references therein):
\begin{eqnarray}
g_{\mu\nu} & \rightarrow & \overline{g}_{\mu\nu}=\Omega^{2}g_{\mu\nu} , \\
\psi & \rightarrow & \overline{\psi}=\Omega^{-(d-1)/2}\psi , \\
\gamma^{\mu}\nabla_{\mu}\psi & \rightarrow & \Omega^{(d+1)/2} \overline{\gamma}^{\mu}\overline{\nabla}_{\mu}\overline{\psi} ,
\end{eqnarray}
Now, the metric becomes, after considering $\Omega=1/r$
\begin{align}
\mathrm{d}\overline{s}^{2} = -\frac{1}{r^{2}}f(r)\mathrm{d}t^{2} + \frac{1}{r^{2}}\bigg( 1 - \frac{r_0}{r}\bigg)^{-1}f(r)^{-1}\mathrm{d}r^{2} + \mathrm{d}\Omega^{2}_{d-2} , 
\end{align}
where we have considered $\overline{\psi}=r^{(d-1)/2}\psi$.
Now, let us take advantage of the fact that we can split the $t-r$ part and the $(d-2)$-sphere part which allow us to decouple the equation and solve the $t-r$. Thus, writing down the Dirac equation in the form (for massless particles):
\begin{align}
\overline{\gamma}^{\mu}\overline{\nabla}_{\mu}\overline{\psi}  =  
\bigg[ 
\left(
\overline{\gamma}^{t}\overline{\nabla}_{t} +
\overline{\gamma}^{r}\overline{\nabla}_{r} \right) \otimes 1
+ 
\overline{\gamma}^{5} \otimes
\left( \overline{\gamma}^{a}
\overline{\nabla}_{a}\right)_{S_{d-2}} 
\bigg] 
\overline{\psi} 
= 0 , 
\end{align}
where as usual $(\overline{\gamma}^{5})^{2}=1$. In what follows, let us change our notation by omitting the bars for simplicity.
}

\textcolor{black}{
We shall now let $\chi_{\ell}^{(\pm)}$ be the eigenspinors for the $(d-2)$-sphere (see \cite{Camporesi:1995fb} for further details):
\begin{equation}
\left( \gamma^{a}\nabla_{a} \right)_{S_{d-2}}\chi_{\ell}^{(\pm)} = \pm i \left( \ell + \frac{d-2}{2}\right) \chi_{\ell}^{(\pm)} ,
\end{equation}
Here $\ell = 0, 1, 2, \dots$. Since the eigenspinors are orthogonal, it is always possible to expand $\psi$ as:
\begin{equation}
\psi = \sum_{\ell} \left( \phi_{\ell}^{(+)} \chi_{\ell}^{(+)} + \phi_{\ell}^{(-)} \chi_{\ell}^{(-)} \right) .
\end{equation}
The Dirac equation acquires the simple form:
\begin{equation}\label{eqn:2Ddirac}
\left \{ 
\gamma^{t} \nabla_{t} + \gamma^{r} \nabla_{r} + \gamma^{5} \left[ \pm i \left( \ell + \frac{d-2}{2} \right) \right] \right \} 
\phi_{\ell}^{(\pm)} = 0 ,
\end{equation}
which is precisely the $2$-dimensional Dirac equation.
In order to solve the corresponding differential equations we make the explicit choice of the Dirac matrices, namely:
\begin{align}
\gamma^{t} &= \frac{r}{\sqrt{f(r)}}(-i\sigma^{3}), 
\\
\gamma^{r} &= \sqrt{1-\frac{r_0}{r}} \sqrt{f(r)} 
\ r\sigma^{2} ,
\end{align}
and $\sigma^{i}$ are the so-called Pauli matrices, defined as:
\begin{equation}
\sigma^{1}=\left(
\begin{array}{cc}
0 & 1 \\ 1 & 0
\end{array}
\right)\ \ \ ,\ \ \ \sigma^{2}=\left(
\begin{array}{cc}
0 & -i \\ i & 0
\end{array}
\right)\ \ \ ,\ \ \ \sigma^{3}=\left(
\begin{array}{cc}
1 & 0 \\ 0 & -1
\end{array}
\right) .
\end{equation}
Also, $\gamma^{5}$ is written in term of the Pauli matrices as follows
\begin{equation}
\gamma^{5} = (-i\sigma^{3})(\sigma^{2}) = - \sigma^{1} .
\end{equation}
The spin connections are then found to be:
\begin{align}
\Gamma_{t} &= \sigma^{1} 
\left( 
\frac{1}{4} r^{2} \sqrt{1-\frac{r_0}{r}}
\right) 
\frac{\mathrm{d}}{\mathrm{d}r} \left( \frac{f(r)}{r^{2}} \right) , \\
\Gamma_{r} &= 0 .
\end{align}
At this point should be mentioned that the treatment for $+$
sign solution is completely equivalent to the $-$ sign case, the reason why we will focus on of them, the positive one. Also let us define the parameter $\xi$ in term of the dimension $d$ and the angular number $\ell$:
\begin{align}
    \xi \equiv \ell + \frac{1}{2}(d-2), 
\end{align}
with $\xi$ being $ +1, +2,...$ and $d=1, ..., N.$ 
Taking the last facts into account, the Dirac equation can then be simplified to be:
\begin{align}
& \left\{ \frac{r}{\sqrt{f(r)}}(-i\sigma^{3}) \left[
\frac{\partial}{\partial t} + 
\sigma^{1} 
\left( 
\frac{1}{4} r^{2} \sqrt{1-\frac{r_0}{r}}
\right) 
\frac{\mathrm{d}}{\mathrm{d}r} \left( \frac{f(r)}{r^{2}} \right)
\right] + 
\sqrt{1-\frac{r_0}{r}} \sqrt{f(r)} 
r \sigma^{2}
\frac{\partial}{\partial r} + (-\sigma^{1})(i) \left( \xi \right) \right\} \phi_{\ell}^{(+)} = 0  
\end{align}
Putting spatial and temporal part separately, we finally have a first order partial differential equation for $\phi_{\ell}^{(+)}$, i.e., 
\begin{align}
\sigma^{2} 
\left( 
\sqrt{1-\frac{r_0}{r}} \sqrt{f(r)} \
r
\right) 
\left[ \frac{\partial}{\partial r} + \frac{r}{2\sqrt{f(r)}}
\frac{\mathrm{d}}{\mathrm{d}r} \left( \frac{\sqrt{f(r)}}{r} \right) \right]
\phi_{\ell}^{(+)} - i \sigma^{1} \xi 
\phi_{\ell}^{(+)} = i \sigma^{3} \left( \frac{r}{\sqrt{f(r)}} \right)
\frac{\partial \phi_{\ell}^{(+)}}{\partial t} . &
\end{align}
Now, let us to obtain solutions of the form:
\begin{equation}
\phi_{\ell}^{(+)} = \left( \frac{\sqrt{f(r)}}{r} \right)^{-1/2} e^{-i \omega t} \left(
\begin{array}{c}
iG(r) \\ F(r)
\end{array}
\right) ,
\end{equation}
%
%where $E$ is the energy. 
%
The Dirac equation can then be reduced to:
\begin{equation}
\sigma^{2} \left( \sqrt{1-\frac{r_0}{r}} \sqrt{f(r)} \ r \right) \left(
\begin{array}{c}
i\frac{\mathrm{d}G(r)}{\mathrm{d}r} \\ 
\ \frac{\mathrm{d}F(r)}{\mathrm{d}r}
\end{array}
\right) -i \sigma^{1} \xi \left(
\begin{array}{c}
i G(r) 
\\ 
F(r)
\end{array}
\right) = \sigma^{3} \omega \left( \frac{r}{\sqrt{f(r)}} \right) \left(
\begin{array}{c}
i G(r) 
\\ 
F(r)
\end{array}
\right) .
\end{equation}
By taking the components separately, we finally write a set of coupled first-order differential equations in terms of variables $G \equiv G(r)$ and $F \equiv F(r)$ as follows:
\begin{eqnarray}
\Bigg [ \sqrt{1-\frac{r_0}{r}} f(r) \Bigg ] \frac{\mathrm{d}G(r)}{\mathrm{d}r} 
- 
\Bigg [ \frac{\sqrt{f(r)}}{r} \xi \Bigg ] G(r) & = & + \omega F(r) ,  \label{Eq1}
\\ 
\Bigg [ \sqrt{1-\frac{r_0}{r}} f(r) \Bigg ] \frac{\mathrm{d}F(r)}{\mathrm{d}r} 
+ 
\Bigg [ \frac{\sqrt{f(r)}}{r} \xi \Bigg ] F(r) & = & - \omega G(r) . \label{Eq2}
\end{eqnarray}
To calculate the quasinormal modes for neutral Dirac particles of this background, becomes convenient to introduce the well-known tortoise coordinate
$r_*$, and the auxiliary function $W(r)$ as follows
\begin{align}
r_*(r) &\equiv \int ^{r}\mathrm{d}\bar{r} 
\Bigg[ 
f(\bar{r})\sqrt{1-\frac{r_0}{\bar{r}}}
\Bigg]^{-1} ,
\\
W(r) &= \frac{\xi \sqrt{f(r)}}{r}
\end{align}
Taking advantage of the last two equations, the first-order linear coupled equations \eqref{Eq1} and \eqref{Eq2} can be reduced to
\begin{align}
    \Bigg[
    \frac{\mathrm{d}}{\mathrm{d}r_{*}} - W(r)
    \Bigg] G(r) &= + \omega F(r)
    \\
    \Bigg[
    \frac{\mathrm{d}}{\mathrm{d}r_{*}} + W(r)
    \Bigg] F(r) &= - \omega G(r)
\end{align}
The set of first-order differential equations for $G(r)$ and $F(r)$ can be trivially decoupled to obtain two Schrodinger-like differential equations, for some concrete effective potential, i.e., 
\begin{align}
\frac{\mathrm{d}^2F}{\mathrm{d}{r_{*}}^2} + [\omega^2 - V_{-}] F & =  0 , \label{SL1}
\\
\frac{\mathrm{d}^2G}{\mathrm{d}{r_{*}}^2} + [\omega^2 - V_{+}] G & =  0 , \label{SL2}
\end{align}
where the potentials are given by 
\begin{equation}
V_{\pm} = W^2 \pm \frac{\mathrm{d}W}{\mathrm{d}r_{*}} .
\end{equation}
It should be mentioned that the effective potentials $V_{+}$ and $V_{-}$
are supersymmetric to each other, the reason why the two functions $F$ and $G$ share the same spectra (i.e., they are isospectral), both for scattering and quasi-normal. In addition, for $\phi_{\ell}^{(-)}$, we have these two potentials.
In this respect, in what follows we will consider using the positive sign, having clarified that the potential is the same in this case. Now, utilizing the standard radial coordinate we have the concrete form of the effective potential according to:
\begin{align}
V_{+} &= f(r) 
\Bigg[
\frac{\xi ^2}{r^2} + 
\sqrt{1-\frac{r_0}{r}}
\Bigg(
\frac{\xi  f'(r)}{2 r \sqrt{f(r)}}-\frac{\xi  \sqrt{f(r)}}{r^2}
\Bigg)
\Bigg]
\end{align}
or, replacing the metric potential explicitly, we have 
\begin{align}
    V_{+} &= 
    \left(1-\frac{2 M}{r}\right) 
    \Bigg[
    \frac{\xi ^2}{r^2} + 
    \sqrt{1-\frac{r_0}{r}} \left(\frac{M \xi }{r^3 \sqrt{1-\frac{2 M}{r}}}-\frac{\xi}{r^2}\sqrt{1-\frac{2 M}{r}} \right) 
    \Bigg].
\end{align}
At this point, we should choose appropriated outgoing boundary conditions at the horizon and spatial infinity i.e., nothing should come in from asymptotic infinity to disturb the system and nothing should come out of the horizon.
In other words, the boundary conditions for Schrodinger-like equations \eqref{SL1} and \eqref{SL2} which are
\begin{align}
   \{F, G\} \rightarrow \: &\exp(+i \omega r_*), \; \; \; \; \; \;  r_* \rightarrow - \infty ,
   \\
   \{F, G\} \rightarrow \: &\exp(-i \omega r_*), \; \; \; \; \; \; r_* \rightarrow + \infty .
\end{align}
Note that, as $\psi \sim \exp(-i \omega t)$, a frequency with a negative imaginary part implies a decaying (stable) mode. Conversely, a frequency with a positive imaginary part means an increasing (and therefore unstable) mode.
The effective potential $V(r)$ is shown for the Dirac case, assuming $d=4$ (for simplicity) against the radial coordinate for different values of the set of parameters $\{ \xi, M, \alpha \}$. We include the Schwarzschild case with $\alpha=0$ for comparison purposes.
To get some insights into how the effective potential $V(r)$ looks like in the Dirac case, we show, in Figs.~\eqref{fig:Veffec_1} and \eqref{fig:Veffec_2}, its behavior against the radial coordinate for different values of the set of parameters $\{ \xi, M, \alpha \}$. For comparison reasons, we include the classical Schwarzschild case.
We have considered in Figs.~\eqref{fig:Veffec_1} and \eqref{fig:Veffec_2} the following cases:
\begin{itemize}
    \item Fig.~\eqref{fig:Veffec_1} shows the evolution of the effective potential for the classical case $(\alpha = 0)$ varying the two free parameters $\xi$ and $M$. Such figures are included for comparison.
    From the left panel we can see that as $\xi$ increases, the maximum of the potential increases, at which point it shifts to the right. All the solutions converge at small radii because their associated horizons are equal in contract to the other cases shown.
    Similarly, the right panel shows that as $M$ increases (for a fixing $\xi$), the maximum of the potential decreases, now shifting to the right.
    \item Fig.~\eqref{fig:Veffec_2} shows the evolution of the effective potential for the quantum-inspired case $(\alpha \neq 0)$ varying the two free parameters $\xi$ and $M$. The case $\alpha=0$ is included for comparison.
    The left panel shows how the effective potential varies for fixed $\{ \xi, M\}$ and different values of the parameter $\alpha$. The figure confirms that as $\alpha$ increases, the maximum of the potential decreases, at the same time as it shifts to the right. In addition, the effective potential tends to overlap quickly between small and large radii.
    The middle panel shows the potential $V(r)$ for fixed $\{ \xi \}$ and different values of the BH mass $M$ and the parameter $\alpha$.  
    The right panel shows the potential $V(r)$ for fixed $\{ M \}$ and different values of the parameters $\xi$ and $\alpha$.
    Note that in the middle and right panels, the quantum corrections are compared with the quantum-inspired case and the modifications, although present, are small.
\end{itemize}
}

\begin{figure*}
    \centering
    \includegraphics[width=0.48\textwidth]{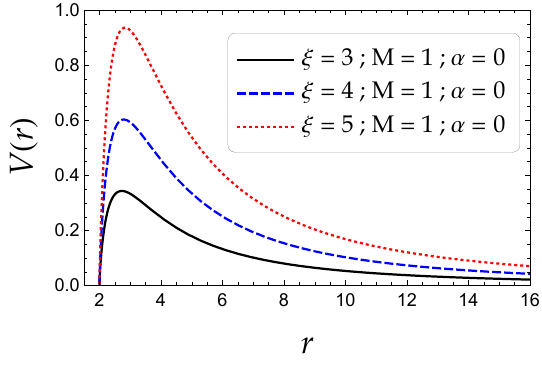}
    \includegraphics[width=0.48\textwidth]{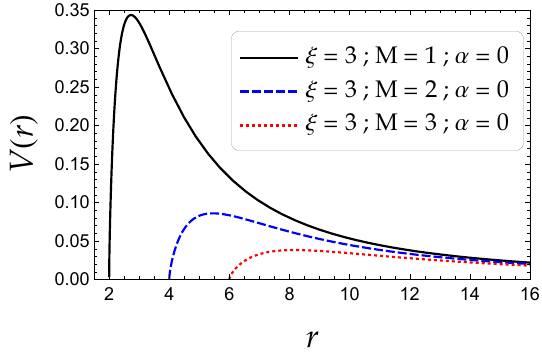}
    \caption{
    \textcolor{black}{
    Effective potential for this quantum-inspired Schwarzschild black hole. For simplicity, we show $V(r)$ varying $\xi$ (proportional to angular number $\ell$) and $M$ (the black hole mass), assuming the classical case, i.e., $\alpha = 0$.
    {\bf{Left panel:}} Effective potential with a fixed mass and varying $\xi$, for $\alpha = 0$.
    {\bf{Right panel:}} Effective potential with a fixed value $\xi$ and varying the mass $M$, for $\alpha = 0$.
    }
    }
    \label{fig:Veffec_1}
\end{figure*}

\begin{figure*}
    \centering
    \includegraphics[width=0.325\textwidth]{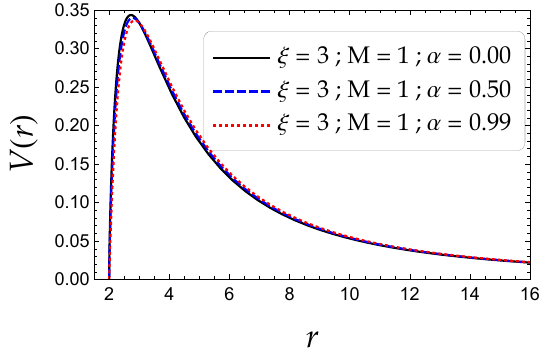}
    \includegraphics[width=0.325\textwidth]{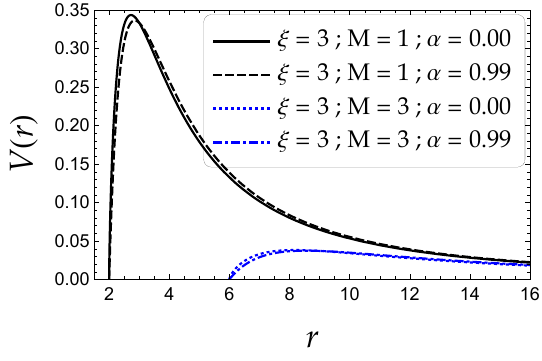}
    \includegraphics[width=0.325\textwidth]{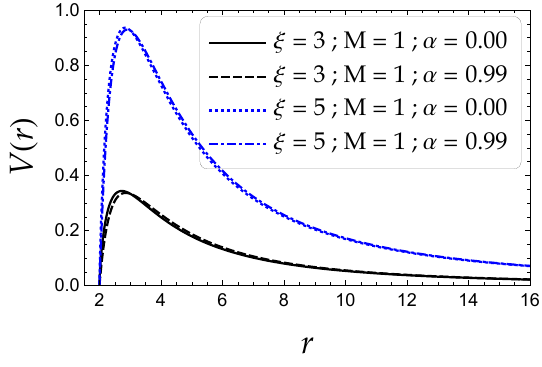}
    \caption{
    \textcolor{black}{
    Effective potential for this quantum-inspired Schwarzschild black hole. We show $V(r)$ varying $\xi$  and $M$, for several different parameter values $\alpha$.
    {\bf{Left panel:}} Effective potential fixing $\xi$ and $M$, varying $\alpha$.
    {\bf{Middle panel:}} Effective potential for a fixed parameter $\xi$, varying $M$ and $\alpha$.
    {\bf{Right panel:}} Effective potential for a fixed mass $M$, varying $\xi$ and $\alpha$.
    All figures show that the peak of the potential is slightly shifted with respect to its classical counterpart $(\alpha = 0)$.
    }
    }
    \label{fig:Veffec_2}
\end{figure*}

%%%%%%%%%%%%%%%%%%%%%%%%%%%%%%%%%%%%5
\section{Quasinormal spectrum: WKB approximation}
%%%%%%%%%%%%%%%%%%%%%%%%%%%%%%%%%%%%%

\textcolor{black}{The characteristic oscillations of black holes, known as quasinormal modes (QNMs), reveal how the black hole responds to external perturbations. Analogous to the oscillations of a struck bell, these modes represent the "ringing" of black holes. Quasinormal modes are defined by complex frequencies, where the real component is the oscillation frequency and the imaginary component is the damping rate induced by gravitational wave emission.
The quasinormal spectra of BHs can only be obtained accurately in a few cases, e.g:
i) when the induced differential equation (for the radial part of the wave function) can be transformed into the Gauss' hypergeometric function (see \cite{Birmingham:2001hc,Fernando:2003ai,Fernando:2008hb,Gonzalez:2010vv,Destounis:2018utr,Ovgun:2018gwt,Rincon:2018ktz} and references therein), or
ii) when the potential barrier takes the form of the P{\"o}schl-Teller potential (see \cite{Poschl:1933zz,Ferrari:1984zz,Cardoso:2001hn,Cardoso:2003sw,Molina:2003ff,Panotopoulos:2018hua} and references therein).
Given the complexity and non-trivial nature of the underlying differential equation, it is important to use numerical or semi-analytical methods to compute the relevant quasinormal frequencies. As a result, several strategies have been developed for this purpose, some of which are widely used. 
}

\textcolor{black}{
Specifically:
i) the method of continued fraction, together with its modifications, originally introduced in \cite{Leaver:1985ax,Nollert:1993zz, Daghigh:2022uws},
ii) the Frobenius method and its generalization (see the references in \cite{Destounis:2020pjk, Fontana:2022whx,Hatsuda:2021gtn}),
iii) the asymptotic iteration method (see \cite{Cho:2011sf,2003JPhA...3611807C,Ciftci:2005xn} and references therein).
among other possibilities.
Additional details and other methods commonly used to compute QNMs can be found in \cite{Konoplya:2011qq}.
Since the effective potential has a well-behaved form, in the present work we will use the WKB semiclassical method to obtain the corresponding quasinormal frequencies (see \cite{Schutz:1985km,Iyer:1986np,Iyer:1986nq,Kokkotas:1988fm,Seidel:1989bp} and references therein).
Schutz and Will obtained the first-order calculation using the WKB method \cite{Schutz:1985km}, followed by improvements made by Iyer and Will \cite{Iyer:1986np}, who developed a semi-analytic expression including second- and third-order corrections. 
The method and its lower-order corrections are quite efficient for determining the lowest overtones among the complex frequencies of, for example, an oscillating Schwarzschild BH.  The accuracy of the approximation improves as the angular harmonic index $\ell$ ($\ell \propto \xi$) increases but deteriorates as the overtone index increases.
After all these works and improvements, R.A. Konoplya generalized the method up to the 6th order \cite{Konoplya:2003ii}, while J. Matyjasek and M. Opala improved the formulas from the 7th to the 13th order \cite{Matyjasek:2017psv}.
}

\textcolor{black}{
Be aware and note that the higher-order equations of the WKB method have not been mathematically proven to consistently converge to the theoretical resolution. 
In this sense, Konoplya proposed a "preliminary" criterion for identifying errors by simply subtracting frequencies at successive orders using the WKB method. 
The idea is reasonable, but it does not provide a strict criterion for selecting the order of the WKB approximation that minimizes the errors. 
It is important to note that the WKB approximation 
gives the best results when the sixth/seventh order is used.
Note that it depends strongly on the background, the reasons why we cannot ensure that a certain order is the best regardless of the background.
Since the metric potentials are not complicated and the effective potential has the appropriate form, the 
the WKB method is sufficient to obtain accurate results. 
The method takes advantage of the one-dimensional Schrodinger-like equation corresponding to a potential barrier.
In short, the WKB formula uses the matching of asymptotic solutions, i.e. a combination of incoming and outgoing waves, together with a Taylor expansion centered around the peak of the potential barrier at $x=x_0$. 
More precisely, the expansion covers the region between the two turning points (the roots of the effective potential $U(x,\omega) \equiv V(x) - \omega^2$).
In this paper, we will consider the WKB method for computing 6th-order QN spectra, using the following generalized expression
\begin{equation}
\omega_n^2 = V_0+(-2V_0'')^{1/2} \Lambda(n) - i \nu (-2V_0'')^{1/2} [1+\Omega(n)]\,,
\end{equation}
}
\textcolor{black}{
where 
i) the second derivative of the potential at maximum
is represented by $V_0''$, 
ii) $\nu = n+1/2$, 
iii) the maximum of the effective barrier is represented by $V_0$, and finally 
iv) $n=0,1,2...$ is the overtone number.
The functions $\Lambda(n), \Omega(n)$ are well-known and also quite long, which is the reason why we bypass the deduction and write down the concrete expression for them. Instead, they can be consulted in \cite{Kokkotas:1988fm}. 
Finally, we have used a Wolfram Mathematica \cite{wolfram} notebook using the WKB method on any order from one to six \cite{Konoplya:2019hlu}.
In our calculations, we will consider only values $n < \xi$. For higher order WKB corrections, the interested reader may consult \cite{Konoplya:2019hlu,Hatsuda:2019eoj}. 
}

\begin{table}[ph!]
\centering
\caption{Dirac Quasinormal frequencies (varying $\xi$, $n$ and $\alpha$) with $M=1$ for the  model considered in this work.}
{
\textcolor{black}{
\begin{tabular}{c|c|ccccc} 
\toprule
$\alpha$ & $\xi$ &  $\omega(n=0)$ & $\omega(n=1)$ & $\omega(n=2)$  & $\omega(n=3)$  & $\omega(n=4)$ 
\\ 
\colrule
     & 3 &  0.574094\, -0.0963070 i & 0.557016\, -0.292717 i & 0.526533\, -0.499713 i & \text{}                & \text{}                \\
0.0  & 4 &  0.767354\, -0.0962705 i & 0.754300\, -0.290969 i & 0.729754\, -0.491909 i & 0.696728\, -0.702338 i & \text{}                \\
     & 5 &  0.960293\, -0.0962539 i & 0.949759\, -0.290148 i & 0.929490\, -0.488114 i & 0.901072\, -0.692514 i & 0.866728\, -0.905117 i \\
%\botrule
%\\ 
\colrule
     & 3 &  0.573408\, -0.0965237 i & 0.556274\, -0.293431 i & 0.525630\, -0.501204 i & \text{}                & \text{}                \\
0.1  & 4 &  0.766865\, -0.0964614 i & 0.753807\, -0.291556 i & 0.729273\, -0.492929 i & 0.696313\, -0.703830 i & \text{}                \\
     & 5 &  0.959913\, -0.0964185 i & 0.949375\, -0.290650 i & 0.929106\, -0.488973 i & 0.900706\, -0.693759 i & 0.866416\, -0.906779 i \\ 
%\botrule
\colrule
     & 3 &  0.572797\, -0.0967595 i & 0.556361\, -0.293799 i & 0.528672\, -0.499092 i & \text{} & \text{} \\
0.2  & 4 &  0.766378\, -0.0966740 i & 0.753327\, -0.292204 i & 0.728869\, -0.494002 i & 0.696161\, -0.705193 i & \text{} \\
     & 5 &  0.959536\, -0.0965998 i & 0.948994\, -0.291201 i & 0.928734\, -0.489907 i & 0.900390\, -0.695076 i & 0.866242\, -0.908429 i \\
%\botrule
\colrule
     & 3 &  0.571947\, -0.0970766 i & 0.554041\, -0.295592 i & 0.521144\, -0.507491 i & \text{} & \text{} \\
0.3  & 4 &  0.765870\, -0.0969152 i & 0.752597\, -0.293023 i & 0.727173\, -0.496070 i & 0.691785\, -0.711001 i & \text{} \\
     & 5 &  0.959172\, -0.0967994 i & 0.948773\, -0.291761 i & 0.929268\, -0.490457 i & 0.903371\, -0.693969 i & 0.874961\, -0.900944 i \\ 
%\botrule
\colrule
     & 3 &  0.579336\, -0.0960210 i & 0.639963\, -0.256187 i & 0.957207\, -0.275872 i & \text{} & \text{} \\
0.4  & 4 &  0.765428\, -0.0971803 i & 0.752802\, -0.293585 i & 0.731124\, -0.494522 i & 0.707856\, -0.696649 i & \text{} \\
     & 5 &  0.958758\, -0.0970275 i & 0.948014\, -0.292555 i & 0.927102\, -0.492540 i & 0.897371\, -0.699918 i & 0.86099\, -0.917216 i \\ 
%\botrule
\colrule
     & 3 &  0.585579\, -0.0951529 i & 0.719173\, -0.227667 i & 1.321910\, -0.197548 i & \text{} & \text{} \\
0.5  & 4 &  0.765004\, -0.0975141 i & 0.757256\, -0.292975 i & 0.772716\, -0.471159 i & 0.898277\, -0.557122 i & \text{} \\
     & 5 &  0.952345\, -0.0979062 i & 0.873636\, -0.318269 i & 0.559106\, -0.819435 i & 0.312456\, -2.020740 i & 0.20724\, -3.84638 i \\ 
\botrule
\end{tabular}
}
\label{table:First set}
}
\end{table}
%}

\textcolor{black}{
For Dirac perturbations, we summarize our results in table \eqref{table:First set} varying $n$, $\alpha$ and $\xi$. We used the QN frequencies using the WKB method of {\bf{6th order}}. 
Based on the frequencies obtained, all modes are found to be stable (given the negative value of the quasinormal frequencies). 
In order to guaranties the accuracy of the method, we consider the limited range $n<\xi$, a sector in which the WKB method works perfectly and agrees
with the other papers (we became aware of these papers during the revision process)
\cite{Bolokhov:2023bwm,Gingrich:2024tuf,Yang:2024ofe,Moreira:2023cxy}}.

\begin{figure*}
    \centering
    \includegraphics[width=0.48\textwidth]{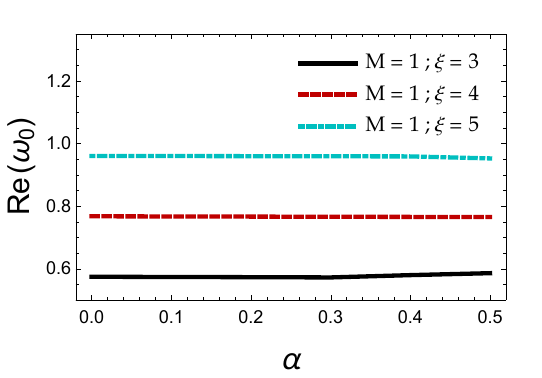}
    \includegraphics[width=0.48\textwidth]{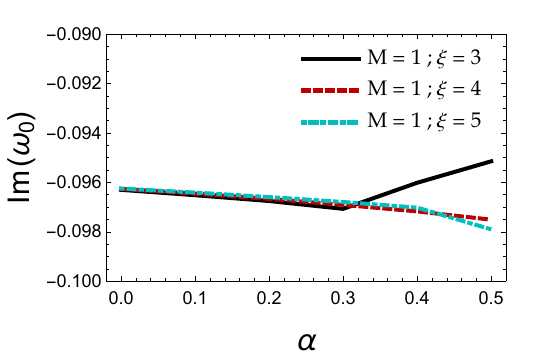}
    \caption{
    \textcolor{black}{
    Dirac Quasinormal modes for this quantum-inspired Schwarzschild black hole. We show $\omega_0$ varying $\xi$, $M$ and $n$.
    {\bf{Left panel:}} Real part of quasinormal modes against the parameter $\alpha$, assuming a fixed mass and varying $\xi$, and $n$.
    {\bf{Right panel:}} Imaginary part of quasinormal modes against the parameter $\alpha$, assuming a fixed mass and varying $\xi$, and $n$.
    }
    }
    \label{fig:QNMs_1}
\end{figure*}

\section{Conclusion and future perspectives}

In the present paper, we have mainly investigated i) the weak deflection angle and ii) the rigorous Greybody bounds of a recent black hole solution in four-dimensional spacetime inspired by loop quantum gravity. After a brief and compact review of the relevant expressions, such as the metric potentials and the alternative mass definitions, we have used the Gauss-Bonnet theorem to obtain a closed expression for computing the weak deflection angle. We have thus obtained an analytical formula to find $\hat{\alpha}$, using as a background the recently discovered quantum Schwarzschild black hole in terms of the black hole mass, or equivalently the Schwarzschild radius $r_H$, plus the auxiliary parameter $\alpha$, which is always less than one.
It should be emphasised that the weak field deflection provides an alternative angle to study quantum effects when light rays are scattered with very large impact parameters. 
The weak deflection angle (which includes both time-like particles and finite distance) $\Theta$ (and also $\Theta_{\text{null}}$) is obtained and plotted in Fig.\eqref{fig_wda} for different values of the parameter $\alpha$ and compared with the Schwarzschild black hole. We observe a qualitatively similar behaviour, although higher than the classical counterpart.
A closed quantity related to the weak deflection angle is also computed with the manuscript. This is the case of the so-called Einstein ring, $\theta_{E}$, which is computed explicitly and shown in Fig.\eqref{fig_einsring} for several values of the parameter $\alpha$. We can conclude that the Einstein ring is significantly lower than the Schwarzschild case (using M87* parameters).
In addition, we have obtained the rigorous Greybody bounds for this black hole solution by considering the Klein-Gordon equation for massless scalar fields, writing the corresponding Schrodinger-like equation, and identifying the effective potential. Thus we have well-defined cases: the first when $s=0$ and the second when $s=1$, where $s$ is the spin of the perturbation. Having obtained $V_{l}(r)$ (also plotted in Fig.\eqref{V01}), we have obtained $T_{b}$ (i.e. the greybody transmission probability) for both cases. The exact expressions for $T_{b}$ are found and the numerical solution is plotted in Fig.\eqref{T01}. Our results show that as $\alpha$ increases, the transmission probability tends to the Schwarzschild bound.

\textcolor{black}{
Finally, we compute the Dirac quasinormal modes for this quantum Schwarzschild black hole. We first study the effective potential (see Figs.~\eqref{fig:Veffec_1} and \eqref{fig:Veffec_2}), as well as the real and imaginary parts of the quasinormal frequencies (see also Fig.~\eqref{fig:QNMs_1}), for different values of the parameter $\alpha$. In the light of our calculations, we conclude that this black hole is stable to perturbations (given the negative sign of the imaginary part of the quasinormal frequency).
}
%%%%%%%%%%%%%%%%%%%%%%%%%%%%%%
\section{ACKNOWLEDGMENTS}
%%%%%%%%%%%%%%%%%%%%%%%%%%%%%%
\textcolor{black}{
The authors of this work wish to thank the reviewers for their valuable and crucial comments, which have significantly improved the quality of this manuscript. 
}
A. R. acknowledges financial support from Conselleria d'Educació, Cultura, Universitats i Ocupació de la Generalitat Valenciana thorugh Prometeo Project CIPROM/2022/13.
A. R. is funded by the María Zambrano contract ZAMBRANO 21-25 (Spain) (with funding from NextGenerationEU).
A. {\"O}. and R. P. would like to acknowledge networking support of the COST Action CA18108 - Quantum gravity phenomenology in the multi-messenger approach (QG-MM), COST Action CA21106 - COSMIC WISPers in the Dark Universe: Theory, astrophysics and experiments (CosmicWISPers), the COST Action CA22113 - Fundamental challenges in theoretical physics (THEORY-CHALLENGES), and the COST Action CA21136 - Addressing observational tensions in cosmology with systematics and fundamental physics (CosmoVerse).
\bibliography{reference.bib}

%merlin.mbs apsrev4-1.bst 2010-07-25 4.21a (PWD, AO, DPC) hacked
%Control: key (0)
%Control: author (0) dotless jnrlst
%Control: editor formatted (1) identically to author
%Control: production of article title (0) allowed
%Control: page (1) range
%Control: year (0) verbatim
%Control: production of eprint (0) enabled
\begin{thebibliography}{159}%
\makeatletter
\providecommand \@ifxundefined [1]{%
 \@ifx{#1\undefined}
}%
\providecommand \@ifnum [1]{%
 \ifnum #1\expandafter \@firstoftwo
 \else \expandafter \@secondoftwo
 \fi
}%
\providecommand \@ifx [1]{%
 \ifx #1\expandafter \@firstoftwo
 \else \expandafter \@secondoftwo
 \fi
}%
\providecommand \natexlab [1]{#1}%
\providecommand \enquote  [1]{``#1''}%
\providecommand \bibnamefont  [1]{#1}%
\providecommand \bibfnamefont [1]{#1}%
\providecommand \citenamefont [1]{#1}%
\providecommand \href@noop [0]{\@secondoftwo}%
\providecommand \href [0]{\begingroup \@sanitize@url \@href}%
\providecommand \@href[1]{\@@startlink{#1}\@@href}%
\providecommand \@@href[1]{\endgroup#1\@@endlink}%
\providecommand \@sanitize@url [0]{\catcode `\\12\catcode `\$12\catcode
  `\&12\catcode `\#12\catcode `\^12\catcode `\_12\catcode `\%12\relax}%
\providecommand \@@startlink[1]{}%
\providecommand \@@endlink[0]{}%
\providecommand \url  [0]{\begingroup\@sanitize@url \@url }%
\providecommand \@url [1]{\endgroup\@href {#1}{\urlprefix }}%
\providecommand \urlprefix  [0]{URL }%
\providecommand \Eprint [0]{\href }%
\providecommand \doibase [0]{http://dx.doi.org/}%
\providecommand \selectlanguage [0]{\@gobble}%
\providecommand \bibinfo  [0]{\@secondoftwo}%
\providecommand \bibfield  [0]{\@secondoftwo}%
\providecommand \translation [1]{[#1]}%
\providecommand \BibitemOpen [0]{}%
\providecommand \bibitemStop [0]{}%
\providecommand \bibitemNoStop [0]{.\EOS\space}%
\providecommand \EOS [0]{\spacefactor3000\relax}%
\providecommand \BibitemShut  [1]{\csname bibitem#1\endcsname}%
\let\auto@bib@innerbib\@empty
%</preamble>
\bibitem [{\citenamefont {Hawking}(1974)}]{hawking1}%
  \BibitemOpen
  \bibfield  {author} {\bibinfo {author} {\bibfnamefont {S.~W.}\ \bibnamefont
  {Hawking}},\ }\bibfield  {title} {\enquote {\bibinfo {title} {{Black hole
  explosions}},}\ }\href {\doibase 10.1038/248030a0} {\bibfield  {journal}
  {\bibinfo  {journal} {Nature}\ }\textbf {\bibinfo {volume} {248}},\ \bibinfo
  {pages} {30--31} (\bibinfo {year} {1974})}\BibitemShut {NoStop}%
%%CITATION = NATUA,248,30;%%
\bibitem [{\citenamefont {Hawking}(1975)}]{hawking2}%
  \BibitemOpen
  \bibfield  {author} {\bibinfo {author} {\bibfnamefont {S.~W.}\ \bibnamefont
  {Hawking}},\ }\bibfield  {title} {\enquote {\bibinfo {title} {{Particle
  Creation by Black Holes}},}\ }\bibfield  {booktitle} {\emph {\bibinfo
  {booktitle} {{Euclidean quantum gravity}}},\ }\href {\doibase
  10.1007/BF02345020, 10.1007/BF01608497} {\bibfield  {journal} {\bibinfo
  {journal} {Commun. Math. Phys.}\ }\textbf {\bibinfo {volume} {43}},\ \bibinfo
  {pages} {199--220} (\bibinfo {year} {1975})},\ \bibinfo {note}
  {[,167(1975)]}\BibitemShut {NoStop}%
%%CITATION = CMPHA,43,199;%%
\bibitem [{\citenamefont {Eichhorn}\ and\ \citenamefont
  {Held}(2022)}]{Eichhorn:2022bgu}%
  \BibitemOpen
  \bibfield  {author} {\bibinfo {author} {\bibfnamefont {Astrid}\ \bibnamefont
  {Eichhorn}}\ and\ \bibinfo {author} {\bibfnamefont {Aaron}\ \bibnamefont
  {Held}},\ }\href@noop {} {\enquote {\bibinfo {title} {{Black holes in
  asymptotically safe gravity and beyond}},}\ } (\bibinfo {year} {2022}),\
  \Eprint {http://arxiv.org/abs/2212.09495} {arXiv:2212.09495 [gr-qc]}
  \BibitemShut {NoStop}%
\bibitem [{\citenamefont {Abbott}\ \emph
  {et~al.}(2016{\natexlab{a}})\citenamefont {Abbott} \emph {et~al.}}]{ligo1}%
  \BibitemOpen
  \bibfield  {author} {\bibinfo {author} {\bibfnamefont {B.~P.}\ \bibnamefont
  {Abbott}} \emph {et~al.} (\bibinfo {collaboration} {LIGO Scientific,
  Virgo}),\ }\bibfield  {title} {\enquote {\bibinfo {title} {{Observation of
  Gravitational Waves from a Binary Black Hole Merger}},}\ }\href {\doibase
  10.1103/PhysRevLett.116.061102} {\bibfield  {journal} {\bibinfo  {journal}
  {Phys. Rev. Lett.}\ }\textbf {\bibinfo {volume} {116}},\ \bibinfo {pages}
  {061102} (\bibinfo {year} {2016}{\natexlab{a}})},\ \Eprint
  {http://arxiv.org/abs/1602.03837} {arXiv:1602.03837 [gr-qc]} \BibitemShut
  {NoStop}%
%%CITATION = ARXIV:1602.03837;%%
\bibitem [{\citenamefont {Abbott}\ \emph
  {et~al.}(2016{\natexlab{b}})\citenamefont {Abbott} \emph {et~al.}}]{ligo2}%
  \BibitemOpen
  \bibfield  {author} {\bibinfo {author} {\bibfnamefont {B.~P.}\ \bibnamefont
  {Abbott}} \emph {et~al.} (\bibinfo {collaboration} {LIGO Scientific,
  Virgo}),\ }\bibfield  {title} {\enquote {\bibinfo {title} {{GW151226:
  Observation of Gravitational Waves from a 22-Solar-Mass Binary Black Hole
  Coalescence}},}\ }\href {\doibase 10.1103/PhysRevLett.116.241103} {\bibfield
  {journal} {\bibinfo  {journal} {Phys. Rev. Lett.}\ }\textbf {\bibinfo
  {volume} {116}},\ \bibinfo {pages} {241103} (\bibinfo {year}
  {2016}{\natexlab{b}})},\ \Eprint {http://arxiv.org/abs/1606.04855}
  {arXiv:1606.04855 [gr-qc]} \BibitemShut {NoStop}%
%%CITATION = ARXIV:1606.04855;%%
\bibitem [{\citenamefont {Abbott}\ \emph
  {et~al.}(2017{\natexlab{a}})\citenamefont {Abbott} \emph {et~al.}}]{ligo3}%
  \BibitemOpen
  \bibfield  {author} {\bibinfo {author} {\bibfnamefont {Benjamin~P.}\
  \bibnamefont {Abbott}} \emph {et~al.} (\bibinfo {collaboration} {LIGO
  Scientific, VIRGO}),\ }\bibfield  {title} {\enquote {\bibinfo {title}
  {{GW170104: Observation of a 50-Solar-Mass Binary Black Hole Coalescence at
  Redshift 0.2}},}\ }\href {\doibase 10.1103/PhysRevLett.118.221101,
  10.1103/PhysRevLett.121.129901} {\bibfield  {journal} {\bibinfo  {journal}
  {Phys. Rev. Lett.}\ }\textbf {\bibinfo {volume} {118}},\ \bibinfo {pages}
  {221101} (\bibinfo {year} {2017}{\natexlab{a}})},\ \bibinfo {note} {[Erratum:
  Phys. Rev. Lett.121,no.12,129901(2018)]},\ \Eprint
  {http://arxiv.org/abs/1706.01812} {arXiv:1706.01812 [gr-qc]} \BibitemShut
  {NoStop}%
%%CITATION = ARXIV:1706.01812;%%
\bibitem [{\citenamefont {Abbott}\ \emph
  {et~al.}(2017{\natexlab{b}})\citenamefont {Abbott} \emph {et~al.}}]{ligo4}%
  \BibitemOpen
  \bibfield  {author} {\bibinfo {author} {\bibfnamefont {B.~P.}\ \bibnamefont
  {Abbott}} \emph {et~al.} (\bibinfo {collaboration} {LIGO Scientific,
  Virgo}),\ }\bibfield  {title} {\enquote {\bibinfo {title} {{GW170814: A
  Three-Detector Observation of Gravitational Waves from a Binary Black Hole
  Coalescence}},}\ }\href {\doibase 10.1103/PhysRevLett.119.141101} {\bibfield
  {journal} {\bibinfo  {journal} {Phys. Rev. Lett.}\ }\textbf {\bibinfo
  {volume} {119}},\ \bibinfo {pages} {141101} (\bibinfo {year}
  {2017}{\natexlab{b}})},\ \Eprint {http://arxiv.org/abs/1709.09660}
  {arXiv:1709.09660 [gr-qc]} \BibitemShut {NoStop}%
%%CITATION = ARXIV:1709.09660;%%
\bibitem [{\citenamefont {Abbott}\ \emph
  {et~al.}(2017{\natexlab{c}})\citenamefont {Abbott} \emph {et~al.}}]{ligo5}%
  \BibitemOpen
  \bibfield  {author} {\bibinfo {author} {\bibfnamefont {B..~P..}\ \bibnamefont
  {Abbott}} \emph {et~al.} (\bibinfo {collaboration} {LIGO Scientific,
  Virgo}),\ }\bibfield  {title} {\enquote {\bibinfo {title} {{GW170608:
  Observation of a 19-solar-mass Binary Black Hole Coalescence}},}\ }\href
  {\doibase 10.3847/2041-8213/aa9f0c} {\bibfield  {journal} {\bibinfo
  {journal} {Astrophys. J.}\ }\textbf {\bibinfo {volume} {851}},\ \bibinfo
  {pages} {L35} (\bibinfo {year} {2017}{\natexlab{c}})},\ \Eprint
  {http://arxiv.org/abs/1711.05578} {arXiv:1711.05578 [astro-ph.HE]}
  \BibitemShut {NoStop}%
%%CITATION = ARXIV:1711.05578;%%
\bibitem [{\citenamefont {Akiyama}\ \emph
  {et~al.}(2019{\natexlab{a}})\citenamefont {Akiyama} \emph {et~al.}}]{L1}%
  \BibitemOpen
  \bibfield  {author} {\bibinfo {author} {\bibfnamefont {Kazunori}\
  \bibnamefont {Akiyama}} \emph {et~al.} (\bibinfo {collaboration} {Event
  Horizon Telescope}),\ }\bibfield  {title} {\enquote {\bibinfo {title} {{First
  M87 Event Horizon Telescope Results. I. The Shadow of the Supermassive Black
  Hole}},}\ }\href {\doibase 10.3847/2041-8213/ab0ec7} {\bibfield  {journal}
  {\bibinfo  {journal} {Astrophys. J.}\ }\textbf {\bibinfo {volume} {875}},\
  \bibinfo {pages} {L1} (\bibinfo {year} {2019}{\natexlab{a}})},\ \Eprint
  {http://arxiv.org/abs/1906.11238} {arXiv:1906.11238 [astro-ph.GA]}
  \BibitemShut {NoStop}%
%%CITATION = ARXIV:1906.11238;%%
\bibitem [{\citenamefont {Akiyama}\ \emph
  {et~al.}(2019{\natexlab{b}})\citenamefont {Akiyama} \emph {et~al.}}]{L2}%
  \BibitemOpen
  \bibfield  {author} {\bibinfo {author} {\bibfnamefont {Kazunori}\
  \bibnamefont {Akiyama}} \emph {et~al.} (\bibinfo {collaboration} {Event
  Horizon Telescope}),\ }\bibfield  {title} {\enquote {\bibinfo {title} {{First
  M87 Event Horizon Telescope Results. II. Array and Instrumentation}},}\
  }\href {\doibase 10.3847/2041-8213/ab0c96} {\bibfield  {journal} {\bibinfo
  {journal} {Astrophys. J.}\ }\textbf {\bibinfo {volume} {875}},\ \bibinfo
  {pages} {L2} (\bibinfo {year} {2019}{\natexlab{b}})},\ \Eprint
  {http://arxiv.org/abs/1906.11239} {arXiv:1906.11239 [astro-ph.IM]}
  \BibitemShut {NoStop}%
%%CITATION = ARXIV:1906.11239;%%
\bibitem [{\citenamefont {Akiyama}\ \emph
  {et~al.}(2019{\natexlab{c}})\citenamefont {Akiyama} \emph {et~al.}}]{L3}%
  \BibitemOpen
  \bibfield  {author} {\bibinfo {author} {\bibfnamefont {Kazunori}\
  \bibnamefont {Akiyama}} \emph {et~al.} (\bibinfo {collaboration} {Event
  Horizon Telescope}),\ }\bibfield  {title} {\enquote {\bibinfo {title} {{First
  M87 Event Horizon Telescope Results. III. Data Processing and
  Calibration}},}\ }\href {\doibase 10.3847/2041-8213/ab0c57} {\bibfield
  {journal} {\bibinfo  {journal} {Astrophys. J.}\ }\textbf {\bibinfo {volume}
  {875}},\ \bibinfo {pages} {L3} (\bibinfo {year} {2019}{\natexlab{c}})},\
  \Eprint {http://arxiv.org/abs/1906.11240} {arXiv:1906.11240 [astro-ph.GA]}
  \BibitemShut {NoStop}%
%%CITATION = ARXIV:1906.11240;%%
\bibitem [{\citenamefont {Akiyama}\ \emph
  {et~al.}(2019{\natexlab{d}})\citenamefont {Akiyama} \emph {et~al.}}]{L4}%
  \BibitemOpen
  \bibfield  {author} {\bibinfo {author} {\bibfnamefont {Kazunori}\
  \bibnamefont {Akiyama}} \emph {et~al.} (\bibinfo {collaboration} {Event
  Horizon Telescope}),\ }\bibfield  {title} {\enquote {\bibinfo {title} {{First
  M87 Event Horizon Telescope Results. IV. Imaging the Central Supermassive
  Black Hole}},}\ }\href {\doibase 10.3847/2041-8213/ab0e85} {\bibfield
  {journal} {\bibinfo  {journal} {Astrophys. J.}\ }\textbf {\bibinfo {volume}
  {875}},\ \bibinfo {pages} {L4} (\bibinfo {year} {2019}{\natexlab{d}})},\
  \Eprint {http://arxiv.org/abs/1906.11241} {arXiv:1906.11241 [astro-ph.GA]}
  \BibitemShut {NoStop}%
%%CITATION = ARXIV:1906.11241;%%
\bibitem [{\citenamefont {Akiyama}\ \emph
  {et~al.}(2019{\natexlab{e}})\citenamefont {Akiyama} \emph {et~al.}}]{L5}%
  \BibitemOpen
  \bibfield  {author} {\bibinfo {author} {\bibfnamefont {Kazunori}\
  \bibnamefont {Akiyama}} \emph {et~al.} (\bibinfo {collaboration} {Event
  Horizon Telescope}),\ }\bibfield  {title} {\enquote {\bibinfo {title} {{First
  M87 Event Horizon Telescope Results. V. Physical Origin of the Asymmetric
  Ring}},}\ }\href {\doibase 10.3847/2041-8213/ab0f43} {\bibfield  {journal}
  {\bibinfo  {journal} {Astrophys. J.}\ }\textbf {\bibinfo {volume} {875}},\
  \bibinfo {pages} {L5} (\bibinfo {year} {2019}{\natexlab{e}})},\ \Eprint
  {http://arxiv.org/abs/1906.11242} {arXiv:1906.11242 [astro-ph.GA]}
  \BibitemShut {NoStop}%
%%CITATION = ARXIV:1906.11242;%%
\bibitem [{\citenamefont {Akiyama}\ \emph
  {et~al.}(2019{\natexlab{f}})\citenamefont {Akiyama} \emph {et~al.}}]{L6}%
  \BibitemOpen
  \bibfield  {author} {\bibinfo {author} {\bibfnamefont {Kazunori}\
  \bibnamefont {Akiyama}} \emph {et~al.} (\bibinfo {collaboration} {Event
  Horizon Telescope}),\ }\bibfield  {title} {\enquote {\bibinfo {title} {{First
  M87 Event Horizon Telescope Results. VI. The Shadow and Mass of the Central
  Black Hole}},}\ }\href {\doibase 10.3847/2041-8213/ab1141} {\bibfield
  {journal} {\bibinfo  {journal} {Astrophys. J.}\ }\textbf {\bibinfo {volume}
  {875}},\ \bibinfo {pages} {L6} (\bibinfo {year} {2019}{\natexlab{f}})},\
  \Eprint {http://arxiv.org/abs/1906.11243} {arXiv:1906.11243 [astro-ph.GA]}
  \BibitemShut {NoStop}%
%%CITATION = ARXIV:1906.11243;%%
\bibitem [{pro()}]{project}%
  \BibitemOpen
  \href@noop {} {\enquote {\bibinfo {title} {The event horizon telescope},}\
  }\bibinfo {howpublished} {\url{https://eventhorizontelescope.org}},\ \bibinfo
  {note} {accessed: 2019-07-12}\BibitemShut {NoStop}%
\bibitem [{\citenamefont {Heusler}(1996)}]{heusler_1996}%
  \BibitemOpen
  \bibfield  {author} {\bibinfo {author} {\bibfnamefont {Markus}\ \bibnamefont
  {Heusler}},\ }\href {\doibase 10.1017/CBO9780511661396} {\emph {\bibinfo
  {title} {Black Hole Uniqueness Theorems}}},\ Cambridge Lecture Notes in
  Physics\ (\bibinfo  {publisher} {Cambridge University Press},\ \bibinfo
  {year} {1996})\BibitemShut {NoStop}%
\bibitem [{\citenamefont {Hawking}\ \emph {et~al.}(2016)\citenamefont
  {Hawking}, \citenamefont {Perry},\ and\ \citenamefont
  {Strominger}}]{Hawking:2016msc}%
  \BibitemOpen
  \bibfield  {author} {\bibinfo {author} {\bibfnamefont {Stephen~W.}\
  \bibnamefont {Hawking}}, \bibinfo {author} {\bibfnamefont {Malcolm~J.}\
  \bibnamefont {Perry}}, \ and\ \bibinfo {author} {\bibfnamefont {Andrew}\
  \bibnamefont {Strominger}},\ }\bibfield  {title} {\enquote {\bibinfo {title}
  {{Soft Hair on Black Holes}},}\ }\href {\doibase
  10.1103/PhysRevLett.116.231301} {\bibfield  {journal} {\bibinfo  {journal}
  {Phys. Rev. Lett.}\ }\textbf {\bibinfo {volume} {116}},\ \bibinfo {pages}
  {231301} (\bibinfo {year} {2016})},\ \Eprint
  {http://arxiv.org/abs/1601.00921} {arXiv:1601.00921 [hep-th]} \BibitemShut
  {NoStop}%
\bibitem [{\citenamefont {Babichev}\ and\ \citenamefont
  {Charmousis}(2014)}]{Babichev:2013cya}%
  \BibitemOpen
  \bibfield  {author} {\bibinfo {author} {\bibfnamefont {Eugeny}\ \bibnamefont
  {Babichev}}\ and\ \bibinfo {author} {\bibfnamefont {Christos}\ \bibnamefont
  {Charmousis}},\ }\bibfield  {title} {\enquote {\bibinfo {title} {{Dressing a
  black hole with a time-dependent Galileon}},}\ }\href {\doibase
  10.1007/JHEP08(2014)106} {\bibfield  {journal} {\bibinfo  {journal} {JHEP}\
  }\textbf {\bibinfo {volume} {08}},\ \bibinfo {pages} {106} (\bibinfo {year}
  {2014})},\ \Eprint {http://arxiv.org/abs/1312.3204} {arXiv:1312.3204 [gr-qc]}
  \BibitemShut {NoStop}%
\bibitem [{\citenamefont {Antoniou}\ \emph {et~al.}(2018)\citenamefont
  {Antoniou}, \citenamefont {Bakopoulos},\ and\ \citenamefont
  {Kanti}}]{Antoniou:2017acq}%
  \BibitemOpen
  \bibfield  {author} {\bibinfo {author} {\bibfnamefont {G.}~\bibnamefont
  {Antoniou}}, \bibinfo {author} {\bibfnamefont {A.}~\bibnamefont
  {Bakopoulos}}, \ and\ \bibinfo {author} {\bibfnamefont {P.}~\bibnamefont
  {Kanti}},\ }\bibfield  {title} {\enquote {\bibinfo {title} {{Evasion of
  No-Hair Theorems and Novel Black-Hole Solutions in Gauss-Bonnet Theories}},}\
  }\href {\doibase 10.1103/PhysRevLett.120.131102} {\bibfield  {journal}
  {\bibinfo  {journal} {Phys. Rev. Lett.}\ }\textbf {\bibinfo {volume} {120}},\
  \bibinfo {pages} {131102} (\bibinfo {year} {2018})},\ \Eprint
  {http://arxiv.org/abs/1711.03390} {arXiv:1711.03390 [hep-th]} \BibitemShut
  {NoStop}%
\bibitem [{\citenamefont {Sotiriou}\ and\ \citenamefont
  {Zhou}(2014)}]{Sotiriou:2013qea}%
  \BibitemOpen
  \bibfield  {author} {\bibinfo {author} {\bibfnamefont {Thomas~P.}\
  \bibnamefont {Sotiriou}}\ and\ \bibinfo {author} {\bibfnamefont
  {Shuang-Yong}\ \bibnamefont {Zhou}},\ }\bibfield  {title} {\enquote {\bibinfo
  {title} {{Black hole hair in generalized scalar-tensor gravity}},}\ }\href
  {\doibase 10.1103/PhysRevLett.112.251102} {\bibfield  {journal} {\bibinfo
  {journal} {Phys. Rev. Lett.}\ }\textbf {\bibinfo {volume} {112}},\ \bibinfo
  {pages} {251102} (\bibinfo {year} {2014})},\ \Eprint
  {http://arxiv.org/abs/1312.3622} {arXiv:1312.3622 [gr-qc]} \BibitemShut
  {NoStop}%
\bibitem [{\citenamefont {Schwarzschild}(1916)}]{Schwarzschild:1916uq}%
  \BibitemOpen
  \bibfield  {author} {\bibinfo {author} {\bibfnamefont {Karl}\ \bibnamefont
  {Schwarzschild}},\ }\bibfield  {title} {\enquote {\bibinfo {title} {{On the
  gravitational field of a mass point according to Einstein's theory}},}\
  }\href@noop {} {\bibfield  {journal} {\bibinfo  {journal} {Sitzungsber.
  Preuss. Akad. Wiss. Berlin (Math. Phys. )}\ }\textbf {\bibinfo {volume}
  {1916}},\ \bibinfo {pages} {189--196} (\bibinfo {year} {1916})},\ \Eprint
  {http://arxiv.org/abs/physics/9905030} {arXiv:physics/9905030} \BibitemShut
  {NoStop}%
\bibitem [{\citenamefont {{Reissner}}(1916)}]{1916AnP...355..106R}%
  \BibitemOpen
  \bibfield  {author} {\bibinfo {author} {\bibfnamefont {H.}~\bibnamefont
  {{Reissner}}},\ }\bibfield  {title} {\enquote {\bibinfo {title} {{{\"U}ber
  die Eigengravitation des elektrischen Feldes nach der Einsteinschen
  Theorie}},}\ }\href {\doibase 10.1002/andp.19163550905} {\bibfield  {journal}
  {\bibinfo  {journal} {Annalen der Physik}\ }\textbf {\bibinfo {volume}
  {355}},\ \bibinfo {pages} {106--120} (\bibinfo {year} {1916})}\BibitemShut
  {NoStop}%
\bibitem [{\citenamefont {{Nordstr{\"o}m}}(1918)}]{1918KNAB...20.1238N}%
  \BibitemOpen
  \bibfield  {author} {\bibinfo {author} {\bibfnamefont {G.}~\bibnamefont
  {{Nordstr{\"o}m}}},\ }\bibfield  {title} {\enquote {\bibinfo {title} {{On the
  Energy of the Gravitation field in Einstein's Theory}},}\ }\href@noop {}
  {\bibfield  {journal} {\bibinfo  {journal} {Koninklijke Nederlandse Akademie
  van Wetenschappen Proceedings Series B Physical Sciences}\ }\textbf {\bibinfo
  {volume} {20}},\ \bibinfo {pages} {1238--1245} (\bibinfo {year}
  {1918})}\BibitemShut {NoStop}%
\bibitem [{\citenamefont {Kerr}(1963)}]{Kerr:1963ud}%
  \BibitemOpen
  \bibfield  {author} {\bibinfo {author} {\bibfnamefont {Roy~P.}\ \bibnamefont
  {Kerr}},\ }\bibfield  {title} {\enquote {\bibinfo {title} {{Gravitational
  field of a spinning mass as an example of algebraically special metrics}},}\
  }\href {\doibase 10.1103/PhysRevLett.11.237} {\bibfield  {journal} {\bibinfo
  {journal} {Phys. Rev. Lett.}\ }\textbf {\bibinfo {volume} {11}},\ \bibinfo
  {pages} {237--238} (\bibinfo {year} {1963})}\BibitemShut {NoStop}%
\bibitem [{\citenamefont {Newman}\ \emph {et~al.}(1965)\citenamefont {Newman},
  \citenamefont {Couch}, \citenamefont {Chinnapared}, \citenamefont {Exton},
  \citenamefont {Prakash},\ and\ \citenamefont {Torrence}}]{Newman:1965my}%
  \BibitemOpen
  \bibfield  {author} {\bibinfo {author} {\bibfnamefont {E~T.}\ \bibnamefont
  {Newman}}, \bibinfo {author} {\bibfnamefont {R.}~\bibnamefont {Couch}},
  \bibinfo {author} {\bibfnamefont {K.}~\bibnamefont {Chinnapared}}, \bibinfo
  {author} {\bibfnamefont {A.}~\bibnamefont {Exton}}, \bibinfo {author}
  {\bibfnamefont {A.}~\bibnamefont {Prakash}}, \ and\ \bibinfo {author}
  {\bibfnamefont {R.}~\bibnamefont {Torrence}},\ }\bibfield  {title} {\enquote
  {\bibinfo {title} {{Metric of a Rotating, Charged Mass}},}\ }\href {\doibase
  10.1063/1.1704351} {\bibfield  {journal} {\bibinfo  {journal} {J. Math.
  Phys.}\ }\textbf {\bibinfo {volume} {6}},\ \bibinfo {pages} {918--919}
  (\bibinfo {year} {1965})}\BibitemShut {NoStop}%
\bibitem [{\citenamefont {Jacobson}(1995)}]{Jacobson:1995ab}%
  \BibitemOpen
  \bibfield  {author} {\bibinfo {author} {\bibfnamefont {Ted}\ \bibnamefont
  {Jacobson}},\ }\bibfield  {title} {\enquote {\bibinfo {title}
  {{Thermodynamics of space-time: The Einstein equation of state}},}\ }\href
  {\doibase 10.1103/PhysRevLett.75.1260} {\bibfield  {journal} {\bibinfo
  {journal} {Phys. Rev. Lett.}\ }\textbf {\bibinfo {volume} {75}},\ \bibinfo
  {pages} {1260--1263} (\bibinfo {year} {1995})},\ \Eprint
  {http://arxiv.org/abs/gr-qc/9504004} {arXiv:gr-qc/9504004 [gr-qc]}
  \BibitemShut {NoStop}%
%%CITATION = GR-QC/9504004;%%
\bibitem [{\citenamefont {Connes}(1996)}]{Connes:1996gi}%
  \BibitemOpen
  \bibfield  {author} {\bibinfo {author} {\bibfnamefont {Alain}\ \bibnamefont
  {Connes}},\ }\bibfield  {title} {\enquote {\bibinfo {title} {{Gravity coupled
  with matter and foundation of noncommutative geometry}},}\ }\href {\doibase
  10.1007/BF02506388} {\bibfield  {journal} {\bibinfo  {journal} {Commun. Math.
  Phys.}\ }\textbf {\bibinfo {volume} {182}},\ \bibinfo {pages} {155--176}
  (\bibinfo {year} {1996})},\ \Eprint {http://arxiv.org/abs/hep-th/9603053}
  {arXiv:hep-th/9603053 [hep-th]} \BibitemShut {NoStop}%
%%CITATION = HEP-TH/9603053;%%
\bibitem [{\citenamefont {Reuter}(1998)}]{Reuter:1996cp}%
  \BibitemOpen
  \bibfield  {author} {\bibinfo {author} {\bibfnamefont {M.}~\bibnamefont
  {Reuter}},\ }\bibfield  {title} {\enquote {\bibinfo {title} {{Nonperturbative
  evolution equation for quantum gravity}},}\ }\href {\doibase
  10.1103/PhysRevD.57.971} {\bibfield  {journal} {\bibinfo  {journal} {Phys.
  Rev. D}\ }\textbf {\bibinfo {volume} {57}},\ \bibinfo {pages} {971--985}
  (\bibinfo {year} {1998})},\ \Eprint {http://arxiv.org/abs/hep-th/9605030}
  {arXiv:hep-th/9605030} \BibitemShut {NoStop}%
\bibitem [{\citenamefont {Rovelli}(1998)}]{Rovelli:1997yv}%
  \BibitemOpen
  \bibfield  {author} {\bibinfo {author} {\bibfnamefont {Carlo}\ \bibnamefont
  {Rovelli}},\ }\bibfield  {title} {\enquote {\bibinfo {title} {{Loop quantum
  gravity}},}\ }\href {\doibase 10.12942/lrr-1998-1} {\bibfield  {journal}
  {\bibinfo  {journal} {Living Rev. Rel.}\ }\textbf {\bibinfo {volume} {1}},\
  \bibinfo {pages} {1} (\bibinfo {year} {1998})},\ \Eprint
  {http://arxiv.org/abs/gr-qc/9710008} {arXiv:gr-qc/9710008 [gr-qc]}
  \BibitemShut {NoStop}%
%%CITATION = GR-QC/9710008;%%
\bibitem [{\citenamefont {Gambini}\ and\ \citenamefont
  {Pullin}(2005)}]{Gambini:2004vz}%
  \BibitemOpen
  \bibfield  {author} {\bibinfo {author} {\bibfnamefont {Rodolfo}\ \bibnamefont
  {Gambini}}\ and\ \bibinfo {author} {\bibfnamefont {Jorge}\ \bibnamefont
  {Pullin}},\ }\bibfield  {title} {\enquote {\bibinfo {title} {{Consistent
  discretization and loop quantum geometry}},}\ }\href {\doibase
  10.1103/PhysRevLett.94.101302} {\bibfield  {journal} {\bibinfo  {journal}
  {Phys. Rev. Lett.}\ }\textbf {\bibinfo {volume} {94}},\ \bibinfo {pages}
  {101302} (\bibinfo {year} {2005})},\ \Eprint
  {http://arxiv.org/abs/gr-qc/0409057} {arXiv:gr-qc/0409057 [gr-qc]}
  \BibitemShut {NoStop}%
%%CITATION = GR-QC/0409057;%%
\bibitem [{\citenamefont {Ashtekar}(2005)}]{Ashtekar:2004vs}%
  \BibitemOpen
  \bibfield  {author} {\bibinfo {author} {\bibfnamefont {Abhay}\ \bibnamefont
  {Ashtekar}},\ }\bibfield  {title} {\enquote {\bibinfo {title} {{Gravity and
  the quantum}},}\ }\href {\doibase 10.1088/1367-2630/7/1/198} {\bibfield
  {journal} {\bibinfo  {journal} {New J. Phys.}\ }\textbf {\bibinfo {volume}
  {7}},\ \bibinfo {pages} {198} (\bibinfo {year} {2005})},\ \Eprint
  {http://arxiv.org/abs/gr-qc/0410054} {arXiv:gr-qc/0410054 [gr-qc]}
  \BibitemShut {NoStop}%
%%CITATION = GR-QC/0410054;%%
\bibitem [{\citenamefont {Nicolini}(2009)}]{Nicolini:2008aj}%
  \BibitemOpen
  \bibfield  {author} {\bibinfo {author} {\bibfnamefont {Piero}\ \bibnamefont
  {Nicolini}},\ }\bibfield  {title} {\enquote {\bibinfo {title}
  {{Noncommutative Black Holes, The Final Appeal To Quantum Gravity: A
  Review}},}\ }\href {\doibase 10.1142/S0217751X09043353} {\bibfield  {journal}
  {\bibinfo  {journal} {Int. J. Mod. Phys.}\ }\textbf {\bibinfo {volume}
  {A24}},\ \bibinfo {pages} {1229--1308} (\bibinfo {year} {2009})},\ \Eprint
  {http://arxiv.org/abs/0807.1939} {arXiv:0807.1939 [hep-th]} \BibitemShut
  {NoStop}%
%%CITATION = ARXIV:0807.1939;%%
\bibitem [{\citenamefont {Horava}(2009)}]{Horava:2009uw}%
  \BibitemOpen
  \bibfield  {author} {\bibinfo {author} {\bibfnamefont {Petr}\ \bibnamefont
  {Horava}},\ }\bibfield  {title} {\enquote {\bibinfo {title} {{Quantum Gravity
  at a Lifshitz Point}},}\ }\href {\doibase 10.1103/PhysRevD.79.084008}
  {\bibfield  {journal} {\bibinfo  {journal} {Phys. Rev.}\ }\textbf {\bibinfo
  {volume} {D79}},\ \bibinfo {pages} {084008} (\bibinfo {year} {2009})},\
  \Eprint {http://arxiv.org/abs/0901.3775} {arXiv:0901.3775 [hep-th]}
  \BibitemShut {NoStop}%
%%CITATION = ARXIV:0901.3775;%%
\bibitem [{\citenamefont {Verlinde}(2011)}]{Verlinde:2010hp}%
  \BibitemOpen
  \bibfield  {author} {\bibinfo {author} {\bibfnamefont {Erik~P.}\ \bibnamefont
  {Verlinde}},\ }\bibfield  {title} {\enquote {\bibinfo {title} {{On the Origin
  of Gravity and the Laws of Newton}},}\ }\href {\doibase
  10.1007/JHEP04(2011)029} {\bibfield  {journal} {\bibinfo  {journal} {JHEP}\
  }\textbf {\bibinfo {volume} {04}},\ \bibinfo {pages} {029} (\bibinfo {year}
  {2011})},\ \Eprint {http://arxiv.org/abs/1001.0785} {arXiv:1001.0785
  [hep-th]} \BibitemShut {NoStop}%
%%CITATION = ARXIV:1001.0785;%%
\bibitem [{\citenamefont {Platania}(2023)}]{Platania:2023srt}%
  \BibitemOpen
  \bibfield  {author} {\bibinfo {author} {\bibfnamefont {Alessia}\ \bibnamefont
  {Platania}},\ }\href@noop {} {\enquote {\bibinfo {title} {{Black Holes in
  Asymptotically Safe Gravity}},}\ } (\bibinfo {year} {2023}),\ \Eprint
  {http://arxiv.org/abs/2302.04272} {arXiv:2302.04272 [gr-qc]} \BibitemShut
  {NoStop}%
\bibitem [{\citenamefont {Saueressig}\ \emph {et~al.}(2016)\citenamefont
  {Saueressig}, \citenamefont {Alkofer}, \citenamefont {D'Odorico},\ and\
  \citenamefont {Vidotto}}]{Saueressig:2015xua}%
  \BibitemOpen
  \bibfield  {author} {\bibinfo {author} {\bibfnamefont {Frank}\ \bibnamefont
  {Saueressig}}, \bibinfo {author} {\bibfnamefont {Natalia}\ \bibnamefont
  {Alkofer}}, \bibinfo {author} {\bibfnamefont {Giulio}\ \bibnamefont
  {D'Odorico}}, \ and\ \bibinfo {author} {\bibfnamefont {Francesca}\
  \bibnamefont {Vidotto}},\ }\bibfield  {title} {\enquote {\bibinfo {title}
  {{Black holes in Asymptotically Safe Gravity}},}\ }\href {\doibase
  10.22323/1.224.0174} {\bibfield  {journal} {\bibinfo  {journal} {PoS}\
  }\textbf {\bibinfo {volume} {FFP14}},\ \bibinfo {pages} {174} (\bibinfo
  {year} {2016})},\ \Eprint {http://arxiv.org/abs/1503.06472} {arXiv:1503.06472
  [hep-th]} \BibitemShut {NoStop}%
\bibitem [{\citenamefont {Koch}\ and\ \citenamefont
  {Saueressig}(2014)}]{Koch:2014cqa}%
  \BibitemOpen
  \bibfield  {author} {\bibinfo {author} {\bibfnamefont {Benjamin}\
  \bibnamefont {Koch}}\ and\ \bibinfo {author} {\bibfnamefont {Frank}\
  \bibnamefont {Saueressig}},\ }\bibfield  {title} {\enquote {\bibinfo {title}
  {{Black holes within Asymptotic Safety}},}\ }\href {\doibase
  10.1142/S0217751X14300117} {\bibfield  {journal} {\bibinfo  {journal} {Int.
  J. Mod. Phys. A}\ }\textbf {\bibinfo {volume} {29}},\ \bibinfo {pages}
  {1430011} (\bibinfo {year} {2014})},\ \Eprint
  {http://arxiv.org/abs/1401.4452} {arXiv:1401.4452 [hep-th]} \BibitemShut
  {NoStop}%
\bibitem [{\citenamefont {Borissova}\ and\ \citenamefont
  {Platania}(2023)}]{Borissova:2022mgd}%
  \BibitemOpen
  \bibfield  {author} {\bibinfo {author} {\bibfnamefont {Johanna~N.}\
  \bibnamefont {Borissova}}\ and\ \bibinfo {author} {\bibfnamefont {Alessia}\
  \bibnamefont {Platania}},\ }\bibfield  {title} {\enquote {\bibinfo {title}
  {{Formation and evaporation of quantum black holes from the decoupling
  mechanism in quantum gravity}},}\ }\href {\doibase 10.1007/JHEP03(2023)046}
  {\bibfield  {journal} {\bibinfo  {journal} {JHEP}\ }\textbf {\bibinfo
  {volume} {03}},\ \bibinfo {pages} {046} (\bibinfo {year} {2023})},\ \Eprint
  {http://arxiv.org/abs/2210.01138} {arXiv:2210.01138 [gr-qc]} \BibitemShut
  {NoStop}%
\bibitem [{\citenamefont {Bonanno}\ and\ \citenamefont
  {Reuter}(2000)}]{Bonanno:2000ep}%
  \BibitemOpen
  \bibfield  {author} {\bibinfo {author} {\bibfnamefont {Alfio}\ \bibnamefont
  {Bonanno}}\ and\ \bibinfo {author} {\bibfnamefont {Martin}\ \bibnamefont
  {Reuter}},\ }\bibfield  {title} {\enquote {\bibinfo {title} {{Renormalization
  group improved black hole space-times}},}\ }\href {\doibase
  10.1103/PhysRevD.62.043008} {\bibfield  {journal} {\bibinfo  {journal} {Phys.
  Rev. D}\ }\textbf {\bibinfo {volume} {62}},\ \bibinfo {pages} {043008}
  (\bibinfo {year} {2000})},\ \Eprint {http://arxiv.org/abs/hep-th/0002196}
  {arXiv:hep-th/0002196} \BibitemShut {NoStop}%
\bibitem [{\citenamefont {Ishibashi}\ \emph {et~al.}(2021)\citenamefont
  {Ishibashi}, \citenamefont {Ohta},\ and\ \citenamefont
  {Yamaguchi}}]{Ishibashi:2021kmf}%
  \BibitemOpen
  \bibfield  {author} {\bibinfo {author} {\bibfnamefont {Akihiro}\ \bibnamefont
  {Ishibashi}}, \bibinfo {author} {\bibfnamefont {Nobuyoshi}\ \bibnamefont
  {Ohta}}, \ and\ \bibinfo {author} {\bibfnamefont {Daiki}\ \bibnamefont
  {Yamaguchi}},\ }\bibfield  {title} {\enquote {\bibinfo {title} {{Quantum
  improved charged black holes}},}\ }\href {\doibase
  10.1103/PhysRevD.104.066016} {\bibfield  {journal} {\bibinfo  {journal}
  {Phys. Rev. D}\ }\textbf {\bibinfo {volume} {104}},\ \bibinfo {pages}
  {066016} (\bibinfo {year} {2021})},\ \Eprint
  {http://arxiv.org/abs/2106.05015} {arXiv:2106.05015 [hep-th]} \BibitemShut
  {NoStop}%
\bibitem [{\citenamefont {Chen}\ \emph {et~al.}(2023)\citenamefont {Chen},
  \citenamefont {Chen}, \citenamefont {Ishibashi},\ and\ \citenamefont
  {Ohta}}]{Chen:2023wdg}%
  \BibitemOpen
  \bibfield  {author} {\bibinfo {author} {\bibfnamefont {Chiang-Mei}\
  \bibnamefont {Chen}}, \bibinfo {author} {\bibfnamefont {Yi}~\bibnamefont
  {Chen}}, \bibinfo {author} {\bibfnamefont {Akihiro}\ \bibnamefont
  {Ishibashi}}, \ and\ \bibinfo {author} {\bibfnamefont {Nobuyoshi}\
  \bibnamefont {Ohta}},\ }\href@noop {} {\enquote {\bibinfo {title} {{Quantum
  Improved Regular Kerr Black Holes}},}\ } (\bibinfo {year} {2023}),\ \Eprint
  {http://arxiv.org/abs/2308.16356} {arXiv:2308.16356 [hep-th]} \BibitemShut
  {NoStop}%
\bibitem [{\citenamefont {Konoplya}(2023)}]{Konoplya:2023bpf}%
  \BibitemOpen
  \bibfield  {author} {\bibinfo {author} {\bibfnamefont {R.~A.}\ \bibnamefont
  {Konoplya}},\ }\href@noop {} {\enquote {\bibinfo {title} {{Hawking radiation
  of renormalization group improved regular black holes}},}\ } (\bibinfo {year}
  {2023}),\ \Eprint {http://arxiv.org/abs/2308.02850} {arXiv:2308.02850
  [gr-qc]} \BibitemShut {NoStop}%
\bibitem [{\citenamefont {Ruiz}\ and\ \citenamefont
  {Tuiran}(2023)}]{Ruiz:2021qfp}%
  \BibitemOpen
  \bibfield  {author} {\bibinfo {author} {\bibfnamefont {O.}~\bibnamefont
  {Ruiz}}\ and\ \bibinfo {author} {\bibfnamefont {E.}~\bibnamefont {Tuiran}},\
  }\bibfield  {title} {\enquote {\bibinfo {title} {{Nonperturbative quantum
  correction to the Reissner-Nordstr\"om spacetime with running
  Newton\textquoteright{}s constant}},}\ }\href {\doibase
  10.1103/PhysRevD.107.066003} {\bibfield  {journal} {\bibinfo  {journal}
  {Phys. Rev. D}\ }\textbf {\bibinfo {volume} {107}},\ \bibinfo {pages}
  {066003} (\bibinfo {year} {2023})},\ \Eprint
  {http://arxiv.org/abs/2112.12519} {arXiv:2112.12519 [gr-qc]} \BibitemShut
  {NoStop}%
\bibitem [{\citenamefont {Rinc\'on}\ and\ \citenamefont
  {Panotopoulos}(2020{\natexlab{a}})}]{Rincon:2020iwy}%
  \BibitemOpen
  \bibfield  {author} {\bibinfo {author} {\bibfnamefont {\'Angel}\ \bibnamefont
  {Rinc\'on}}\ and\ \bibinfo {author} {\bibfnamefont {Grigoris}\ \bibnamefont
  {Panotopoulos}},\ }\bibfield  {title} {\enquote {\bibinfo {title}
  {{Quasinormal modes of an improved Schwarzschild black hole}},}\ }\href
  {\doibase 10.1016/j.dark.2020.100639} {\bibfield  {journal} {\bibinfo
  {journal} {Phys. Dark Univ.}\ }\textbf {\bibinfo {volume} {30}},\ \bibinfo
  {pages} {100639} (\bibinfo {year} {2020}{\natexlab{a}})},\ \Eprint
  {http://arxiv.org/abs/2006.11889} {arXiv:2006.11889 [gr-qc]} \BibitemShut
  {NoStop}%
\bibitem [{\citenamefont {Rinc\'on}\ and\ \citenamefont
  {Panotopoulos}(2018{\natexlab{a}})}]{Rincon:2018sgd}%
  \BibitemOpen
  \bibfield  {author} {\bibinfo {author} {\bibfnamefont {\'Angel}\ \bibnamefont
  {Rinc\'on}}\ and\ \bibinfo {author} {\bibfnamefont {Grigoris}\ \bibnamefont
  {Panotopoulos}},\ }\bibfield  {title} {\enquote {\bibinfo {title}
  {{Quasinormal modes of scale dependent black holes in ( 1+2 )-dimensional
  Einstein-power-Maxwell theory}},}\ }\href {\doibase
  10.1103/PhysRevD.97.024027} {\bibfield  {journal} {\bibinfo  {journal} {Phys.
  Rev. D}\ }\textbf {\bibinfo {volume} {97}},\ \bibinfo {pages} {024027}
  (\bibinfo {year} {2018}{\natexlab{a}})},\ \Eprint
  {http://arxiv.org/abs/1801.03248} {arXiv:1801.03248 [hep-th]} \BibitemShut
  {NoStop}%
\bibitem [{\citenamefont {Rinc\'on}\ \emph {et~al.}(2017)\citenamefont
  {Rinc\'on}, \citenamefont {Contreras}, \citenamefont {Bargue\~no},
  \citenamefont {Koch}, \citenamefont {Panotopoulos},\ and\ \citenamefont
  {Hern\'andez-Arboleda}}]{Rincon:2017goj}%
  \BibitemOpen
  \bibfield  {author} {\bibinfo {author} {\bibfnamefont {\'Angel}\ \bibnamefont
  {Rinc\'on}}, \bibinfo {author} {\bibfnamefont {Ernesto}\ \bibnamefont
  {Contreras}}, \bibinfo {author} {\bibfnamefont {Pedro}\ \bibnamefont
  {Bargue\~no}}, \bibinfo {author} {\bibfnamefont {Benjamin}\ \bibnamefont
  {Koch}}, \bibinfo {author} {\bibfnamefont {Grigorios}\ \bibnamefont
  {Panotopoulos}}, \ and\ \bibinfo {author} {\bibfnamefont {Alejandro}\
  \bibnamefont {Hern\'andez-Arboleda}},\ }\bibfield  {title} {\enquote
  {\bibinfo {title} {{Scale dependent three-dimensional charged black holes in
  linear and non-linear electrodynamics}},}\ }\href {\doibase
  10.1140/epjc/s10052-017-5045-9} {\bibfield  {journal} {\bibinfo  {journal}
  {Eur. Phys. J. C}\ }\textbf {\bibinfo {volume} {77}},\ \bibinfo {pages} {494}
  (\bibinfo {year} {2017})},\ \Eprint {http://arxiv.org/abs/1704.04845}
  {arXiv:1704.04845 [hep-th]} \BibitemShut {NoStop}%
\bibitem [{\citenamefont {Koch}\ \emph {et~al.}(2016)\citenamefont {Koch},
  \citenamefont {Reyes},\ and\ \citenamefont {Rincón}}]{Koch:2016uso}%
  \BibitemOpen
  \bibfield  {author} {\bibinfo {author} {\bibfnamefont {Benjamin}\
  \bibnamefont {Koch}}, \bibinfo {author} {\bibfnamefont {Ignacio~A.}\
  \bibnamefont {Reyes}}, \ and\ \bibinfo {author} {\bibfnamefont {\'Angel}\
  \bibnamefont {Rincón}},\ }\bibfield  {title} {\enquote {\bibinfo {title} {{A
  scale dependent black hole in three-dimensional space–time}},}\ }\href
  {\doibase 10.1088/0264-9381/33/22/225010} {\bibfield  {journal} {\bibinfo
  {journal} {Class. Quant. Grav.}\ }\textbf {\bibinfo {volume} {33}},\ \bibinfo
  {pages} {225010} (\bibinfo {year} {2016})},\ \Eprint
  {http://arxiv.org/abs/1606.04123} {arXiv:1606.04123 [hep-th]} \BibitemShut
  {NoStop}%
%%CITATION = ARXIV:1606.04123;%%
\bibitem [{\citenamefont {Contreras}\ \emph {et~al.}(2019)\citenamefont
  {Contreras}, \citenamefont {Rincón}, \citenamefont {Panotopoulos},
  \citenamefont {Bargueño},\ and\ \citenamefont {Koch}}]{Contreras:2019cmf}%
  \BibitemOpen
  \bibfield  {author} {\bibinfo {author} {\bibfnamefont {Ernesto}\ \bibnamefont
  {Contreras}}, \bibinfo {author} {\bibfnamefont {\'Angel}\ \bibnamefont
  {Rincón}}, \bibinfo {author} {\bibfnamefont {Grigoris}\ \bibnamefont
  {Panotopoulos}}, \bibinfo {author} {\bibfnamefont {Pedro}\ \bibnamefont
  {Bargueño}}, \ and\ \bibinfo {author} {\bibfnamefont {Benjamin}\
  \bibnamefont {Koch}},\ }\href@noop {} {\enquote {\bibinfo {title} {{Black
  hole shadow of a rotating scale--dependent black hole}},}\ } (\bibinfo {year}
  {2019}),\ \Eprint {http://arxiv.org/abs/1906.06990} {arXiv:1906.06990
  [gr-qc]} \BibitemShut {NoStop}%
%%CITATION = ARXIV:1906.06990;%%
\bibitem [{\citenamefont {Rincón}\ \emph {et~al.}(2018)\citenamefont
  {Rincón}, \citenamefont {Contreras}, \citenamefont {Bargueño},
  \citenamefont {Koch},\ and\ \citenamefont {Panotopoulos}}]{Rincon:2018dsq}%
  \BibitemOpen
  \bibfield  {author} {\bibinfo {author} {\bibfnamefont {\'Angel}\ \bibnamefont
  {Rincón}}, \bibinfo {author} {\bibfnamefont {Ernesto}\ \bibnamefont
  {Contreras}}, \bibinfo {author} {\bibfnamefont {Pedro}\ \bibnamefont
  {Bargueño}}, \bibinfo {author} {\bibfnamefont {Benjamin}\ \bibnamefont
  {Koch}}, \ and\ \bibinfo {author} {\bibfnamefont {Grigorios}\ \bibnamefont
  {Panotopoulos}},\ }\bibfield  {title} {\enquote {\bibinfo {title}
  {{Scale-dependent ( $2+1$ )-dimensional electrically charged black holes in
  Einstein-power-Maxwell theory}},}\ }\href {\doibase
  10.1140/epjc/s10052-018-6106-4} {\bibfield  {journal} {\bibinfo  {journal}
  {Eur. Phys. J.}\ }\textbf {\bibinfo {volume} {C78}},\ \bibinfo {pages} {641}
  (\bibinfo {year} {2018})},\ \Eprint {http://arxiv.org/abs/1807.08047}
  {arXiv:1807.08047 [hep-th]} \BibitemShut {NoStop}%
%%CITATION = ARXIV:1807.08047;%%
\bibitem [{\citenamefont {Contreras}\ \emph {et~al.}(2017)\citenamefont
  {Contreras}, \citenamefont {Rinc\'on}, \citenamefont {Koch},\ and\
  \citenamefont {Bargue\~no}}]{Contreras:2017eza}%
  \BibitemOpen
  \bibfield  {author} {\bibinfo {author} {\bibfnamefont {Ernesto}\ \bibnamefont
  {Contreras}}, \bibinfo {author} {\bibfnamefont {\'Angel}\ \bibnamefont
  {Rinc\'on}}, \bibinfo {author} {\bibfnamefont {Benjamin}\ \bibnamefont
  {Koch}}, \ and\ \bibinfo {author} {\bibfnamefont {Pedro}\ \bibnamefont
  {Bargue\~no}},\ }\bibfield  {title} {\enquote {\bibinfo {title} {{A regular
  scale-dependent black hole solution}},}\ }\href {\doibase
  10.1142/S0218271818500323} {\bibfield  {journal} {\bibinfo  {journal} {Int.
  J. Mod. Phys. D}\ }\textbf {\bibinfo {volume} {27}},\ \bibinfo {pages}
  {1850032} (\bibinfo {year} {2017})},\ \Eprint
  {http://arxiv.org/abs/1711.08400} {arXiv:1711.08400 [gr-qc]} \BibitemShut
  {NoStop}%
\bibitem [{\citenamefont {Rincón}\ and\ \citenamefont
  {Koch}(2018)}]{Rincon:2018lyd}%
  \BibitemOpen
  \bibfield  {author} {\bibinfo {author} {\bibfnamefont {\'Angel}\ \bibnamefont
  {Rincón}}\ and\ \bibinfo {author} {\bibfnamefont {Benjamin}\ \bibnamefont
  {Koch}},\ }\bibfield  {title} {\enquote {\bibinfo {title} {{Scale-dependent
  rotating BTZ black hole}},}\ }\href {\doibase 10.1140/epjc/s10052-018-6488-3}
  {\bibfield  {journal} {\bibinfo  {journal} {Eur. Phys. J.}\ }\textbf
  {\bibinfo {volume} {C78}},\ \bibinfo {pages} {1022} (\bibinfo {year}
  {2018})},\ \Eprint {http://arxiv.org/abs/1806.03024} {arXiv:1806.03024
  [hep-th]} \BibitemShut {NoStop}%
%%CITATION = ARXIV:1806.03024;%%
\bibitem [{\citenamefont {Panotopoulos}\ and\ \citenamefont
  {Rinc\'on}(2021)}]{Panotopoulos:2020mii}%
  \BibitemOpen
  \bibfield  {author} {\bibinfo {author} {\bibfnamefont {Grigoris}\
  \bibnamefont {Panotopoulos}}\ and\ \bibinfo {author} {\bibfnamefont
  {\'Angel}\ \bibnamefont {Rinc\'on}},\ }\bibfield  {title} {\enquote {\bibinfo
  {title} {{Quasinormal spectra of scale-dependent Schwarzschild\textendash{}de
  Sitter black holes}},}\ }\href {\doibase 10.1016/j.dark.2020.100743}
  {\bibfield  {journal} {\bibinfo  {journal} {Phys. Dark Univ.}\ }\textbf
  {\bibinfo {volume} {31}},\ \bibinfo {pages} {100743} (\bibinfo {year}
  {2021})},\ \Eprint {http://arxiv.org/abs/2011.02860} {arXiv:2011.02860
  [gr-qc]} \BibitemShut {NoStop}%
\bibitem [{\citenamefont {Fathi}\ \emph {et~al.}(2019)\citenamefont {Fathi},
  \citenamefont {Rinc\'on},\ and\ \citenamefont {Villanueva}}]{Fathi:2019jid}%
  \BibitemOpen
  \bibfield  {author} {\bibinfo {author} {\bibfnamefont {Mohsen}\ \bibnamefont
  {Fathi}}, \bibinfo {author} {\bibfnamefont {\'Angel}\ \bibnamefont
  {Rinc\'on}}, \ and\ \bibinfo {author} {\bibfnamefont {J.~R.}\ \bibnamefont
  {Villanueva}},\ }\href@noop {} {\enquote {\bibinfo {title} {{Photons
  trajectories on a first order scale-dependent static BTZ black hole}},}\ }
  (\bibinfo {year} {2019}),\ \Eprint {http://arxiv.org/abs/1903.09037}
  {arXiv:1903.09037 [gr-qc]} \BibitemShut {NoStop}%
%%CITATION = ARXIV:1903.09037;%%
\bibitem [{\citenamefont {Panotopoulos}\ \emph {et~al.}(2020)\citenamefont
  {Panotopoulos}, \citenamefont {Rinc\'on},\ and\ \citenamefont
  {Lopes}}]{Panotopoulos:2020zqa}%
  \BibitemOpen
  \bibfield  {author} {\bibinfo {author} {\bibfnamefont {Grigoris}\
  \bibnamefont {Panotopoulos}}, \bibinfo {author} {\bibfnamefont {\'Angel}\
  \bibnamefont {Rinc\'on}}, \ and\ \bibinfo {author} {\bibfnamefont
  {Il\'\i{}dio}\ \bibnamefont {Lopes}},\ }\bibfield  {title} {\enquote
  {\bibinfo {title} {{Interior solutions of relativistic stars in the
  scale-dependent scenario}},}\ }\href {\doibase
  10.1140/epjc/s10052-020-7900-3} {\bibfield  {journal} {\bibinfo  {journal}
  {Eur. Phys. J. C}\ }\textbf {\bibinfo {volume} {80}},\ \bibinfo {pages} {318}
  (\bibinfo {year} {2020})},\ \Eprint {http://arxiv.org/abs/2004.02627}
  {arXiv:2004.02627 [gr-qc]} \BibitemShut {NoStop}%
\bibitem [{\citenamefont {\"Ovg\"un}\ \emph {et~al.}(2023)\citenamefont
  {\"Ovg\"un}, \citenamefont {Pantig},\ and\ \citenamefont
  {Rinc\'on}}]{Ovgun:2023ego}%
  \BibitemOpen
  \bibfield  {author} {\bibinfo {author} {\bibfnamefont {Ali}\ \bibnamefont
  {\"Ovg\"un}}, \bibinfo {author} {\bibfnamefont {Reggie~C.}\ \bibnamefont
  {Pantig}}, \ and\ \bibinfo {author} {\bibfnamefont {\'Angel}\ \bibnamefont
  {Rinc\'on}},\ }\bibfield  {title} {\enquote {\bibinfo {title} {{4D
  scale-dependent Schwarzschild-AdS/dS black holes: study of shadow and weak
  deflection angle and greybody bounding}},}\ }\href {\doibase
  10.1140/epjp/s13360-023-03793-w} {\bibfield  {journal} {\bibinfo  {journal}
  {Eur. Phys. J. Plus}\ }\textbf {\bibinfo {volume} {138}},\ \bibinfo {pages}
  {192} (\bibinfo {year} {2023})},\ \Eprint {http://arxiv.org/abs/2303.01696}
  {arXiv:2303.01696 [gr-qc]} \BibitemShut {NoStop}%
\bibitem [{\citenamefont {Rinc\'on}\ \emph {et~al.}(2021)\citenamefont
  {Rinc\'on}, \citenamefont {Contreras}, \citenamefont {Bargue\~no},
  \citenamefont {Koch},\ and\ \citenamefont {Panotopoulos}}]{Rincon:2021hjj}%
  \BibitemOpen
  \bibfield  {author} {\bibinfo {author} {\bibfnamefont {\'Angel}\ \bibnamefont
  {Rinc\'on}}, \bibinfo {author} {\bibfnamefont {Ernesto}\ \bibnamefont
  {Contreras}}, \bibinfo {author} {\bibfnamefont {Pedro}\ \bibnamefont
  {Bargue\~no}}, \bibinfo {author} {\bibfnamefont {Benjamin}\ \bibnamefont
  {Koch}}, \ and\ \bibinfo {author} {\bibfnamefont {Grigoris}\ \bibnamefont
  {Panotopoulos}},\ }\bibfield  {title} {\enquote {\bibinfo {title} {{Four
  dimensional Einstein-power-Maxwell black hole solutions in scale-dependent
  gravity}},}\ }\href {\doibase 10.1016/j.dark.2021.100783} {\bibfield
  {journal} {\bibinfo  {journal} {Phys. Dark Univ.}\ }\textbf {\bibinfo
  {volume} {31}},\ \bibinfo {pages} {100783} (\bibinfo {year} {2021})},\
  \Eprint {http://arxiv.org/abs/2102.02426} {arXiv:2102.02426 [gr-qc]}
  \BibitemShut {NoStop}%
\bibitem [{\citenamefont {Rinc\'on}\ and\ \citenamefont
  {Panotopoulos}(2020{\natexlab{b}})}]{Rincon:2020cpz}%
  \BibitemOpen
  \bibfield  {author} {\bibinfo {author} {\bibfnamefont {\'Angel}\ \bibnamefont
  {Rinc\'on}}\ and\ \bibinfo {author} {\bibfnamefont {Grigoris}\ \bibnamefont
  {Panotopoulos}},\ }\bibfield  {title} {\enquote {\bibinfo {title}
  {{Scale-dependent slowly rotating black holes with flat horizon
  structure}},}\ }\href {\doibase 10.1016/j.dark.2020.100725} {\bibfield
  {journal} {\bibinfo  {journal} {Phys. Dark Univ.}\ }\textbf {\bibinfo
  {volume} {30}},\ \bibinfo {pages} {100725} (\bibinfo {year}
  {2020}{\natexlab{b}})},\ \Eprint {http://arxiv.org/abs/2009.14678}
  {arXiv:2009.14678 [gr-qc]} \BibitemShut {NoStop}%
\bibitem [{\citenamefont {Alvarez}\ \emph {et~al.}(2022)\citenamefont
  {Alvarez}, \citenamefont {Koch}, \citenamefont {Laporte}, \citenamefont
  {Canales},\ and\ \citenamefont {Rinc\'on}}]{Alvarez:2022mlf}%
  \BibitemOpen
  \bibfield  {author} {\bibinfo {author} {\bibfnamefont {Pedro~D.}\
  \bibnamefont {Alvarez}}, \bibinfo {author} {\bibfnamefont {Benjamin}\
  \bibnamefont {Koch}}, \bibinfo {author} {\bibfnamefont {Cristobal}\
  \bibnamefont {Laporte}}, \bibinfo {author} {\bibfnamefont {Felipe}\
  \bibnamefont {Canales}}, \ and\ \bibinfo {author} {\bibfnamefont {\'Angel}\
  \bibnamefont {Rinc\'on}},\ }\bibfield  {title} {\enquote {\bibinfo {title}
  {{Statefinder analysis of scale-dependent cosmology}},}\ }\href {\doibase
  10.1088/1475-7516/2022/10/071} {\bibfield  {journal} {\bibinfo  {journal}
  {JCAP}\ }\textbf {\bibinfo {volume} {10}},\ \bibinfo {pages} {071} (\bibinfo
  {year} {2022})},\ \Eprint {http://arxiv.org/abs/2205.05592} {arXiv:2205.05592
  [gr-qc]} \BibitemShut {NoStop}%
\bibitem [{\citenamefont {Rinc\'on}\ \emph {et~al.}(2023)\citenamefont
  {Rinc\'on}, \citenamefont {Koch}, \citenamefont {Laporte}, \citenamefont
  {Canales},\ and\ \citenamefont {Cruz}}]{Rincon:2022hpy}%
  \BibitemOpen
  \bibfield  {author} {\bibinfo {author} {\bibfnamefont {\'Angel}\ \bibnamefont
  {Rinc\'on}}, \bibinfo {author} {\bibfnamefont {Benjamin}\ \bibnamefont
  {Koch}}, \bibinfo {author} {\bibfnamefont {Cristobal}\ \bibnamefont
  {Laporte}}, \bibinfo {author} {\bibfnamefont {Felipe}\ \bibnamefont
  {Canales}}, \ and\ \bibinfo {author} {\bibfnamefont {Norman}\ \bibnamefont
  {Cruz}},\ }\bibfield  {title} {\enquote {\bibinfo {title} {{The effects of
  running gravitational coupling on three dimensional black holes}},}\ }\href
  {\doibase 10.1140/epjc/s10052-023-11169-8} {\bibfield  {journal} {\bibinfo
  {journal} {Eur. Phys. J. C}\ }\textbf {\bibinfo {volume} {83}},\ \bibinfo
  {pages} {105} (\bibinfo {year} {2023})},\ \Eprint
  {http://arxiv.org/abs/2212.13623} {arXiv:2212.13623 [gr-qc]} \BibitemShut
  {NoStop}%
\bibitem [{\citenamefont {Rinc\'on}\ and\ \citenamefont
  {Bargue\~no}(2023)}]{Rincon:2023kun}%
  \BibitemOpen
  \bibfield  {author} {\bibinfo {author} {\bibfnamefont {\'Angel}\ \bibnamefont
  {Rinc\'on}}\ and\ \bibinfo {author} {\bibfnamefont {Pedro}\ \bibnamefont
  {Bargue\~no}},\ }\bibfield  {title} {\enquote {\bibinfo {title} {{Nariai-like
  black holes in light of scale-dependent gravity}},}\ }\href {\doibase
  10.1140/epjc/s10052-023-12004-w} {\bibfield  {journal} {\bibinfo  {journal}
  {Eur. Phys. J. C}\ }\textbf {\bibinfo {volume} {83}},\ \bibinfo {pages} {836}
  (\bibinfo {year} {2023})}\BibitemShut {NoStop}%
\bibitem [{\citenamefont {Fathi}(2023)}]{Fathi:2023qyl}%
  \BibitemOpen
  \bibfield  {author} {\bibinfo {author} {\bibfnamefont {Mohsen}\ \bibnamefont
  {Fathi}},\ }\bibfield  {title} {\enquote {\bibinfo {title} {{Analytical study
  of particle geodesics around a scale-dependent de Sitter black hole}},}\
  }\href {\doibase 10.1016/j.aop.2023.169401} {\bibfield  {journal} {\bibinfo
  {journal} {Annals Phys.}\ }\textbf {\bibinfo {volume} {457}},\ \bibinfo
  {pages} {169401} (\bibinfo {year} {2023})},\ \Eprint
  {http://arxiv.org/abs/2305.09797} {arXiv:2305.09797 [gr-qc]} \BibitemShut
  {NoStop}%
\bibitem [{\citenamefont {Perez}(2017)}]{Perez:2017cmj}%
  \BibitemOpen
  \bibfield  {author} {\bibinfo {author} {\bibfnamefont {Alejandro}\
  \bibnamefont {Perez}},\ }\bibfield  {title} {\enquote {\bibinfo {title}
  {{Black Holes in Loop Quantum Gravity}},}\ }\href {\doibase
  10.1088/1361-6633/aa7e14} {\bibfield  {journal} {\bibinfo  {journal} {Rept.
  Prog. Phys.}\ }\textbf {\bibinfo {volume} {80}},\ \bibinfo {pages} {126901}
  (\bibinfo {year} {2017})},\ \Eprint {http://arxiv.org/abs/1703.09149}
  {arXiv:1703.09149 [gr-qc]} \BibitemShut {NoStop}%
\bibitem [{\citenamefont {Ashtekar}\ and\ \citenamefont
  {Bianchi}(2021)}]{Ashtekar:2021kfp}%
  \BibitemOpen
  \bibfield  {author} {\bibinfo {author} {\bibfnamefont {Abhay}\ \bibnamefont
  {Ashtekar}}\ and\ \bibinfo {author} {\bibfnamefont {Eugenio}\ \bibnamefont
  {Bianchi}},\ }\bibfield  {title} {\enquote {\bibinfo {title} {{A short review
  of loop quantum gravity}},}\ }\href {\doibase 10.1088/1361-6633/abed91}
  {\bibfield  {journal} {\bibinfo  {journal} {Rept. Prog. Phys.}\ }\textbf
  {\bibinfo {volume} {84}},\ \bibinfo {pages} {042001} (\bibinfo {year}
  {2021})},\ \Eprint {http://arxiv.org/abs/2104.04394} {arXiv:2104.04394
  [gr-qc]} \BibitemShut {NoStop}%
\bibitem [{\citenamefont {Abbott}\ \emph
  {et~al.}(2016{\natexlab{c}})\citenamefont {Abbott} \emph
  {et~al.}}]{LIGOScientific:2016aoc}%
  \BibitemOpen
  \bibfield  {author} {\bibinfo {author} {\bibfnamefont {B.~P.}\ \bibnamefont
  {Abbott}} \emph {et~al.} (\bibinfo {collaboration} {LIGO Scientific,
  Virgo}),\ }\bibfield  {title} {\enquote {\bibinfo {title} {{Observation of
  Gravitational Waves from a Binary Black Hole Merger}},}\ }\href {\doibase
  10.1103/PhysRevLett.116.061102} {\bibfield  {journal} {\bibinfo  {journal}
  {Phys. Rev. Lett.}\ }\textbf {\bibinfo {volume} {116}},\ \bibinfo {pages}
  {061102} (\bibinfo {year} {2016}{\natexlab{c}})},\ \Eprint
  {http://arxiv.org/abs/1602.03837} {arXiv:1602.03837 [gr-qc]} \BibitemShut
  {NoStop}%
\bibitem [{\citenamefont {Kokkotas}\ and\ \citenamefont
  {Schmidt}(1999)}]{Kokkotas:1999bd}%
  \BibitemOpen
  \bibfield  {author} {\bibinfo {author} {\bibfnamefont {Kostas~D.}\
  \bibnamefont {Kokkotas}}\ and\ \bibinfo {author} {\bibfnamefont {Bernd~G.}\
  \bibnamefont {Schmidt}},\ }\bibfield  {title} {\enquote {\bibinfo {title}
  {{Quasinormal modes of stars and black holes}},}\ }\href {\doibase
  10.12942/lrr-1999-2} {\bibfield  {journal} {\bibinfo  {journal} {Living Rev.
  Rel.}\ }\textbf {\bibinfo {volume} {2}},\ \bibinfo {pages} {2} (\bibinfo
  {year} {1999})},\ \Eprint {http://arxiv.org/abs/gr-qc/9909058}
  {arXiv:gr-qc/9909058} \BibitemShut {NoStop}%
\bibitem [{\citenamefont {Alonso-Bardaji}\ \emph
  {et~al.}(2022{\natexlab{a}})\citenamefont {Alonso-Bardaji}, \citenamefont
  {Brizuela},\ and\ \citenamefont {Vera}}]{Alonso-Bardaji:2021yls}%
  \BibitemOpen
  \bibfield  {author} {\bibinfo {author} {\bibfnamefont {Asier}\ \bibnamefont
  {Alonso-Bardaji}}, \bibinfo {author} {\bibfnamefont {David}\ \bibnamefont
  {Brizuela}}, \ and\ \bibinfo {author} {\bibfnamefont {Ra\"ul}\ \bibnamefont
  {Vera}},\ }\bibfield  {title} {\enquote {\bibinfo {title} {{An effective
  model for the quantum Schwarzschild black hole}},}\ }\href {\doibase
  10.1016/j.physletb.2022.137075} {\bibfield  {journal} {\bibinfo  {journal}
  {Phys. Lett. B}\ }\textbf {\bibinfo {volume} {829}},\ \bibinfo {pages}
  {137075} (\bibinfo {year} {2022}{\natexlab{a}})},\ \Eprint
  {http://arxiv.org/abs/2112.12110} {arXiv:2112.12110 [gr-qc]} \BibitemShut
  {NoStop}%
\bibitem [{\citenamefont {Alonso-Bardaji}\ \emph {et~al.}(2023)\citenamefont
  {Alonso-Bardaji}, \citenamefont {Brizuela},\ and\ \citenamefont
  {Vera}}]{Alonso-Bardaji:2023niu}%
  \BibitemOpen
  \bibfield  {author} {\bibinfo {author} {\bibfnamefont {Asier}\ \bibnamefont
  {Alonso-Bardaji}}, \bibinfo {author} {\bibfnamefont {David}\ \bibnamefont
  {Brizuela}}, \ and\ \bibinfo {author} {\bibfnamefont {Ra\"ul}\ \bibnamefont
  {Vera}},\ }\href@noop {} {\enquote {\bibinfo {title} {{Singularity resolution
  by holonomy corrections: Spherical charged black holes in cosmological
  backgrounds}},}\ } (\bibinfo {year} {2023}),\ \Eprint
  {http://arxiv.org/abs/2302.10619} {arXiv:2302.10619 [gr-qc]} \BibitemShut
  {NoStop}%
\bibitem [{\citenamefont {Vagnozzi}\ \emph {et~al.}(2023)\citenamefont
  {Vagnozzi} \emph {et~al.}}]{Vagnozzi:2022moj}%
  \BibitemOpen
  \bibfield  {author} {\bibinfo {author} {\bibfnamefont {Sunny}\ \bibnamefont
  {Vagnozzi}} \emph {et~al.},\ }\bibfield  {title} {\enquote {\bibinfo {title}
  {{Horizon-scale tests of gravity theories and fundamental physics from the
  Event Horizon Telescope image of Sagittarius A$^*$}},}\ }\href {\doibase
  10.1088/1361-6382/acd97b} {\bibfield  {journal} {\bibinfo  {journal} {Class.
  Quant. Grav.}\ }\textbf {\bibinfo {volume} {40}},\ \bibinfo {pages} {165007}
  (\bibinfo {year} {2023})},\ \Eprint {http://arxiv.org/abs/2205.07787}
  {arXiv:2205.07787 [gr-qc]} \BibitemShut {NoStop}%
\bibitem [{\citenamefont {Boehmer}\ and\ \citenamefont
  {Vandersloot}(2007)}]{Boehmer:2007ket}%
  \BibitemOpen
  \bibfield  {author} {\bibinfo {author} {\bibfnamefont {Christian~G.}\
  \bibnamefont {Boehmer}}\ and\ \bibinfo {author} {\bibfnamefont {Kevin}\
  \bibnamefont {Vandersloot}},\ }\bibfield  {title} {\enquote {\bibinfo {title}
  {{Loop Quantum Dynamics of the Schwarzschild Interior}},}\ }\href {\doibase
  10.1103/PhysRevD.76.104030} {\bibfield  {journal} {\bibinfo  {journal} {Phys.
  Rev. D}\ }\textbf {\bibinfo {volume} {76}},\ \bibinfo {pages} {104030}
  (\bibinfo {year} {2007})},\ \Eprint {http://arxiv.org/abs/0709.2129}
  {arXiv:0709.2129 [gr-qc]} \BibitemShut {NoStop}%
\bibitem [{\citenamefont {Gambini}\ and\ \citenamefont
  {Pullin}(2008)}]{Gambini:2008dy}%
  \BibitemOpen
  \bibfield  {author} {\bibinfo {author} {\bibfnamefont {Rodolfo}\ \bibnamefont
  {Gambini}}\ and\ \bibinfo {author} {\bibfnamefont {Jorge}\ \bibnamefont
  {Pullin}},\ }\bibfield  {title} {\enquote {\bibinfo {title} {{Black holes in
  loop quantum gravity: The Complete space-time}},}\ }\href {\doibase
  10.1103/PhysRevLett.101.161301} {\bibfield  {journal} {\bibinfo  {journal}
  {Phys. Rev. Lett.}\ }\textbf {\bibinfo {volume} {101}},\ \bibinfo {pages}
  {161301} (\bibinfo {year} {2008})},\ \Eprint {http://arxiv.org/abs/0805.1187}
  {arXiv:0805.1187 [gr-qc]} \BibitemShut {NoStop}%
\bibitem [{\citenamefont {Gambini}\ \emph {et~al.}(2022)\citenamefont
  {Gambini}, \citenamefont {Ben\'\i{}tez},\ and\ \citenamefont
  {Pullin}}]{Gambini:2021uzf}%
  \BibitemOpen
  \bibfield  {author} {\bibinfo {author} {\bibfnamefont {Rodolfo}\ \bibnamefont
  {Gambini}}, \bibinfo {author} {\bibfnamefont {Florencia}\ \bibnamefont
  {Ben\'\i{}tez}}, \ and\ \bibinfo {author} {\bibfnamefont {Jorge}\
  \bibnamefont {Pullin}},\ }\bibfield  {title} {\enquote {\bibinfo {title} {{A
  Covariant Polymerized Scalar Field in Semi-Classical Loop Quantum
  Gravity}},}\ }\href {\doibase 10.3390/universe8100526} {\bibfield  {journal}
  {\bibinfo  {journal} {Universe}\ }\textbf {\bibinfo {volume} {8}},\ \bibinfo
  {pages} {526} (\bibinfo {year} {2022})},\ \Eprint
  {http://arxiv.org/abs/2102.09501} {arXiv:2102.09501 [gr-qc]} \BibitemShut
  {NoStop}%
\bibitem [{\citenamefont {Alonso-Bardaji}\ \emph
  {et~al.}(2022{\natexlab{b}})\citenamefont {Alonso-Bardaji}, \citenamefont
  {Brizuela},\ and\ \citenamefont {Vera}}]{Alonso-Bardaji:2022ear}%
  \BibitemOpen
  \bibfield  {author} {\bibinfo {author} {\bibfnamefont {Asier}\ \bibnamefont
  {Alonso-Bardaji}}, \bibinfo {author} {\bibfnamefont {David}\ \bibnamefont
  {Brizuela}}, \ and\ \bibinfo {author} {\bibfnamefont {Ra\"ul}\ \bibnamefont
  {Vera}},\ }\bibfield  {title} {\enquote {\bibinfo {title} {{Nonsingular
  spherically symmetric black-hole model with holonomy corrections}},}\ }\href
  {\doibase 10.1103/PhysRevD.106.024035} {\bibfield  {journal} {\bibinfo
  {journal} {Phys. Rev. D}\ }\textbf {\bibinfo {volume} {106}},\ \bibinfo
  {pages} {024035} (\bibinfo {year} {2022}{\natexlab{b}})},\ \Eprint
  {http://arxiv.org/abs/2205.02098} {arXiv:2205.02098 [gr-qc]} \BibitemShut
  {NoStop}%
\bibitem [{\citenamefont {Soares}\ \emph {et~al.}(2023)\citenamefont {Soares},
  \citenamefont {Pereira}, \citenamefont {Vit\'oria},\ and\ \citenamefont
  {Rocha}}]{Soares:2023uup}%
  \BibitemOpen
  \bibfield  {author} {\bibinfo {author} {\bibfnamefont {A.~R.}\ \bibnamefont
  {Soares}}, \bibinfo {author} {\bibfnamefont {C.~F.~S.}\ \bibnamefont
  {Pereira}}, \bibinfo {author} {\bibfnamefont {R.~L.~L.}\ \bibnamefont
  {Vit\'oria}}, \ and\ \bibinfo {author} {\bibfnamefont {Erick~Melo}\
  \bibnamefont {Rocha}},\ }\bibfield  {title} {\enquote {\bibinfo {title}
  {{Holonomy corrected Schwarzschild black hole lensing}},}\ }\href {\doibase
  10.1103/PhysRevD.108.124024} {\bibfield  {journal} {\bibinfo  {journal}
  {Phys. Rev. D}\ }\textbf {\bibinfo {volume} {108}},\ \bibinfo {pages}
  {124024} (\bibinfo {year} {2023})},\ \Eprint
  {http://arxiv.org/abs/2309.05106} {arXiv:2309.05106 [gr-qc]} \BibitemShut
  {NoStop}%
\bibitem [{\citenamefont {Junior}\ \emph {et~al.}(2024)\citenamefont {Junior},
  \citenamefont {Lobo}, \citenamefont {Rodrigues},\ and\ \citenamefont
  {Vieira}}]{Junior:2023xgl}%
  \BibitemOpen
  \bibfield  {author} {\bibinfo {author} {\bibfnamefont {Ednaldo L.~B.}\
  \bibnamefont {Junior}}, \bibinfo {author} {\bibfnamefont {Francisco S.~N.}\
  \bibnamefont {Lobo}}, \bibinfo {author} {\bibfnamefont {Manuel~E.}\
  \bibnamefont {Rodrigues}}, \ and\ \bibinfo {author} {\bibfnamefont
  {Henrique~A.}\ \bibnamefont {Vieira}},\ }\bibfield  {title} {\enquote
  {\bibinfo {title} {{Gravitational lens effect of a holonomy corrected
  Schwarzschild black hole}},}\ }\href {\doibase 10.1103/PhysRevD.109.024004}
  {\bibfield  {journal} {\bibinfo  {journal} {Phys. Rev. D}\ }\textbf {\bibinfo
  {volume} {109}},\ \bibinfo {pages} {024004} (\bibinfo {year} {2024})},\
  \Eprint {http://arxiv.org/abs/2309.02658} {arXiv:2309.02658 [gr-qc]}
  \BibitemShut {NoStop}%
\bibitem [{\citenamefont {Jha}(2023)}]{Jha:2023rem}%
  \BibitemOpen
  \bibfield  {author} {\bibinfo {author} {\bibfnamefont {Sohan~Kumar}\
  \bibnamefont {Jha}},\ }\bibfield  {title} {\enquote {\bibinfo {title}
  {{Shadow, quasinormal modes, greybody bounds, and Hawking sparsity of loop
  quantum gravity motivated non-rotating black hole}},}\ }\href {\doibase
  10.1140/epjc/s10052-023-12123-4} {\bibfield  {journal} {\bibinfo  {journal}
  {Eur. Phys. J. C}\ }\textbf {\bibinfo {volume} {83}},\ \bibinfo {pages} {952}
  (\bibinfo {year} {2023})},\ \Eprint {http://arxiv.org/abs/2310.04759}
  {arXiv:2310.04759 [gr-qc]} \BibitemShut {NoStop}%
\bibitem [{\citenamefont {Moreira}\ \emph {et~al.}(2023)\citenamefont
  {Moreira}, \citenamefont {Lima~Junior}, \citenamefont {Crispino},\ and\
  \citenamefont {Herdeiro}}]{Moreira:2023cxy}%
  \BibitemOpen
  \bibfield  {author} {\bibinfo {author} {\bibfnamefont {Zeus~S.}\ \bibnamefont
  {Moreira}}, \bibinfo {author} {\bibfnamefont {Haroldo C.~D.}\ \bibnamefont
  {Lima~Junior}}, \bibinfo {author} {\bibfnamefont {Lu\'\i{}s C.~B.}\
  \bibnamefont {Crispino}}, \ and\ \bibinfo {author} {\bibfnamefont {Carlos
  A.~R.}\ \bibnamefont {Herdeiro}},\ }\bibfield  {title} {\enquote {\bibinfo
  {title} {{Quasinormal modes of a holonomy corrected Schwarzschild black
  hole}},}\ }\href {\doibase 10.1103/PhysRevD.107.104016} {\bibfield  {journal}
  {\bibinfo  {journal} {Phys. Rev. D}\ }\textbf {\bibinfo {volume} {107}},\
  \bibinfo {pages} {104016} (\bibinfo {year} {2023})},\ \Eprint
  {http://arxiv.org/abs/2302.14722} {arXiv:2302.14722 [gr-qc]} \BibitemShut
  {NoStop}%
\bibitem [{\citenamefont {Beig}(1978)}]{BEIG1978153}%
  \BibitemOpen
  \bibfield  {author} {\bibinfo {author} {\bibfnamefont {R.}~\bibnamefont
  {Beig}},\ }\bibfield  {title} {\enquote {\bibinfo {title}
  {Arnowitt-deser-misner energy and g00},}\ }\href {\doibase
  https://doi.org/10.1016/0375-9601(78)90198-6} {\bibfield  {journal} {\bibinfo
   {journal} {Physics Letters A}\ }\textbf {\bibinfo {volume} {69}},\ \bibinfo
  {pages} {153--155} (\bibinfo {year} {1978})}\BibitemShut {NoStop}%
\bibitem [{\citenamefont {Hayward}(1996)}]{Hayward:1994bu}%
  \BibitemOpen
  \bibfield  {author} {\bibinfo {author} {\bibfnamefont {Sean~A.}\ \bibnamefont
  {Hayward}},\ }\bibfield  {title} {\enquote {\bibinfo {title} {{Gravitational
  energy in spherical symmetry}},}\ }\href {\doibase 10.1103/PhysRevD.53.1938}
  {\bibfield  {journal} {\bibinfo  {journal} {Phys. Rev. D}\ }\textbf {\bibinfo
  {volume} {53}},\ \bibinfo {pages} {1938--1949} (\bibinfo {year} {1996})},\
  \Eprint {http://arxiv.org/abs/gr-qc/9408002} {arXiv:gr-qc/9408002}
  \BibitemShut {NoStop}%
\bibitem [{\citenamefont {Li}\ \emph {et~al.}(2020)\citenamefont {Li},
  \citenamefont {Zhang},\ and\ \citenamefont {\"Ovg\"un}}]{Li:2020wvn}%
  \BibitemOpen
  \bibfield  {author} {\bibinfo {author} {\bibfnamefont {Zonghai}\ \bibnamefont
  {Li}}, \bibinfo {author} {\bibfnamefont {Guodong}\ \bibnamefont {Zhang}}, \
  and\ \bibinfo {author} {\bibfnamefont {Ali}\ \bibnamefont {\"Ovg\"un}},\
  }\bibfield  {title} {\enquote {\bibinfo {title} {{Circular Orbit of a
  Particle and Weak Gravitational Lensing}},}\ }\href {\doibase
  10.1103/PhysRevD.101.124058} {\bibfield  {journal} {\bibinfo  {journal}
  {Phys. Rev. D}\ }\textbf {\bibinfo {volume} {101}},\ \bibinfo {pages}
  {124058} (\bibinfo {year} {2020})},\ \Eprint
  {http://arxiv.org/abs/2006.13047} {arXiv:2006.13047 [gr-qc]} \BibitemShut
  {NoStop}%
\bibitem [{\citenamefont {Gibbons}\ and\ \citenamefont
  {Werner}(2008)}]{Gibbons:2008rj}%
  \BibitemOpen
  \bibfield  {author} {\bibinfo {author} {\bibfnamefont {G.~W.}\ \bibnamefont
  {Gibbons}}\ and\ \bibinfo {author} {\bibfnamefont {M.~C.}\ \bibnamefont
  {Werner}},\ }\bibfield  {title} {\enquote {\bibinfo {title} {{Applications of
  the Gauss-Bonnet theorem to gravitational lensing}},}\ }\href {\doibase
  10.1088/0264-9381/25/23/235009} {\bibfield  {journal} {\bibinfo  {journal}
  {Class. Quant. Grav.}\ }\textbf {\bibinfo {volume} {25}},\ \bibinfo {pages}
  {235009} (\bibinfo {year} {2008})},\ \Eprint {http://arxiv.org/abs/0807.0854}
  {arXiv:0807.0854 [gr-qc]} \BibitemShut {NoStop}%
\bibitem [{\citenamefont {Ishihara}\ \emph {et~al.}(2016)\citenamefont
  {Ishihara}, \citenamefont {Suzuki}, \citenamefont {Ono}, \citenamefont
  {Kitamura},\ and\ \citenamefont {Asada}}]{Ishihara:2016vdc}%
  \BibitemOpen
  \bibfield  {author} {\bibinfo {author} {\bibfnamefont {Asahi}\ \bibnamefont
  {Ishihara}}, \bibinfo {author} {\bibfnamefont {Yusuke}\ \bibnamefont
  {Suzuki}}, \bibinfo {author} {\bibfnamefont {Toshiaki}\ \bibnamefont {Ono}},
  \bibinfo {author} {\bibfnamefont {Takao}\ \bibnamefont {Kitamura}}, \ and\
  \bibinfo {author} {\bibfnamefont {Hideki}\ \bibnamefont {Asada}},\ }\bibfield
   {title} {\enquote {\bibinfo {title} {{Gravitational bending angle of light
  for finite distance and the Gauss-Bonnet theorem}},}\ }\href {\doibase
  10.1103/PhysRevD.94.084015} {\bibfield  {journal} {\bibinfo  {journal} {Phys.
  Rev. D}\ }\textbf {\bibinfo {volume} {94}},\ \bibinfo {pages} {084015}
  (\bibinfo {year} {2016})},\ \Eprint {http://arxiv.org/abs/1604.08308}
  {arXiv:1604.08308 [gr-qc]} \BibitemShut {NoStop}%
\bibitem [{\citenamefont {Bozza}\ and\ \citenamefont
  {Mancini}(2009)}]{Bozza:2008zr}%
  \BibitemOpen
  \bibfield  {author} {\bibinfo {author} {\bibfnamefont {Valerio}\ \bibnamefont
  {Bozza}}\ and\ \bibinfo {author} {\bibfnamefont {Luigi}\ \bibnamefont
  {Mancini}},\ }\bibfield  {title} {\enquote {\bibinfo {title} {{Gravitational
  lensing of stars orbiting the Massive Black Hole in the Galactic Center}},}\
  }\href {\doibase 10.1088/0004-637X/696/1/701} {\bibfield  {journal} {\bibinfo
   {journal} {Astrophys. J.}\ }\textbf {\bibinfo {volume} {696}},\ \bibinfo
  {pages} {701--705} (\bibinfo {year} {2009})},\ \Eprint
  {http://arxiv.org/abs/0812.3853} {arXiv:0812.3853 [astro-ph]} \BibitemShut
  {NoStop}%
\bibitem [{\citenamefont {Virbhadra}\ and\ \citenamefont
  {Ellis}(2000)}]{Virbhadra:1999nm}%
  \BibitemOpen
  \bibfield  {author} {\bibinfo {author} {\bibfnamefont {K.~S.}\ \bibnamefont
  {Virbhadra}}\ and\ \bibinfo {author} {\bibfnamefont {George F.~R.}\
  \bibnamefont {Ellis}},\ }\bibfield  {title} {\enquote {\bibinfo {title}
  {{Schwarzschild black hole lensing}},}\ }\href {\doibase
  10.1103/PhysRevD.62.084003} {\bibfield  {journal} {\bibinfo  {journal} {Phys.
  Rev. D}\ }\textbf {\bibinfo {volume} {62}},\ \bibinfo {pages} {084003}
  (\bibinfo {year} {2000})},\ \Eprint {http://arxiv.org/abs/astro-ph/9904193}
  {arXiv:astro-ph/9904193} \BibitemShut {NoStop}%
\bibitem [{\citenamefont {Virbhadra}\ and\ \citenamefont
  {Ellis}(2002)}]{Virbhadra:2002ju}%
  \BibitemOpen
  \bibfield  {author} {\bibinfo {author} {\bibfnamefont {K.~S.}\ \bibnamefont
  {Virbhadra}}\ and\ \bibinfo {author} {\bibfnamefont {G.~F.~R.}\ \bibnamefont
  {Ellis}},\ }\bibfield  {title} {\enquote {\bibinfo {title} {{Gravitational
  lensing by naked singularities}},}\ }\href {\doibase
  10.1103/PhysRevD.65.103004} {\bibfield  {journal} {\bibinfo  {journal} {Phys.
  Rev. D}\ }\textbf {\bibinfo {volume} {65}},\ \bibinfo {pages} {103004}
  (\bibinfo {year} {2002})}\BibitemShut {NoStop}%
\bibitem [{\citenamefont {Bozza}\ \emph {et~al.}(2001)\citenamefont {Bozza},
  \citenamefont {Capozziello}, \citenamefont {Iovane},\ and\ \citenamefont
  {Scarpetta}}]{Bozza:2001xd}%
  \BibitemOpen
  \bibfield  {author} {\bibinfo {author} {\bibfnamefont {V.}~\bibnamefont
  {Bozza}}, \bibinfo {author} {\bibfnamefont {S.}~\bibnamefont {Capozziello}},
  \bibinfo {author} {\bibfnamefont {G.}~\bibnamefont {Iovane}}, \ and\ \bibinfo
  {author} {\bibfnamefont {G.}~\bibnamefont {Scarpetta}},\ }\bibfield  {title}
  {\enquote {\bibinfo {title} {{Strong field limit of black hole gravitational
  lensing}},}\ }\href {\doibase 10.1023/A:1012292927358} {\bibfield  {journal}
  {\bibinfo  {journal} {Gen. Rel. Grav.}\ }\textbf {\bibinfo {volume} {33}},\
  \bibinfo {pages} {1535--1548} (\bibinfo {year} {2001})},\ \Eprint
  {http://arxiv.org/abs/gr-qc/0102068} {arXiv:gr-qc/0102068} \BibitemShut
  {NoStop}%
\bibitem [{\citenamefont {Bozza}(2002)}]{Bozza:2002zj}%
  \BibitemOpen
  \bibfield  {author} {\bibinfo {author} {\bibfnamefont {V.}~\bibnamefont
  {Bozza}},\ }\bibfield  {title} {\enquote {\bibinfo {title} {{Gravitational
  lensing in the strong field limit}},}\ }\href {\doibase
  10.1103/PhysRevD.66.103001} {\bibfield  {journal} {\bibinfo  {journal} {Phys.
  Rev. D}\ }\textbf {\bibinfo {volume} {66}},\ \bibinfo {pages} {103001}
  (\bibinfo {year} {2002})},\ \Eprint {http://arxiv.org/abs/gr-qc/0208075}
  {arXiv:gr-qc/0208075} \BibitemShut {NoStop}%
\bibitem [{\citenamefont {Hasse}\ and\ \citenamefont
  {Perlick}(2002)}]{Hasse:2001by}%
  \BibitemOpen
  \bibfield  {author} {\bibinfo {author} {\bibfnamefont {Wolfgang}\
  \bibnamefont {Hasse}}\ and\ \bibinfo {author} {\bibfnamefont {Volker}\
  \bibnamefont {Perlick}},\ }\bibfield  {title} {\enquote {\bibinfo {title}
  {{Gravitational lensing in spherically symmetric static space-times with
  centrifugal force reversal}},}\ }\href {\doibase 10.1023/A:1015384604371}
  {\bibfield  {journal} {\bibinfo  {journal} {Gen. Rel. Grav.}\ }\textbf
  {\bibinfo {volume} {34}},\ \bibinfo {pages} {415--433} (\bibinfo {year}
  {2002})},\ \Eprint {http://arxiv.org/abs/gr-qc/0108002} {arXiv:gr-qc/0108002}
  \BibitemShut {NoStop}%
\bibitem [{\citenamefont {Perlick}(2004)}]{Perlick:2003vg}%
  \BibitemOpen
  \bibfield  {author} {\bibinfo {author} {\bibfnamefont {Volker}\ \bibnamefont
  {Perlick}},\ }\bibfield  {title} {\enquote {\bibinfo {title} {{On the Exact
  gravitational lens equation in spherically symmetric and static
  space-times}},}\ }\href {\doibase 10.1103/PhysRevD.69.064017} {\bibfield
  {journal} {\bibinfo  {journal} {Phys. Rev. D}\ }\textbf {\bibinfo {volume}
  {69}},\ \bibinfo {pages} {064017} (\bibinfo {year} {2004})},\ \Eprint
  {http://arxiv.org/abs/gr-qc/0307072} {arXiv:gr-qc/0307072} \BibitemShut
  {NoStop}%
\bibitem [{\citenamefont {Atamurotov}\ \emph {et~al.}(2024)\citenamefont
  {Atamurotov}, \citenamefont {Yunusov}, \citenamefont {Abdujabbarov},\ and\
  \citenamefont {Mustafa}}]{Atamurotov:2023rye}%
  \BibitemOpen
  \bibfield  {author} {\bibinfo {author} {\bibfnamefont {Farruh}\ \bibnamefont
  {Atamurotov}}, \bibinfo {author} {\bibfnamefont {Odil}\ \bibnamefont
  {Yunusov}}, \bibinfo {author} {\bibfnamefont {Ahmadjon}\ \bibnamefont
  {Abdujabbarov}}, \ and\ \bibinfo {author} {\bibfnamefont {G.}~\bibnamefont
  {Mustafa}},\ }\bibfield  {title} {\enquote {\bibinfo {title} {{Gravitational
  weak lensing of hairy black hole in presence of plasma}},}\ }\href {\doibase
  10.1016/j.newast.2023.102098} {\bibfield  {journal} {\bibinfo  {journal} {New
  Astron.}\ }\textbf {\bibinfo {volume} {105}},\ \bibinfo {pages} {102098}
  (\bibinfo {year} {2024})}\BibitemShut {NoStop}%
\bibitem [{\citenamefont {Abdujabbarov}\ \emph {et~al.}(2017)\citenamefont
  {Abdujabbarov}, \citenamefont {Ahmedov}, \citenamefont {Dadhich},\ and\
  \citenamefont {Atamurotov}}]{Abdujabbarov:2017pfw}%
  \BibitemOpen
  \bibfield  {author} {\bibinfo {author} {\bibfnamefont {Ahmadjon}\
  \bibnamefont {Abdujabbarov}}, \bibinfo {author} {\bibfnamefont {Bobomurat}\
  \bibnamefont {Ahmedov}}, \bibinfo {author} {\bibfnamefont {Naresh}\
  \bibnamefont {Dadhich}}, \ and\ \bibinfo {author} {\bibfnamefont {Farruh}\
  \bibnamefont {Atamurotov}},\ }\bibfield  {title} {\enquote {\bibinfo {title}
  {{Optical properties of a braneworld black hole: Gravitational lensing and
  retrolensing}},}\ }\href {\doibase 10.1103/PhysRevD.96.084017} {\bibfield
  {journal} {\bibinfo  {journal} {Phys. Rev. D}\ }\textbf {\bibinfo {volume}
  {96}},\ \bibinfo {pages} {084017} (\bibinfo {year} {2017})}\BibitemShut
  {NoStop}%
\bibitem [{\citenamefont {Atamurotov}\ \emph {et~al.}(2022)\citenamefont
  {Atamurotov}, \citenamefont {Alloqulov}, \citenamefont {Abdujabbarov},\ and\
  \citenamefont {Ahmedov}}]{Atamurotov:2022ahu}%
  \BibitemOpen
  \bibfield  {author} {\bibinfo {author} {\bibfnamefont {Farruh}\ \bibnamefont
  {Atamurotov}}, \bibinfo {author} {\bibfnamefont {Mirzabek}\ \bibnamefont
  {Alloqulov}}, \bibinfo {author} {\bibfnamefont {Ahmadjon}\ \bibnamefont
  {Abdujabbarov}}, \ and\ \bibinfo {author} {\bibfnamefont {Bobomurat}\
  \bibnamefont {Ahmedov}},\ }\bibfield  {title} {\enquote {\bibinfo {title}
  {{Testing the Einstein-\AE{}ther gravity: particle dynamics and gravitational
  lensing}},}\ }\href {\doibase 10.1140/epjp/s13360-022-02846-w} {\bibfield
  {journal} {\bibinfo  {journal} {Eur. Phys. J. Plus}\ }\textbf {\bibinfo
  {volume} {137}},\ \bibinfo {pages} {634} (\bibinfo {year}
  {2022})}\BibitemShut {NoStop}%
\bibitem [{\citenamefont {Atamurotov}\ \emph {et~al.}(2023)\citenamefont
  {Atamurotov}, \citenamefont {Alibekov}, \citenamefont {Abdujabbarov},
  \citenamefont {Mustafa},\ and\ \citenamefont {Aripov}}]{Atamurotov:2023tff}%
  \BibitemOpen
  \bibfield  {author} {\bibinfo {author} {\bibfnamefont {Farruh}\ \bibnamefont
  {Atamurotov}}, \bibinfo {author} {\bibfnamefont {Husan}\ \bibnamefont
  {Alibekov}}, \bibinfo {author} {\bibfnamefont {Ahmadjon}\ \bibnamefont
  {Abdujabbarov}}, \bibinfo {author} {\bibfnamefont {Ghulam}\ \bibnamefont
  {Mustafa}}, \ and\ \bibinfo {author} {\bibfnamefont {Mersaid~M.}\
  \bibnamefont {Aripov}},\ }\bibfield  {title} {\enquote {\bibinfo {title}
  {{Weak Gravitational Lensing around Bardeen Black Hole with a String Cloud in
  the Presence of Plasma}},}\ }\href {\doibase 10.3390/sym15040848} {\bibfield
  {journal} {\bibinfo  {journal} {Symmetry}\ }\textbf {\bibinfo {volume}
  {15}},\ \bibinfo {pages} {848} (\bibinfo {year} {2023})}\BibitemShut
  {NoStop}%
\bibitem [{\citenamefont {Hoshimov}\ \emph {et~al.}(2024)\citenamefont
  {Hoshimov}, \citenamefont {Yunusov}, \citenamefont {Atamurotov},
  \citenamefont {Jamil},\ and\ \citenamefont
  {Abdujabbarov}}]{Hoshimov:2023tlz}%
  \BibitemOpen
  \bibfield  {author} {\bibinfo {author} {\bibfnamefont {Husanboy}\
  \bibnamefont {Hoshimov}}, \bibinfo {author} {\bibfnamefont {Odil}\
  \bibnamefont {Yunusov}}, \bibinfo {author} {\bibfnamefont {Farruh}\
  \bibnamefont {Atamurotov}}, \bibinfo {author} {\bibfnamefont {Mubasher}\
  \bibnamefont {Jamil}}, \ and\ \bibinfo {author} {\bibfnamefont {Ahmadjon}\
  \bibnamefont {Abdujabbarov}},\ }\bibfield  {title} {\enquote {\bibinfo
  {title} {{Weak gravitational lensing and shadow of a GUP-modified
  Schwarzschild black hole in the presence of plasma}},}\ }\href {\doibase
  10.1016/j.dark.2023.101392} {\bibfield  {journal} {\bibinfo  {journal} {Phys.
  Dark Univ.}\ }\textbf {\bibinfo {volume} {43}},\ \bibinfo {pages} {101392}
  (\bibinfo {year} {2024})},\ \Eprint {http://arxiv.org/abs/2312.10678}
  {arXiv:2312.10678 [gr-qc]} \BibitemShut {NoStop}%
\bibitem [{\citenamefont {Parbin}\ \emph {et~al.}(2023)\citenamefont {Parbin},
  \citenamefont {Gogoi},\ and\ \citenamefont {Goswami}}]{Parbin:2023zik}%
  \BibitemOpen
  \bibfield  {author} {\bibinfo {author} {\bibfnamefont {Nashiba}\ \bibnamefont
  {Parbin}}, \bibinfo {author} {\bibfnamefont {Dhruba~Jyoti}\ \bibnamefont
  {Gogoi}}, \ and\ \bibinfo {author} {\bibfnamefont {Umananda~Dev}\
  \bibnamefont {Goswami}},\ }\bibfield  {title} {\enquote {\bibinfo {title}
  {{Weak gravitational lensing and shadow cast by rotating black holes in
  axionic Chern\textendash{}Simons theory}},}\ }\href {\doibase
  10.1016/j.dark.2023.101265} {\bibfield  {journal} {\bibinfo  {journal} {Phys.
  Dark Univ.}\ }\textbf {\bibinfo {volume} {41}},\ \bibinfo {pages} {101265}
  (\bibinfo {year} {2023})},\ \Eprint {http://arxiv.org/abs/2305.09157}
  {arXiv:2305.09157 [gr-qc]} \BibitemShut {NoStop}%
\bibitem [{\citenamefont {Mushtaq}\ \emph {et~al.}(2024)\citenamefont
  {Mushtaq}, \citenamefont {Tiecheng}, \citenamefont {Yasir},\ and\
  \citenamefont {Ahsan}}]{Mushtaq:2024utq}%
  \BibitemOpen
  \bibfield  {author} {\bibinfo {author} {\bibfnamefont {Farzan}\ \bibnamefont
  {Mushtaq}}, \bibinfo {author} {\bibfnamefont {Xia}\ \bibnamefont {Tiecheng}},
  \bibinfo {author} {\bibfnamefont {Muhammad}\ \bibnamefont {Yasir}}, \ and\
  \bibinfo {author} {\bibfnamefont {Aitazaz}\ \bibnamefont {Ahsan}},\
  }\bibfield  {title} {\enquote {\bibinfo {title} {{Weak gravitational lensing
  by multi-horizons black hole}},}\ }\href {\doibase 10.1209/0295-5075/ad2729}
  {\bibfield  {journal} {\bibinfo  {journal} {EPL}\ }\textbf {\bibinfo {volume}
  {145}},\ \bibinfo {pages} {59002} (\bibinfo {year} {2024})}\BibitemShut
  {NoStop}%
\bibitem [{\citenamefont {Al-Badawi}\ \emph {et~al.}(2024)\citenamefont
  {Al-Badawi}, \citenamefont {Alloqulov}, \citenamefont {Shaymatov},\ and\
  \citenamefont {Ahmedov}}]{Al-Badawi:2024dzc}%
  \BibitemOpen
  \bibfield  {author} {\bibinfo {author} {\bibfnamefont {Ahmad}\ \bibnamefont
  {Al-Badawi}}, \bibinfo {author} {\bibfnamefont {Mirzabek}\ \bibnamefont
  {Alloqulov}}, \bibinfo {author} {\bibfnamefont {Sanjar}\ \bibnamefont
  {Shaymatov}}, \ and\ \bibinfo {author} {\bibfnamefont {Bobomurat}\
  \bibnamefont {Ahmedov}},\ }\bibfield  {title} {\enquote {\bibinfo {title}
  {{Shadows and weak gravitational lensing for black holes within
  Einstein-Maxwell-scalar theory*}},}\ }\href {\doibase
  10.1088/1674-1137/ad5a70} {\bibfield  {journal} {\bibinfo  {journal} {Chin.
  Phys. C}\ }\textbf {\bibinfo {volume} {48}},\ \bibinfo {pages} {095105}
  (\bibinfo {year} {2024})},\ \Eprint {http://arxiv.org/abs/2401.04584}
  {arXiv:2401.04584 [gr-qc]} \BibitemShut {NoStop}%
\bibitem [{\citenamefont {Qi}\ \emph {et~al.}(2023)\citenamefont {Qi},
  \citenamefont {Meng}, \citenamefont {Wang},\ and\ \citenamefont
  {Kuang}}]{QiQi:2023nex}%
  \BibitemOpen
  \bibfield  {author} {\bibinfo {author} {\bibfnamefont {Qi}~\bibnamefont
  {Qi}}, \bibinfo {author} {\bibfnamefont {Yuan}\ \bibnamefont {Meng}},
  \bibinfo {author} {\bibfnamefont {Xi-Jing}\ \bibnamefont {Wang}}, \ and\
  \bibinfo {author} {\bibfnamefont {Xiao-Mei}\ \bibnamefont {Kuang}},\
  }\bibfield  {title} {\enquote {\bibinfo {title} {{Gravitational lensing
  effects of black hole with conformally coupled scalar hair}},}\ }\href
  {\doibase 10.1140/epjc/s10052-023-12233-z} {\bibfield  {journal} {\bibinfo
  {journal} {Eur. Phys. J. C}\ }\textbf {\bibinfo {volume} {83}},\ \bibinfo
  {pages} {1043} (\bibinfo {year} {2023})}\BibitemShut {NoStop}%
\bibitem [{\citenamefont {Boonserm}\ \emph {et~al.}(2018)\citenamefont
  {Boonserm}, \citenamefont {Ngampitipan},\ and\ \citenamefont
  {Wongjun}}]{Boonserm:2017qcq}%
  \BibitemOpen
  \bibfield  {author} {\bibinfo {author} {\bibfnamefont {Petarpa}\ \bibnamefont
  {Boonserm}}, \bibinfo {author} {\bibfnamefont {Tritos}\ \bibnamefont
  {Ngampitipan}}, \ and\ \bibinfo {author} {\bibfnamefont {Pitayuth}\
  \bibnamefont {Wongjun}},\ }\bibfield  {title} {\enquote {\bibinfo {title}
  {{Greybody factor for black holes in dRGT massive gravity}},}\ }\href
  {\doibase 10.1140/epjc/s10052-018-5975-x} {\bibfield  {journal} {\bibinfo
  {journal} {Eur. Phys. J. C}\ }\textbf {\bibinfo {volume} {78}},\ \bibinfo
  {pages} {492} (\bibinfo {year} {2018})},\ \Eprint
  {http://arxiv.org/abs/1705.03278} {arXiv:1705.03278 [gr-qc]} \BibitemShut
  {NoStop}%
\bibitem [{\citenamefont {Boonserm}\ \emph {et~al.}(2019)\citenamefont
  {Boonserm}, \citenamefont {Ngampitipan},\ and\ \citenamefont
  {Wongjun}}]{Boonserm:2019mon}%
  \BibitemOpen
  \bibfield  {author} {\bibinfo {author} {\bibfnamefont {P.}~\bibnamefont
  {Boonserm}}, \bibinfo {author} {\bibfnamefont {T.}~\bibnamefont
  {Ngampitipan}}, \ and\ \bibinfo {author} {\bibfnamefont {P.}~\bibnamefont
  {Wongjun}},\ }\bibfield  {title} {\enquote {\bibinfo {title} {{Greybody
  factor for black string in dRGT massive gravity}},}\ }\href {\doibase
  10.1140/epjc/s10052-019-6827-z} {\bibfield  {journal} {\bibinfo  {journal}
  {Eur. Phys. J. C}\ }\textbf {\bibinfo {volume} {79}},\ \bibinfo {pages} {330}
  (\bibinfo {year} {2019})},\ \Eprint {http://arxiv.org/abs/1902.05215}
  {arXiv:1902.05215 [gr-qc]} \BibitemShut {NoStop}%
\bibitem [{\citenamefont {Liu}\ \emph {et~al.}(2023)\citenamefont {Liu},
  \citenamefont {Yang}, \citenamefont {\"Ovg\"un}, \citenamefont {Long},\ and\
  \citenamefont {Xu}}]{Liu:2022ygf}%
  \BibitemOpen
  \bibfield  {author} {\bibinfo {author} {\bibfnamefont {Dong}\ \bibnamefont
  {Liu}}, \bibinfo {author} {\bibfnamefont {Yi}~\bibnamefont {Yang}}, \bibinfo
  {author} {\bibfnamefont {Ali}\ \bibnamefont {\"Ovg\"un}}, \bibinfo {author}
  {\bibfnamefont {Zheng-Wen}\ \bibnamefont {Long}}, \ and\ \bibinfo {author}
  {\bibfnamefont {Zhaoyi}\ \bibnamefont {Xu}},\ }\bibfield  {title} {\enquote
  {\bibinfo {title} {{Gravitational ringing and superradiant instabilities of
  the Kerr-like black holes in a dark matter halo}},}\ }\href {\doibase
  10.1140/epjc/s10052-023-11739-w} {\bibfield  {journal} {\bibinfo  {journal}
  {Eur. Phys. J. C}\ }\textbf {\bibinfo {volume} {83}},\ \bibinfo {pages} {565}
  (\bibinfo {year} {2023})},\ \Eprint {http://arxiv.org/abs/2204.11563}
  {arXiv:2204.11563 [gr-qc]} \BibitemShut {NoStop}%
\bibitem [{\citenamefont {Yang}\ \emph {et~al.}(2023)\citenamefont {Yang},
  \citenamefont {Liu}, \citenamefont {\"Ovg\"un}, \citenamefont {Long},\ and\
  \citenamefont {Xu}}]{Yang:2022ifo}%
  \BibitemOpen
  \bibfield  {author} {\bibinfo {author} {\bibfnamefont {Yi}~\bibnamefont
  {Yang}}, \bibinfo {author} {\bibfnamefont {Dong}\ \bibnamefont {Liu}},
  \bibinfo {author} {\bibfnamefont {Ali}\ \bibnamefont {\"Ovg\"un}}, \bibinfo
  {author} {\bibfnamefont {Zheng-Wen}\ \bibnamefont {Long}}, \ and\ \bibinfo
  {author} {\bibfnamefont {Zhaoyi}\ \bibnamefont {Xu}},\ }\bibfield  {title}
  {\enquote {\bibinfo {title} {{Probing hairy black holes caused by
  gravitational decoupling using quasinormal modes and greybody bounds}},}\
  }\href {\doibase 10.1103/PhysRevD.107.064042} {\bibfield  {journal} {\bibinfo
   {journal} {Phys. Rev. D}\ }\textbf {\bibinfo {volume} {107}},\ \bibinfo
  {pages} {064042} (\bibinfo {year} {2023})},\ \Eprint
  {http://arxiv.org/abs/2203.11551} {arXiv:2203.11551 [gr-qc]} \BibitemShut
  {NoStop}%
\bibitem [{\citenamefont {Gray}\ and\ \citenamefont
  {Visser}(2018)}]{Gray:2015xig}%
  \BibitemOpen
  \bibfield  {author} {\bibinfo {author} {\bibfnamefont {Finnian}\ \bibnamefont
  {Gray}}\ and\ \bibinfo {author} {\bibfnamefont {Matt}\ \bibnamefont
  {Visser}},\ }\bibfield  {title} {\enquote {\bibinfo {title} {{Greybody
  Factors for Schwarzschild Black Holes: Path-Ordered Exponentials and Product
  Integrals}},}\ }\href {\doibase 10.3390/universe4090093} {\bibfield
  {journal} {\bibinfo  {journal} {Universe}\ }\textbf {\bibinfo {volume} {4}},\
  \bibinfo {pages} {93} (\bibinfo {year} {2018})},\ \Eprint
  {http://arxiv.org/abs/1512.05018} {arXiv:1512.05018 [gr-qc]} \BibitemShut
  {NoStop}%
\bibitem [{\citenamefont {Boonserm}\ \emph
  {et~al.}(2014{\natexlab{a}})\citenamefont {Boonserm}, \citenamefont
  {Chatrabhuti}, \citenamefont {Ngampitipan},\ and\ \citenamefont
  {Visser}}]{Boonserm:2014fja}%
  \BibitemOpen
  \bibfield  {author} {\bibinfo {author} {\bibfnamefont {Petarpa}\ \bibnamefont
  {Boonserm}}, \bibinfo {author} {\bibfnamefont {Auttakit}\ \bibnamefont
  {Chatrabhuti}}, \bibinfo {author} {\bibfnamefont {Tritos}\ \bibnamefont
  {Ngampitipan}}, \ and\ \bibinfo {author} {\bibfnamefont {Matt}\ \bibnamefont
  {Visser}},\ }\bibfield  {title} {\enquote {\bibinfo {title} {{Greybody
  factors for Myers-Perry black holes}},}\ }\href {\doibase 10.1063/1.4901127}
  {\bibfield  {journal} {\bibinfo  {journal} {J. Math. Phys.}\ }\textbf
  {\bibinfo {volume} {55}},\ \bibinfo {pages} {112502} (\bibinfo {year}
  {2014}{\natexlab{a}})},\ \Eprint {http://arxiv.org/abs/1405.5678}
  {arXiv:1405.5678 [gr-qc]} \BibitemShut {NoStop}%
\bibitem [{\citenamefont {Boonserm}\ \emph
  {et~al.}(2014{\natexlab{b}})\citenamefont {Boonserm}, \citenamefont
  {Ngampitipan},\ and\ \citenamefont {Visser}}]{Boonserm:2014rma}%
  \BibitemOpen
  \bibfield  {author} {\bibinfo {author} {\bibfnamefont {Petarpa}\ \bibnamefont
  {Boonserm}}, \bibinfo {author} {\bibfnamefont {Tritos}\ \bibnamefont
  {Ngampitipan}}, \ and\ \bibinfo {author} {\bibfnamefont {Matt}\ \bibnamefont
  {Visser}},\ }\bibfield  {title} {\enquote {\bibinfo {title} {{Bounding the
  greybody factors for scalar field excitations on the Kerr-Newman
  spacetime}},}\ }\href {\doibase 10.1007/JHEP03(2014)113} {\bibfield
  {journal} {\bibinfo  {journal} {JHEP}\ }\textbf {\bibinfo {volume} {03}},\
  \bibinfo {pages} {113} (\bibinfo {year} {2014}{\natexlab{b}})},\ \Eprint
  {http://arxiv.org/abs/1401.0568} {arXiv:1401.0568 [gr-qc]} \BibitemShut
  {NoStop}%
\bibitem [{\citenamefont {Boonserm}\ \emph {et~al.}(2013)\citenamefont
  {Boonserm}, \citenamefont {Ngampitipan},\ and\ \citenamefont
  {Visser}}]{Boonserm:2013dua}%
  \BibitemOpen
  \bibfield  {author} {\bibinfo {author} {\bibfnamefont {Petarpa}\ \bibnamefont
  {Boonserm}}, \bibinfo {author} {\bibfnamefont {Tritos}\ \bibnamefont
  {Ngampitipan}}, \ and\ \bibinfo {author} {\bibfnamefont {Matt}\ \bibnamefont
  {Visser}},\ }\bibfield  {title} {\enquote {\bibinfo {title} {{Regge-Wheeler
  equation, linear stability, and greybody factors for dirty black holes}},}\
  }\href {\doibase 10.1103/PhysRevD.88.041502} {\bibfield  {journal} {\bibinfo
  {journal} {Phys. Rev. D}\ }\textbf {\bibinfo {volume} {88}},\ \bibinfo
  {pages} {041502} (\bibinfo {year} {2013})},\ \Eprint
  {http://arxiv.org/abs/1305.1416} {arXiv:1305.1416 [gr-qc]} \BibitemShut
  {NoStop}%
\bibitem [{\citenamefont {Ngampitipan}\ and\ \citenamefont
  {Boonserm}(2013)}]{Ngampitipan:2012dq}%
  \BibitemOpen
  \bibfield  {author} {\bibinfo {author} {\bibfnamefont {Tritos}\ \bibnamefont
  {Ngampitipan}}\ and\ \bibinfo {author} {\bibfnamefont {Petarpa}\ \bibnamefont
  {Boonserm}},\ }\bibfield  {title} {\enquote {\bibinfo {title} {{Bounding the
  Greybody Factors for Non-rotating Black Holes}},}\ }\href {\doibase
  10.1142/S0218271813500582} {\bibfield  {journal} {\bibinfo  {journal} {Int.
  J. Mod. Phys. D}\ }\textbf {\bibinfo {volume} {22}},\ \bibinfo {pages}
  {1350058} (\bibinfo {year} {2013})},\ \Eprint
  {http://arxiv.org/abs/1211.4070} {arXiv:1211.4070 [math-ph]} \BibitemShut
  {NoStop}%
\bibitem [{\citenamefont {Boonserm}(2009)}]{Boonserm:2009zba}%
  \BibitemOpen
  \bibfield  {author} {\bibinfo {author} {\bibfnamefont {Petarpa}\ \bibnamefont
  {Boonserm}},\ }\href@noop {} {\enquote {\bibinfo {title} {{Rigorous bounds on
  Transmission, Reflection, and Bogoliubov coefficients}},}\ } (\bibinfo {year}
  {2009}),\ \Eprint {http://arxiv.org/abs/0907.0045} {arXiv:0907.0045
  [math-ph]} \BibitemShut {NoStop}%
\bibitem [{\citenamefont {Boonserm}\ and\ \citenamefont
  {Visser}(2010)}]{Boonserm:2009mi}%
  \BibitemOpen
  \bibfield  {author} {\bibinfo {author} {\bibfnamefont {Petarpa}\ \bibnamefont
  {Boonserm}}\ and\ \bibinfo {author} {\bibfnamefont {Matt}\ \bibnamefont
  {Visser}},\ }\bibfield  {title} {\enquote {\bibinfo {title} {{Analytic bounds
  on transmission probabilities}},}\ }\href {\doibase
  10.1016/j.aop.2010.02.005} {\bibfield  {journal} {\bibinfo  {journal} {Annals
  Phys.}\ }\textbf {\bibinfo {volume} {325}},\ \bibinfo {pages} {1328--1339}
  (\bibinfo {year} {2010})},\ \Eprint {http://arxiv.org/abs/0901.0944}
  {arXiv:0901.0944 [math-ph]} \BibitemShut {NoStop}%
\bibitem [{\citenamefont {Kumaran}\ and\ \citenamefont
  {\"Ovg\"un}(2023)}]{Kumaran:2023brp}%
  \BibitemOpen
  \bibfield  {author} {\bibinfo {author} {\bibfnamefont {Yashmitha}\
  \bibnamefont {Kumaran}}\ and\ \bibinfo {author} {\bibfnamefont {Ali}\
  \bibnamefont {\"Ovg\"un}},\ }\bibfield  {title} {\enquote {\bibinfo {title}
  {{Shadow and deflection angle of asymptotic, magnetically-charged,
  non-singular black hole}},}\ }\href {\doibase
  10.1140/epjc/s10052-023-12001-z} {\bibfield  {journal} {\bibinfo  {journal}
  {Eur. Phys. J. C}\ }\textbf {\bibinfo {volume} {83}},\ \bibinfo {pages} {812}
  (\bibinfo {year} {2023})},\ \Eprint {http://arxiv.org/abs/2306.04705}
  {arXiv:2306.04705 [gr-qc]} \BibitemShut {NoStop}%
\bibitem [{\citenamefont {Baruah}\ \emph {et~al.}(2023)\citenamefont {Baruah},
  \citenamefont {\"Ovg\"un},\ and\ \citenamefont
  {Deshamukhya}}]{Baruah:2023rhd}%
  \BibitemOpen
  \bibfield  {author} {\bibinfo {author} {\bibfnamefont {Anshuman}\
  \bibnamefont {Baruah}}, \bibinfo {author} {\bibfnamefont {Ali}\ \bibnamefont
  {\"Ovg\"un}}, \ and\ \bibinfo {author} {\bibfnamefont {Atri}\ \bibnamefont
  {Deshamukhya}},\ }\bibfield  {title} {\enquote {\bibinfo {title}
  {{Quasinormal modes and bounding greybody factors of GUP-corrected black
  holes in Kalb\textendash{}Ramond gravity}},}\ }\href {\doibase
  10.1016/j.aop.2023.169393} {\bibfield  {journal} {\bibinfo  {journal} {Annals
  Phys.}\ }\textbf {\bibinfo {volume} {455}},\ \bibinfo {pages} {169393}
  (\bibinfo {year} {2023})},\ \Eprint {http://arxiv.org/abs/2304.07761}
  {arXiv:2304.07761 [gr-qc]} \BibitemShut {NoStop}%
\bibitem [{\citenamefont {Javed}\ \emph {et~al.}(2022)\citenamefont {Javed},
  \citenamefont {Atique},\ and\ \citenamefont {\"Ovg\"un}}]{Javed:2022bdi}%
  \BibitemOpen
  \bibfield  {author} {\bibinfo {author} {\bibfnamefont {Wajiha}\ \bibnamefont
  {Javed}}, \bibinfo {author} {\bibfnamefont {Mehak}\ \bibnamefont {Atique}}, \
  and\ \bibinfo {author} {\bibfnamefont {Ali}\ \bibnamefont {\"Ovg\"un}},\
  }\bibfield  {title} {\enquote {\bibinfo {title} {{Probing effective loop
  quantum gravity on weak gravitational lensing, Hawking radiation and bounding
  greybody factor by black holes}},}\ }\href {\doibase
  10.1007/s10714-022-03028-w} {\bibfield  {journal} {\bibinfo  {journal} {Gen.
  Rel. Grav.}\ }\textbf {\bibinfo {volume} {54}},\ \bibinfo {pages} {135}
  (\bibinfo {year} {2022})},\ \Eprint {http://arxiv.org/abs/2210.17277}
  {arXiv:2210.17277 [gr-qc]} \BibitemShut {NoStop}%
\bibitem [{\citenamefont {Al-Badawi}(2023)}]{Al-Badawi:2023emj}%
  \BibitemOpen
  \bibfield  {author} {\bibinfo {author} {\bibfnamefont {Ahmad}\ \bibnamefont
  {Al-Badawi}},\ }\bibfield  {title} {\enquote {\bibinfo {title} {{Greybody
  factors emitted by a regular black hole in a non-minimally coupled
  Einstein\textendash{}Yang\textendash{}Mills theory}},}\ }\href {\doibase
  10.1140/epjc/s10052-023-11550-7} {\bibfield  {journal} {\bibinfo  {journal}
  {Eur. Phys. J. C}\ }\textbf {\bibinfo {volume} {83}},\ \bibinfo {pages} {380}
  (\bibinfo {year} {2023})},\ \Eprint {http://arxiv.org/abs/2305.07436}
  {arXiv:2305.07436 [gr-qc]} \BibitemShut {NoStop}%
\bibitem [{\citenamefont {Cardoso}\ and\ \citenamefont
  {Lemos}(2001{\natexlab{a}})}]{Cardoso:2001bb}%
  \BibitemOpen
  \bibfield  {author} {\bibinfo {author} {\bibfnamefont {Vitor}\ \bibnamefont
  {Cardoso}}\ and\ \bibinfo {author} {\bibfnamefont {Jose P.~S.}\ \bibnamefont
  {Lemos}},\ }\bibfield  {title} {\enquote {\bibinfo {title} {{Quasinormal
  modes of Schwarzschild anti-de Sitter black holes: Electromagnetic and
  gravitational perturbations}},}\ }\href {\doibase 10.1103/PhysRevD.64.084017}
  {\bibfield  {journal} {\bibinfo  {journal} {Phys. Rev. D}\ }\textbf {\bibinfo
  {volume} {64}},\ \bibinfo {pages} {084017} (\bibinfo {year}
  {2001}{\natexlab{a}})},\ \Eprint {http://arxiv.org/abs/gr-qc/0105103}
  {arXiv:gr-qc/0105103} \BibitemShut {NoStop}%
\bibitem [{\citenamefont {Konoplya}\ and\ \citenamefont
  {Zhidenko}(2007)}]{Konoplya:2006rv}%
  \BibitemOpen
  \bibfield  {author} {\bibinfo {author} {\bibfnamefont {R.~A.}\ \bibnamefont
  {Konoplya}}\ and\ \bibinfo {author} {\bibfnamefont {A.}~\bibnamefont
  {Zhidenko}},\ }\bibfield  {title} {\enquote {\bibinfo {title} {{Perturbations
  and quasi-normal modes of black holes in Einstein-Aether theory}},}\ }\href
  {\doibase 10.1016/j.physletb.2006.11.036} {\bibfield  {journal} {\bibinfo
  {journal} {Phys. Lett. B}\ }\textbf {\bibinfo {volume} {644}},\ \bibinfo
  {pages} {186--191} (\bibinfo {year} {2007})},\ \Eprint
  {http://arxiv.org/abs/gr-qc/0605082} {arXiv:gr-qc/0605082} \BibitemShut
  {NoStop}%
\bibitem [{\citenamefont {Konoplya}\ and\ \citenamefont
  {Zhidenko}(2022)}]{Konoplya:2022tvv}%
  \BibitemOpen
  \bibfield  {author} {\bibinfo {author} {\bibfnamefont {R.~A.}\ \bibnamefont
  {Konoplya}}\ and\ \bibinfo {author} {\bibfnamefont {A.}~\bibnamefont
  {Zhidenko}},\ }\bibfield  {title} {\enquote {\bibinfo {title} {{Quasinormal
  ringing of general spherically symmetric parametrized black holes}},}\ }\href
  {\doibase 10.1103/PhysRevD.105.104032} {\bibfield  {journal} {\bibinfo
  {journal} {Phys. Rev. D}\ }\textbf {\bibinfo {volume} {105}},\ \bibinfo
  {pages} {104032} (\bibinfo {year} {2022})},\ \Eprint
  {http://arxiv.org/abs/2201.12897} {arXiv:2201.12897 [gr-qc]} \BibitemShut
  {NoStop}%
\bibitem [{\citenamefont {Konoplya}\ \emph
  {et~al.}(2019{\natexlab{a}})\citenamefont {Konoplya}, \citenamefont
  {Zinhailo},\ and\ \citenamefont {Stuchl\'\i{}k}}]{Konoplya:2019hml}%
  \BibitemOpen
  \bibfield  {author} {\bibinfo {author} {\bibfnamefont {R.~A.}\ \bibnamefont
  {Konoplya}}, \bibinfo {author} {\bibfnamefont {A.~F.}\ \bibnamefont
  {Zinhailo}}, \ and\ \bibinfo {author} {\bibfnamefont {Z.}~\bibnamefont
  {Stuchl\'\i{}k}},\ }\bibfield  {title} {\enquote {\bibinfo {title}
  {{Quasinormal modes, scattering, and Hawking radiation in the vicinity of an
  Einstein-dilaton-Gauss-Bonnet black hole}},}\ }\href {\doibase
  10.1103/PhysRevD.99.124042} {\bibfield  {journal} {\bibinfo  {journal} {Phys.
  Rev. D}\ }\textbf {\bibinfo {volume} {99}},\ \bibinfo {pages} {124042}
  (\bibinfo {year} {2019}{\natexlab{a}})},\ \Eprint
  {http://arxiv.org/abs/1903.03483} {arXiv:1903.03483 [gr-qc]} \BibitemShut
  {NoStop}%
\bibitem [{\citenamefont {Cai}\ \emph {et~al.}(2016)\citenamefont {Cai},
  \citenamefont {Cheng}, \citenamefont {Liu}, \citenamefont {Wang},\ and\
  \citenamefont {Zhang}}]{Cai:2015fia}%
  \BibitemOpen
  \bibfield  {author} {\bibinfo {author} {\bibfnamefont {Yi-Fu}\ \bibnamefont
  {Cai}}, \bibinfo {author} {\bibfnamefont {Gong}\ \bibnamefont {Cheng}},
  \bibinfo {author} {\bibfnamefont {Junyu}\ \bibnamefont {Liu}}, \bibinfo
  {author} {\bibfnamefont {Min}\ \bibnamefont {Wang}}, \ and\ \bibinfo {author}
  {\bibfnamefont {Hezi}\ \bibnamefont {Zhang}},\ }\bibfield  {title} {\enquote
  {\bibinfo {title} {{Features and stability analysis of non-Schwarzschild
  black hole in quadratic gravity}},}\ }\href {\doibase
  10.1007/JHEP01(2016)108} {\bibfield  {journal} {\bibinfo  {journal} {JHEP}\
  }\textbf {\bibinfo {volume} {01}},\ \bibinfo {pages} {108} (\bibinfo {year}
  {2016})},\ \Eprint {http://arxiv.org/abs/1508.04776} {arXiv:1508.04776
  [hep-th]} \BibitemShut {NoStop}%
\bibitem [{\citenamefont {Visser}(1999)}]{Visser:1998ke}%
  \BibitemOpen
  \bibfield  {author} {\bibinfo {author} {\bibfnamefont {Matt}\ \bibnamefont
  {Visser}},\ }\bibfield  {title} {\enquote {\bibinfo {title} {{Some general
  bounds for 1-D scattering}},}\ }\href {\doibase 10.1103/PhysRevA.59.427}
  {\bibfield  {journal} {\bibinfo  {journal} {Phys. Rev. A}\ }\textbf {\bibinfo
  {volume} {59}},\ \bibinfo {pages} {427--438} (\bibinfo {year} {1999})},\
  \Eprint {http://arxiv.org/abs/quant-ph/9901030} {arXiv:quant-ph/9901030}
  \BibitemShut {NoStop}%
\bibitem [{\citenamefont {Boonserm}\ and\ \citenamefont
  {Visser}(2008)}]{Boonserm:2008zg}%
  \BibitemOpen
  \bibfield  {author} {\bibinfo {author} {\bibfnamefont {Petarpa}\ \bibnamefont
  {Boonserm}}\ and\ \bibinfo {author} {\bibfnamefont {Matt}\ \bibnamefont
  {Visser}},\ }\bibfield  {title} {\enquote {\bibinfo {title} {{Bounding the
  greybody factors for Schwarzschild black holes}},}\ }\href {\doibase
  10.1103/PhysRevD.78.101502} {\bibfield  {journal} {\bibinfo  {journal} {Phys.
  Rev. D}\ }\textbf {\bibinfo {volume} {78}},\ \bibinfo {pages} {101502}
  (\bibinfo {year} {2008})},\ \Eprint {http://arxiv.org/abs/0806.2209}
  {arXiv:0806.2209 [gr-qc]} \BibitemShut {NoStop}%
\bibitem [{\citenamefont {Cho}\ \emph {et~al.}(2007)\citenamefont {Cho},
  \citenamefont {Cornell}, \citenamefont {Doukas},\ and\ \citenamefont
  {Naylor}}]{Cho:2007zi}%
  \BibitemOpen
  \bibfield  {author} {\bibinfo {author} {\bibfnamefont {H.~T.}\ \bibnamefont
  {Cho}}, \bibinfo {author} {\bibfnamefont {Alan~S.}\ \bibnamefont {Cornell}},
  \bibinfo {author} {\bibfnamefont {Jason}\ \bibnamefont {Doukas}}, \ and\
  \bibinfo {author} {\bibfnamefont {Wade}\ \bibnamefont {Naylor}},\ }\bibfield
  {title} {\enquote {\bibinfo {title} {{Split fermion quasi-normal modes}},}\
  }\href {\doibase 10.1103/PhysRevD.75.104005} {\bibfield  {journal} {\bibinfo
  {journal} {Phys. Rev. D}\ }\textbf {\bibinfo {volume} {75}},\ \bibinfo
  {pages} {104005} (\bibinfo {year} {2007})},\ \Eprint
  {http://arxiv.org/abs/hep-th/0701193} {arXiv:hep-th/0701193} \BibitemShut
  {NoStop}%
\bibitem [{\citenamefont {Das}\ \emph {et~al.}(1997)\citenamefont {Das},
  \citenamefont {Gibbons},\ and\ \citenamefont {Mathur}}]{Das:1996we}%
  \BibitemOpen
  \bibfield  {author} {\bibinfo {author} {\bibfnamefont {Sumit~R.}\
  \bibnamefont {Das}}, \bibinfo {author} {\bibfnamefont {Gary~W.}\ \bibnamefont
  {Gibbons}}, \ and\ \bibinfo {author} {\bibfnamefont {Samir~D.}\ \bibnamefont
  {Mathur}},\ }\bibfield  {title} {\enquote {\bibinfo {title} {{Universality of
  low-energy absorption cross-sections for black holes}},}\ }\href {\doibase
  10.1103/PhysRevLett.78.417} {\bibfield  {journal} {\bibinfo  {journal} {Phys.
  Rev. Lett.}\ }\textbf {\bibinfo {volume} {78}},\ \bibinfo {pages} {417--419}
  (\bibinfo {year} {1997})},\ \Eprint {http://arxiv.org/abs/hep-th/9609052}
  {arXiv:hep-th/9609052} \BibitemShut {NoStop}%
\bibitem [{\citenamefont {Gibbons}\ and\ \citenamefont
  {Steif}(1993)}]{Gibbons:1993hg}%
  \BibitemOpen
  \bibfield  {author} {\bibinfo {author} {\bibfnamefont {G.~W.}\ \bibnamefont
  {Gibbons}}\ and\ \bibinfo {author} {\bibfnamefont {Alan~R.}\ \bibnamefont
  {Steif}},\ }\bibfield  {title} {\enquote {\bibinfo {title} {{Anomalous
  fermion production in gravitational collapse}},}\ }\href {\doibase
  10.1016/0370-2693(93)91315-E} {\bibfield  {journal} {\bibinfo  {journal}
  {Phys. Lett. B}\ }\textbf {\bibinfo {volume} {314}},\ \bibinfo {pages}
  {13--20} (\bibinfo {year} {1993})},\ \Eprint
  {http://arxiv.org/abs/gr-qc/9305018} {arXiv:gr-qc/9305018} \BibitemShut
  {NoStop}%
\bibitem [{\citenamefont {Camporesi}\ and\ \citenamefont
  {Higuchi}(1996)}]{Camporesi:1995fb}%
  \BibitemOpen
  \bibfield  {author} {\bibinfo {author} {\bibfnamefont {Roberto}\ \bibnamefont
  {Camporesi}}\ and\ \bibinfo {author} {\bibfnamefont {Atsushi}\ \bibnamefont
  {Higuchi}},\ }\bibfield  {title} {\enquote {\bibinfo {title} {{On the Eigen
  functions of the Dirac operator on spheres and real hyperbolic spaces}},}\
  }\href {\doibase 10.1016/0393-0440(95)00042-9} {\bibfield  {journal}
  {\bibinfo  {journal} {J. Geom. Phys.}\ }\textbf {\bibinfo {volume} {20}},\
  \bibinfo {pages} {1--18} (\bibinfo {year} {1996})},\ \Eprint
  {http://arxiv.org/abs/gr-qc/9505009} {arXiv:gr-qc/9505009} \BibitemShut
  {NoStop}%
\bibitem [{\citenamefont {Birmingham}(2001)}]{Birmingham:2001hc}%
  \BibitemOpen
  \bibfield  {author} {\bibinfo {author} {\bibfnamefont {Danny}\ \bibnamefont
  {Birmingham}},\ }\bibfield  {title} {\enquote {\bibinfo {title} {{Choptuik
  scaling and quasinormal modes in the AdS / CFT correspondence}},}\ }\href
  {\doibase 10.1103/PhysRevD.64.064024} {\bibfield  {journal} {\bibinfo
  {journal} {Phys. Rev. D}\ }\textbf {\bibinfo {volume} {64}},\ \bibinfo
  {pages} {064024} (\bibinfo {year} {2001})},\ \Eprint
  {http://arxiv.org/abs/hep-th/0101194} {arXiv:hep-th/0101194} \BibitemShut
  {NoStop}%
\bibitem [{\citenamefont {Fernando}(2004)}]{Fernando:2003ai}%
  \BibitemOpen
  \bibfield  {author} {\bibinfo {author} {\bibfnamefont {Sharmanthie}\
  \bibnamefont {Fernando}},\ }\bibfield  {title} {\enquote {\bibinfo {title}
  {{Quasinormal modes of charged dilaton black holes in (2+1)-dimensions}},}\
  }\href {\doibase 10.1023/B:GERG.0000006694.68399.c9} {\bibfield  {journal}
  {\bibinfo  {journal} {Gen. Rel. Grav.}\ }\textbf {\bibinfo {volume} {36}},\
  \bibinfo {pages} {71--82} (\bibinfo {year} {2004})},\ \Eprint
  {http://arxiv.org/abs/hep-th/0306214} {arXiv:hep-th/0306214} \BibitemShut
  {NoStop}%
\bibitem [{\citenamefont {Fernando}(2008)}]{Fernando:2008hb}%
  \BibitemOpen
  \bibfield  {author} {\bibinfo {author} {\bibfnamefont {Sharmanthie}\
  \bibnamefont {Fernando}},\ }\bibfield  {title} {\enquote {\bibinfo {title}
  {{Quasinormal modes of charged scalars around dilaton black holes in 2+1
  dimensions: Exact frequencies}},}\ }\href {\doibase
  10.1103/PhysRevD.77.124005} {\bibfield  {journal} {\bibinfo  {journal} {Phys.
  Rev. D}\ }\textbf {\bibinfo {volume} {77}},\ \bibinfo {pages} {124005}
  (\bibinfo {year} {2008})},\ \Eprint {http://arxiv.org/abs/0802.3321}
  {arXiv:0802.3321 [hep-th]} \BibitemShut {NoStop}%
\bibitem [{\citenamefont {Gonzalez}\ \emph {et~al.}(2010)\citenamefont
  {Gonzalez}, \citenamefont {Papantonopoulos},\ and\ \citenamefont
  {Saavedra}}]{Gonzalez:2010vv}%
  \BibitemOpen
  \bibfield  {author} {\bibinfo {author} {\bibfnamefont {Pablo}\ \bibnamefont
  {Gonzalez}}, \bibinfo {author} {\bibfnamefont {Eleftherios}\ \bibnamefont
  {Papantonopoulos}}, \ and\ \bibinfo {author} {\bibfnamefont {Joel}\
  \bibnamefont {Saavedra}},\ }\bibfield  {title} {\enquote {\bibinfo {title}
  {{Chern-Simons black holes: scalar perturbations, mass and area spectrum and
  greybody factors}},}\ }\href {\doibase 10.1007/JHEP08(2010)050} {\bibfield
  {journal} {\bibinfo  {journal} {JHEP}\ }\textbf {\bibinfo {volume} {08}},\
  \bibinfo {pages} {050} (\bibinfo {year} {2010})},\ \Eprint
  {http://arxiv.org/abs/1003.1381} {arXiv:1003.1381 [hep-th]} \BibitemShut
  {NoStop}%
\bibitem [{\citenamefont {Destounis}\ \emph {et~al.}(2018)\citenamefont
  {Destounis}, \citenamefont {Panotopoulos},\ and\ \citenamefont
  {Rinc\'on}}]{Destounis:2018utr}%
  \BibitemOpen
  \bibfield  {author} {\bibinfo {author} {\bibfnamefont {Kyriakos}\
  \bibnamefont {Destounis}}, \bibinfo {author} {\bibfnamefont {Grigoris}\
  \bibnamefont {Panotopoulos}}, \ and\ \bibinfo {author} {\bibfnamefont
  {\'Angel}\ \bibnamefont {Rinc\'on}},\ }\bibfield  {title} {\enquote {\bibinfo
  {title} {{Stability under scalar perturbations and quasinormal modes of 4D
  Einstein\textendash{}Born\textendash{}Infeld dilaton spacetime: exact
  spectrum}},}\ }\href {\doibase 10.1140/epjc/s10052-018-5576-8} {\bibfield
  {journal} {\bibinfo  {journal} {Eur. Phys. J. C}\ }\textbf {\bibinfo {volume}
  {78}},\ \bibinfo {pages} {139} (\bibinfo {year} {2018})},\ \Eprint
  {http://arxiv.org/abs/1801.08955} {arXiv:1801.08955 [gr-qc]} \BibitemShut
  {NoStop}%
\bibitem [{\citenamefont {Ovg\"un}\ and\ \citenamefont
  {Jusufi}(2018)}]{Ovgun:2018gwt}%
  \BibitemOpen
  \bibfield  {author} {\bibinfo {author} {\bibfnamefont {Al\"\i{}}\
  \bibnamefont {Ovg\"un}}\ and\ \bibinfo {author} {\bibfnamefont {Kimet}\
  \bibnamefont {Jusufi}},\ }\bibfield  {title} {\enquote {\bibinfo {title}
  {{Quasinormal Modes and Greybody Factors of $f(R)$ gravity minimally coupled
  to a cloud of strings in $2+1$ Dimensions}},}\ }\href {\doibase
  10.1016/j.aop.2018.05.013} {\bibfield  {journal} {\bibinfo  {journal} {Annals
  Phys.}\ }\textbf {\bibinfo {volume} {395}},\ \bibinfo {pages} {138--151}
  (\bibinfo {year} {2018})},\ \Eprint {http://arxiv.org/abs/1801.02555}
  {arXiv:1801.02555 [gr-qc]} \BibitemShut {NoStop}%
\bibitem [{\citenamefont {Rinc\'on}\ and\ \citenamefont
  {Panotopoulos}(2018{\natexlab{b}})}]{Rincon:2018ktz}%
  \BibitemOpen
  \bibfield  {author} {\bibinfo {author} {\bibfnamefont {\'Angel}\ \bibnamefont
  {Rinc\'on}}\ and\ \bibinfo {author} {\bibfnamefont {Grigoris}\ \bibnamefont
  {Panotopoulos}},\ }\bibfield  {title} {\enquote {\bibinfo {title} {{Greybody
  factors and quasinormal modes for a nonminimally coupled scalar field in a
  cloud of strings in (2+1)-dimensional background}},}\ }\href {\doibase
  10.1140/epjc/s10052-018-6352-5} {\bibfield  {journal} {\bibinfo  {journal}
  {Eur. Phys. J. C}\ }\textbf {\bibinfo {volume} {78}},\ \bibinfo {pages} {858}
  (\bibinfo {year} {2018}{\natexlab{b}})},\ \Eprint
  {http://arxiv.org/abs/1810.08822} {arXiv:1810.08822 [gr-qc]} \BibitemShut
  {NoStop}%
\bibitem [{\citenamefont {Poschl}\ and\ \citenamefont
  {Teller}(1933)}]{Poschl:1933zz}%
  \BibitemOpen
  \bibfield  {author} {\bibinfo {author} {\bibfnamefont {G.}~\bibnamefont
  {Poschl}}\ and\ \bibinfo {author} {\bibfnamefont {E.}~\bibnamefont
  {Teller}},\ }\bibfield  {title} {\enquote {\bibinfo {title} {{Bemerkungen zur
  Quantenmechanik des anharmonischen Oszillators}},}\ }\href {\doibase
  10.1007/BF01331132} {\bibfield  {journal} {\bibinfo  {journal} {Z. Phys.}\
  }\textbf {\bibinfo {volume} {83}},\ \bibinfo {pages} {143--151} (\bibinfo
  {year} {1933})}\BibitemShut {NoStop}%
\bibitem [{\citenamefont {Ferrari}\ and\ \citenamefont
  {Mashhoon}(1984)}]{Ferrari:1984zz}%
  \BibitemOpen
  \bibfield  {author} {\bibinfo {author} {\bibfnamefont {Valeria}\ \bibnamefont
  {Ferrari}}\ and\ \bibinfo {author} {\bibfnamefont {Bahram}\ \bibnamefont
  {Mashhoon}},\ }\bibfield  {title} {\enquote {\bibinfo {title} {{New approach
  to the quasinormal modes of a black hole}},}\ }\href {\doibase
  10.1103/PhysRevD.30.295} {\bibfield  {journal} {\bibinfo  {journal} {Phys.
  Rev. D}\ }\textbf {\bibinfo {volume} {30}},\ \bibinfo {pages} {295--304}
  (\bibinfo {year} {1984})}\BibitemShut {NoStop}%
\bibitem [{\citenamefont {Cardoso}\ and\ \citenamefont
  {Lemos}(2001{\natexlab{b}})}]{Cardoso:2001hn}%
  \BibitemOpen
  \bibfield  {author} {\bibinfo {author} {\bibfnamefont {Vitor}\ \bibnamefont
  {Cardoso}}\ and\ \bibinfo {author} {\bibfnamefont {Jose P.~S.}\ \bibnamefont
  {Lemos}},\ }\bibfield  {title} {\enquote {\bibinfo {title} {{Scalar,
  electromagnetic and Weyl perturbations of BTZ black holes: Quasinormal
  modes}},}\ }\href {\doibase 10.1103/PhysRevD.63.124015} {\bibfield  {journal}
  {\bibinfo  {journal} {Phys. Rev. D}\ }\textbf {\bibinfo {volume} {63}},\
  \bibinfo {pages} {124015} (\bibinfo {year} {2001}{\natexlab{b}})},\ \Eprint
  {http://arxiv.org/abs/gr-qc/0101052} {arXiv:gr-qc/0101052} \BibitemShut
  {NoStop}%
\bibitem [{\citenamefont {Cardoso}\ and\ \citenamefont
  {Lemos}(2003)}]{Cardoso:2003sw}%
  \BibitemOpen
  \bibfield  {author} {\bibinfo {author} {\bibfnamefont {Vitor}\ \bibnamefont
  {Cardoso}}\ and\ \bibinfo {author} {\bibfnamefont {Jose P.~S.}\ \bibnamefont
  {Lemos}},\ }\bibfield  {title} {\enquote {\bibinfo {title} {{Quasinormal
  modes of the near extremal Schwarzschild-de Sitter black hole}},}\ }\href
  {\doibase 10.1103/PhysRevD.67.084020} {\bibfield  {journal} {\bibinfo
  {journal} {Phys. Rev. D}\ }\textbf {\bibinfo {volume} {67}},\ \bibinfo
  {pages} {084020} (\bibinfo {year} {2003})},\ \Eprint
  {http://arxiv.org/abs/gr-qc/0301078} {arXiv:gr-qc/0301078} \BibitemShut
  {NoStop}%
\bibitem [{\citenamefont {Molina}(2003)}]{Molina:2003ff}%
  \BibitemOpen
  \bibfield  {author} {\bibinfo {author} {\bibfnamefont {C.}~\bibnamefont
  {Molina}},\ }\bibfield  {title} {\enquote {\bibinfo {title} {{Quasinormal
  modes of d-dimensional spherical black holes with near extreme cosmological
  constant}},}\ }\href {\doibase 10.1103/PhysRevD.68.064007} {\bibfield
  {journal} {\bibinfo  {journal} {Phys. Rev. D}\ }\textbf {\bibinfo {volume}
  {68}},\ \bibinfo {pages} {064007} (\bibinfo {year} {2003})},\ \Eprint
  {http://arxiv.org/abs/gr-qc/0304053} {arXiv:gr-qc/0304053} \BibitemShut
  {NoStop}%
\bibitem [{\citenamefont {Panotopoulos}(2018)}]{Panotopoulos:2018hua}%
  \BibitemOpen
  \bibfield  {author} {\bibinfo {author} {\bibfnamefont {Grigoris}\
  \bibnamefont {Panotopoulos}},\ }\bibfield  {title} {\enquote {\bibinfo
  {title} {{Electromagnetic quasinormal modes of the nearly-extremal
  higher-dimensional Schwarzschild\textendash{}de Sitter black hole}},}\ }\href
  {\doibase 10.1142/S0217732318501304} {\bibfield  {journal} {\bibinfo
  {journal} {Mod. Phys. Lett. A}\ }\textbf {\bibinfo {volume} {33}},\ \bibinfo
  {pages} {1850130} (\bibinfo {year} {2018})},\ \Eprint
  {http://arxiv.org/abs/1807.03278} {arXiv:1807.03278 [gr-qc]} \BibitemShut
  {NoStop}%
\bibitem [{\citenamefont {Leaver}(1985)}]{Leaver:1985ax}%
  \BibitemOpen
  \bibfield  {author} {\bibinfo {author} {\bibfnamefont {E.~W.}\ \bibnamefont
  {Leaver}},\ }\bibfield  {title} {\enquote {\bibinfo {title} {{An Analytic
  representation for the quasi normal modes of Kerr black holes}},}\ }\href
  {\doibase 10.1098/rspa.1985.0119} {\bibfield  {journal} {\bibinfo  {journal}
  {Proc. Roy. Soc. Lond. A}\ }\textbf {\bibinfo {volume} {402}},\ \bibinfo
  {pages} {285--298} (\bibinfo {year} {1985})}\BibitemShut {NoStop}%
\bibitem [{\citenamefont {Nollert}(1993)}]{Nollert:1993zz}%
  \BibitemOpen
  \bibfield  {author} {\bibinfo {author} {\bibfnamefont {Hans-Peter}\
  \bibnamefont {Nollert}},\ }\bibfield  {title} {\enquote {\bibinfo {title}
  {{Quasinormal modes of Schwarzschild black holes: The determination of
  quasinormal frequencies with very large imaginary parts}},}\ }\href {\doibase
  10.1103/PhysRevD.47.5253} {\bibfield  {journal} {\bibinfo  {journal} {Phys.
  Rev. D}\ }\textbf {\bibinfo {volume} {47}},\ \bibinfo {pages} {5253--5258}
  (\bibinfo {year} {1993})}\BibitemShut {NoStop}%
\bibitem [{\citenamefont {Daghigh}\ \emph {et~al.}(2023)\citenamefont
  {Daghigh}, \citenamefont {Green},\ and\ \citenamefont
  {Morey}}]{Daghigh:2022uws}%
  \BibitemOpen
  \bibfield  {author} {\bibinfo {author} {\bibfnamefont {Ramin~G.}\
  \bibnamefont {Daghigh}}, \bibinfo {author} {\bibfnamefont {Michael~D.}\
  \bibnamefont {Green}}, \ and\ \bibinfo {author} {\bibfnamefont {Jodin~C.}\
  \bibnamefont {Morey}},\ }\bibfield  {title} {\enquote {\bibinfo {title}
  {{Calculating quasinormal modes of Schwarzschild anti-de Sitter black holes
  using the continued fraction method}},}\ }\href {\doibase
  10.1103/PhysRevD.107.024023} {\bibfield  {journal} {\bibinfo  {journal}
  {Phys. Rev. D}\ }\textbf {\bibinfo {volume} {107}},\ \bibinfo {pages}
  {024023} (\bibinfo {year} {2023})},\ \Eprint
  {http://arxiv.org/abs/2209.09324} {arXiv:2209.09324 [gr-qc]} \BibitemShut
  {NoStop}%
\bibitem [{\citenamefont {Destounis}\ \emph {et~al.}(2020)\citenamefont
  {Destounis}, \citenamefont {Fontana},\ and\ \citenamefont
  {Mena}}]{Destounis:2020pjk}%
  \BibitemOpen
  \bibfield  {author} {\bibinfo {author} {\bibfnamefont {Kyriakos}\
  \bibnamefont {Destounis}}, \bibinfo {author} {\bibfnamefont {Rodrigo D.~B.}\
  \bibnamefont {Fontana}}, \ and\ \bibinfo {author} {\bibfnamefont {Filipe~C.}\
  \bibnamefont {Mena}},\ }\bibfield  {title} {\enquote {\bibinfo {title}
  {{Accelerating black holes: quasinormal modes and late-time tails}},}\ }\href
  {\doibase 10.1103/PhysRevD.102.044005} {\bibfield  {journal} {\bibinfo
  {journal} {Phys. Rev. D}\ }\textbf {\bibinfo {volume} {102}},\ \bibinfo
  {pages} {044005} (\bibinfo {year} {2020})},\ \Eprint
  {http://arxiv.org/abs/2005.03028} {arXiv:2005.03028 [gr-qc]} \BibitemShut
  {NoStop}%
\bibitem [{\citenamefont {Fontana}\ and\ \citenamefont
  {Mena}(2022)}]{Fontana:2022whx}%
  \BibitemOpen
  \bibfield  {author} {\bibinfo {author} {\bibfnamefont {Rodrigo D.~B.}\
  \bibnamefont {Fontana}}\ and\ \bibinfo {author} {\bibfnamefont {Filipe~C.}\
  \bibnamefont {Mena}},\ }\bibfield  {title} {\enquote {\bibinfo {title}
  {{Quasinormal modes and stability of accelerating Reissner-Norsdtr\"om AdS
  black holes}},}\ }\href {\doibase 10.1007/JHEP10(2022)047} {\bibfield
  {journal} {\bibinfo  {journal} {JHEP}\ }\textbf {\bibinfo {volume} {10}},\
  \bibinfo {pages} {047} (\bibinfo {year} {2022})},\ \Eprint
  {http://arxiv.org/abs/2203.13933} {arXiv:2203.13933 [gr-qc]} \BibitemShut
  {NoStop}%
\bibitem [{\citenamefont {Hatsuda}\ and\ \citenamefont
  {Kimura}(2021)}]{Hatsuda:2021gtn}%
  \BibitemOpen
  \bibfield  {author} {\bibinfo {author} {\bibfnamefont {Yasuyuki}\
  \bibnamefont {Hatsuda}}\ and\ \bibinfo {author} {\bibfnamefont {Masashi}\
  \bibnamefont {Kimura}},\ }\bibfield  {title} {\enquote {\bibinfo {title}
  {{Spectral Problems for Quasinormal Modes of Black Holes}},}\ }\href
  {\doibase 10.3390/universe7120476} {\bibfield  {journal} {\bibinfo  {journal}
  {Universe}\ }\textbf {\bibinfo {volume} {7}},\ \bibinfo {pages} {476}
  (\bibinfo {year} {2021})},\ \Eprint {http://arxiv.org/abs/2111.15197}
  {arXiv:2111.15197 [gr-qc]} \BibitemShut {NoStop}%
\bibitem [{\citenamefont {Cho}\ \emph {et~al.}(2012)\citenamefont {Cho},
  \citenamefont {Cornell}, \citenamefont {Doukas}, \citenamefont {Huang},\ and\
  \citenamefont {Naylor}}]{Cho:2011sf}%
  \BibitemOpen
  \bibfield  {author} {\bibinfo {author} {\bibfnamefont {H.~T.}\ \bibnamefont
  {Cho}}, \bibinfo {author} {\bibfnamefont {A.~S.}\ \bibnamefont {Cornell}},
  \bibinfo {author} {\bibfnamefont {Jason}\ \bibnamefont {Doukas}}, \bibinfo
  {author} {\bibfnamefont {T.~R.}\ \bibnamefont {Huang}}, \ and\ \bibinfo
  {author} {\bibfnamefont {Wade}\ \bibnamefont {Naylor}},\ }\bibfield  {title}
  {\enquote {\bibinfo {title} {{A New Approach to Black Hole Quasinormal Modes:
  A Review of the Asymptotic Iteration Method}},}\ }\href {\doibase
  10.1155/2012/281705} {\bibfield  {journal} {\bibinfo  {journal} {Adv. Math.
  Phys.}\ }\textbf {\bibinfo {volume} {2012}},\ \bibinfo {pages} {281705}
  (\bibinfo {year} {2012})},\ \Eprint {http://arxiv.org/abs/1111.5024}
  {arXiv:1111.5024 [gr-qc]} \BibitemShut {NoStop}%
\bibitem [{\citenamefont {{Ciftci}}\ \emph {et~al.}(2003)\citenamefont
  {{Ciftci}}, \citenamefont {{Hall}},\ and\ \citenamefont
  {{Saad}}}]{2003JPhA...3611807C}%
  \BibitemOpen
  \bibfield  {author} {\bibinfo {author} {\bibfnamefont {Hakan}\ \bibnamefont
  {{Ciftci}}}, \bibinfo {author} {\bibfnamefont {Richard~L.}\ \bibnamefont
  {{Hall}}}, \ and\ \bibinfo {author} {\bibfnamefont {Nasser}\ \bibnamefont
  {{Saad}}},\ }\bibfield  {title} {\enquote {\bibinfo {title} {{Asymptotic
  iteration method for eigenvalue problems}},}\ }\href {\doibase
  10.1088/0305-4470/36/47/008} {\bibfield  {journal} {\bibinfo  {journal}
  {Journal of Physics A Mathematical General}\ }\textbf {\bibinfo {volume}
  {36}},\ \bibinfo {pages} {11807--11816} (\bibinfo {year} {2003})},\ \Eprint
  {http://arxiv.org/abs/math-ph/0309066} {arXiv:math-ph/0309066 [math-ph]}
  \BibitemShut {NoStop}%
\bibitem [{\citenamefont {Ciftci}\ \emph {et~al.}(2005)\citenamefont {Ciftci},
  \citenamefont {Hall},\ and\ \citenamefont {Saad}}]{Ciftci:2005xn}%
  \BibitemOpen
  \bibfield  {author} {\bibinfo {author} {\bibfnamefont {H.}~\bibnamefont
  {Ciftci}}, \bibinfo {author} {\bibfnamefont {R.~L.}\ \bibnamefont {Hall}}, \
  and\ \bibinfo {author} {\bibfnamefont {N.}~\bibnamefont {Saad}},\ }\bibfield
  {title} {\enquote {\bibinfo {title} {{Perturbation theory in a framework of
  iteration methods}},}\ }\href {\doibase 10.1016/j.physleta.2005.04.030}
  {\bibfield  {journal} {\bibinfo  {journal} {Phys. Lett. A}\ }\textbf
  {\bibinfo {volume} {340}},\ \bibinfo {pages} {388--396} (\bibinfo {year}
  {2005})},\ \Eprint {http://arxiv.org/abs/math-ph/0504056}
  {arXiv:math-ph/0504056} \BibitemShut {NoStop}%
\bibitem [{\citenamefont {Konoplya}\ and\ \citenamefont
  {Zhidenko}(2011)}]{Konoplya:2011qq}%
  \BibitemOpen
  \bibfield  {author} {\bibinfo {author} {\bibfnamefont {R.~A.}\ \bibnamefont
  {Konoplya}}\ and\ \bibinfo {author} {\bibfnamefont {A.}~\bibnamefont
  {Zhidenko}},\ }\bibfield  {title} {\enquote {\bibinfo {title} {{Quasinormal
  modes of black holes: From astrophysics to string theory}},}\ }\href
  {\doibase 10.1103/RevModPhys.83.793} {\bibfield  {journal} {\bibinfo
  {journal} {Rev. Mod. Phys.}\ }\textbf {\bibinfo {volume} {83}},\ \bibinfo
  {pages} {793--836} (\bibinfo {year} {2011})},\ \Eprint
  {http://arxiv.org/abs/1102.4014} {arXiv:1102.4014 [gr-qc]} \BibitemShut
  {NoStop}%
\bibitem [{\citenamefont {Schutz}\ and\ \citenamefont
  {Will}(1985)}]{Schutz:1985km}%
  \BibitemOpen
  \bibfield  {author} {\bibinfo {author} {\bibfnamefont {Bernard~F.}\
  \bibnamefont {Schutz}}\ and\ \bibinfo {author} {\bibfnamefont {Clifford~M.}\
  \bibnamefont {Will}},\ }\bibfield  {title} {\enquote {\bibinfo {title}
  {{BLACK HOLE NORMAL MODES: A SEMIANALYTIC APPROACH}},}\ }\href {\doibase
  10.1086/184453} {\bibfield  {journal} {\bibinfo  {journal} {Astrophys. J.
  Lett.}\ }\textbf {\bibinfo {volume} {291}},\ \bibinfo {pages} {L33--L36}
  (\bibinfo {year} {1985})}\BibitemShut {NoStop}%
\bibitem [{\citenamefont {Iyer}\ and\ \citenamefont
  {Will}(1987)}]{Iyer:1986np}%
  \BibitemOpen
  \bibfield  {author} {\bibinfo {author} {\bibfnamefont {Sai}\ \bibnamefont
  {Iyer}}\ and\ \bibinfo {author} {\bibfnamefont {Clifford~M.}\ \bibnamefont
  {Will}},\ }\bibfield  {title} {\enquote {\bibinfo {title} {{Black Hole Normal
  Modes: A {WKB} Approach. 1. Foundations and Application of a Higher Order
  {WKB} Analysis of Potential Barrier Scattering}},}\ }\href {\doibase
  10.1103/PhysRevD.35.3621} {\bibfield  {journal} {\bibinfo  {journal} {Phys.
  Rev. D}\ }\textbf {\bibinfo {volume} {35}},\ \bibinfo {pages} {3621}
  (\bibinfo {year} {1987})}\BibitemShut {NoStop}%
\bibitem [{\citenamefont {Iyer}(1987)}]{Iyer:1986nq}%
  \BibitemOpen
  \bibfield  {author} {\bibinfo {author} {\bibfnamefont {Sai}\ \bibnamefont
  {Iyer}},\ }\bibfield  {title} {\enquote {\bibinfo {title} {{BLACK HOLE NORMAL
  MODES: A WKB APPROACH. 2. SCHWARZSCHILD BLACK HOLES}},}\ }\href {\doibase
  10.1103/PhysRevD.35.3632} {\bibfield  {journal} {\bibinfo  {journal} {Phys.
  Rev. D}\ }\textbf {\bibinfo {volume} {35}},\ \bibinfo {pages} {3632}
  (\bibinfo {year} {1987})}\BibitemShut {NoStop}%
\bibitem [{\citenamefont {Kokkotas}\ and\ \citenamefont
  {Schutz}(1988)}]{Kokkotas:1988fm}%
  \BibitemOpen
  \bibfield  {author} {\bibinfo {author} {\bibfnamefont {K.~D.}\ \bibnamefont
  {Kokkotas}}\ and\ \bibinfo {author} {\bibfnamefont {Bernard~F.}\ \bibnamefont
  {Schutz}},\ }\bibfield  {title} {\enquote {\bibinfo {title} {{Black Hole
  Normal Modes: A {WKB} Approach. 3. The {Reissner-Nordstrom} Black Hole}},}\
  }\href {\doibase 10.1103/PhysRevD.37.3378} {\bibfield  {journal} {\bibinfo
  {journal} {Phys. Rev. D}\ }\textbf {\bibinfo {volume} {37}},\ \bibinfo
  {pages} {3378--3387} (\bibinfo {year} {1988})}\BibitemShut {NoStop}%
\bibitem [{\citenamefont {Seidel}\ and\ \citenamefont
  {Iyer}(1990)}]{Seidel:1989bp}%
  \BibitemOpen
  \bibfield  {author} {\bibinfo {author} {\bibfnamefont {Edward}\ \bibnamefont
  {Seidel}}\ and\ \bibinfo {author} {\bibfnamefont {Sai}\ \bibnamefont
  {Iyer}},\ }\bibfield  {title} {\enquote {\bibinfo {title} {{BLACK HOLE NORMAL
  MODES: A WKB APPROACH. 4. KERR BLACK HOLES}},}\ }\href {\doibase
  10.1103/PhysRevD.41.374} {\bibfield  {journal} {\bibinfo  {journal} {Phys.
  Rev. D}\ }\textbf {\bibinfo {volume} {41}},\ \bibinfo {pages} {374--382}
  (\bibinfo {year} {1990})}\BibitemShut {NoStop}%
\bibitem [{\citenamefont {Konoplya}(2003)}]{Konoplya:2003ii}%
  \BibitemOpen
  \bibfield  {author} {\bibinfo {author} {\bibfnamefont {R.~A.}\ \bibnamefont
  {Konoplya}},\ }\bibfield  {title} {\enquote {\bibinfo {title} {{Quasinormal
  behavior of the d-dimensional Schwarzschild black hole and higher order WKB
  approach}},}\ }\href {\doibase 10.1103/PhysRevD.68.024018} {\bibfield
  {journal} {\bibinfo  {journal} {Phys. Rev. D}\ }\textbf {\bibinfo {volume}
  {68}},\ \bibinfo {pages} {024018} (\bibinfo {year} {2003})},\ \Eprint
  {http://arxiv.org/abs/gr-qc/0303052} {arXiv:gr-qc/0303052} \BibitemShut
  {NoStop}%
\bibitem [{\citenamefont {Matyjasek}\ and\ \citenamefont
  {Opala}(2017)}]{Matyjasek:2017psv}%
  \BibitemOpen
  \bibfield  {author} {\bibinfo {author} {\bibfnamefont {Jerzy}\ \bibnamefont
  {Matyjasek}}\ and\ \bibinfo {author} {\bibfnamefont {Micha\l{}}\ \bibnamefont
  {Opala}},\ }\bibfield  {title} {\enquote {\bibinfo {title} {{Quasinormal
  modes of black holes. The improved semianalytic approach}},}\ }\href
  {\doibase 10.1103/PhysRevD.96.024011} {\bibfield  {journal} {\bibinfo
  {journal} {Phys. Rev. D}\ }\textbf {\bibinfo {volume} {96}},\ \bibinfo
  {pages} {024011} (\bibinfo {year} {2017})},\ \Eprint
  {http://arxiv.org/abs/1704.00361} {arXiv:1704.00361 [gr-qc]} \BibitemShut
  {NoStop}%
\bibitem [{wol()}]{wolfram}%
  \BibitemOpen
  \href {https://www.wolframalpha.com} {\enquote {\bibinfo {title} {Wolfram
  alpha},}\ }\BibitemShut {NoStop}%
\bibitem [{\citenamefont {Konoplya}\ \emph
  {et~al.}(2019{\natexlab{b}})\citenamefont {Konoplya}, \citenamefont
  {Zhidenko},\ and\ \citenamefont {Zinhailo}}]{Konoplya:2019hlu}%
  \BibitemOpen
  \bibfield  {author} {\bibinfo {author} {\bibfnamefont {R.~A.}\ \bibnamefont
  {Konoplya}}, \bibinfo {author} {\bibfnamefont {A.}~\bibnamefont {Zhidenko}},
  \ and\ \bibinfo {author} {\bibfnamefont {A.~F.}\ \bibnamefont {Zinhailo}},\
  }\bibfield  {title} {\enquote {\bibinfo {title} {{Higher order WKB formula
  for quasinormal modes and grey-body factors: recipes for quick and accurate
  calculations}},}\ }\href {\doibase 10.1088/1361-6382/ab2e25} {\bibfield
  {journal} {\bibinfo  {journal} {Class. Quant. Grav.}\ }\textbf {\bibinfo
  {volume} {36}},\ \bibinfo {pages} {155002} (\bibinfo {year}
  {2019}{\natexlab{b}})},\ \Eprint {http://arxiv.org/abs/1904.10333}
  {arXiv:1904.10333 [gr-qc]} \BibitemShut {NoStop}%
\bibitem [{\citenamefont {Hatsuda}(2020)}]{Hatsuda:2019eoj}%
  \BibitemOpen
  \bibfield  {author} {\bibinfo {author} {\bibfnamefont {Yasuyuki}\
  \bibnamefont {Hatsuda}},\ }\bibfield  {title} {\enquote {\bibinfo {title}
  {{Quasinormal modes of black holes and Borel summation}},}\ }\href {\doibase
  10.1103/PhysRevD.101.024008} {\bibfield  {journal} {\bibinfo  {journal}
  {Phys. Rev. D}\ }\textbf {\bibinfo {volume} {101}},\ \bibinfo {pages}
  {024008} (\bibinfo {year} {2020})},\ \Eprint
  {http://arxiv.org/abs/1906.07232} {arXiv:1906.07232 [gr-qc]} \BibitemShut
  {NoStop}%
\bibitem [{\citenamefont {Bolokhov}(2024)}]{Bolokhov:2023bwm}%
  \BibitemOpen
  \bibfield  {author} {\bibinfo {author} {\bibfnamefont {S.~V.}\ \bibnamefont
  {Bolokhov}},\ }\bibfield  {title} {\enquote {\bibinfo {title} {{Long-lived
  quasinormal modes and overtones\textquoteright{} behavior of
  holonomy-corrected black holes}},}\ }\href {\doibase
  10.1103/PhysRevD.110.024010} {\bibfield  {journal} {\bibinfo  {journal}
  {Phys. Rev. D}\ }\textbf {\bibinfo {volume} {110}},\ \bibinfo {pages}
  {024010} (\bibinfo {year} {2024})},\ \Eprint
  {http://arxiv.org/abs/2311.05503} {arXiv:2311.05503 [gr-qc]} \BibitemShut
  {NoStop}%
\bibitem [{\citenamefont {Gingrich}(2024)}]{Gingrich:2024tuf}%
  \BibitemOpen
  \bibfield  {author} {\bibinfo {author} {\bibfnamefont {Douglas~M.}\
  \bibnamefont {Gingrich}},\ }\href@noop {} {\enquote {\bibinfo {title}
  {{Quasinormal modes of a nonsingular spherically symmetric black hole
  effective model with holonomy corrections}},}\ } (\bibinfo {year} {2024}),\
  \Eprint {http://arxiv.org/abs/2404.04447} {arXiv:2404.04447 [gr-qc]}
  \BibitemShut {NoStop}%
\bibitem [{\citenamefont {Yang}\ \emph {et~al.}(2024)\citenamefont {Yang},
  \citenamefont {Guo}, \citenamefont {Tan}, \citenamefont {Zhao},\ and\
  \citenamefont {Liu}}]{Yang:2024ofe}%
  \BibitemOpen
  \bibfield  {author} {\bibinfo {author} {\bibfnamefont {Sen}\ \bibnamefont
  {Yang}}, \bibinfo {author} {\bibfnamefont {Wen-Di}\ \bibnamefont {Guo}},
  \bibinfo {author} {\bibfnamefont {Qin}\ \bibnamefont {Tan}}, \bibinfo
  {author} {\bibfnamefont {Li}~\bibnamefont {Zhao}}, \ and\ \bibinfo {author}
  {\bibfnamefont {Yu-Xiao}\ \bibnamefont {Liu}},\ }\href@noop {} {\enquote
  {\bibinfo {title} {{Parameterized quasinormal frequencies and Hawking
  radiation for axial gravitational perturbations of a holonomy-corrected black
  hole}},}\ } (\bibinfo {year} {2024}),\ \Eprint
  {http://arxiv.org/abs/2406.15711} {arXiv:2406.15711 [gr-qc]} \BibitemShut
  {NoStop}%
\end{thebibliography}%
\end{document}